\documentclass[fleqn,usenatbib]{mnras}

\usepackage{newtxtext,newtxmath}

\usepackage[T1]{fontenc}

\DeclareRobustCommand{\VAN}[3]{#2}
\let\VANthebibliography\thebibliography
\def\thebibliography{\DeclareRobustCommand{\VAN}[3]{##3}\VANthebibliography}

\usepackage{graphicx}	

\newcommand{\hyperfit}{{\sc HyperFit}}	
\newcommand{\profuse}{{\sc ProFuse}}	
\newcommand{\profound}{{\sc ProFound}}	
\newcommand{\profit}{{\sc ProFit}}
\newcommand{\prospect}{{\sc ProSpect}}
\newcommand{\highlander}{{\sc Highlander}}

\newcommand{\R}{{\sc R}}
\newcommand{\sersic}{S\'{e}rsic}
\newcommand{\msol}{M$_{\odot}$}

\title[ProFuse: Physical Multi-Band Structural Decomposition]{ProFuse: Physical Multi-Band Structural Decomposition of Galaxies and the Mass--Size--Age Plane}

\author[A.~S.~G.~ Robotham et al.]{
A.~S.~G. Robotham,$^{1,2}$\thanks{E-mail: aaron.robotham@uwa.edu.au}
S. Bellstedt,$^{1}$
S.~P. Driver,$^{1}$
\\\\
$^{1}$ICRAR, M468, University of Western Australia, Crawley, WA 6009, Australia\\
$^{2}$ARC Centre of Excellence for Astrophysics in Three Dimensions (ASTRO3D)\\
}

\date{Accepted XXX. Received YYY; in original form ZZZ}

\pubyear{2021}

\begin{document}
\label{firstpage}
\pagerange{\pageref{firstpage}--\pageref{lastpage}}
\maketitle

\begin{abstract}
We present the new \profuse{} \R{} package, a simultaneous spectral (ultraviolet to far infrared) and spatial structural decomposition tool that produces physical models of galaxies and their components. This combines the functionality of the recently released \profound{} (for automatic source extraction), \profit{} (for extended source profiling) and \prospect{} (for stellar population modelling) software packages. The key novelty of \profuse{} is that it generates images using a self-consistent model for the star formation and metallicity history of the bulge and disk separately, and uses target images across a range of wavelengths to define the model likelihood and optimise our physical galaxy reconstruction. The first part of the paper explores the \profuse{} approach in detail, and compares results to published structural and stellar population properties. The latter part of the paper applies \profuse{} to 6,664 $z<0.06$ GAMA galaxies. Using re-processed $ugri$ZYJHKs imaging we extract structural and stellar population properties for bulges and disks in parallel. As well as producing true stellar mass based mass--size relationships, we further extend this correlation to explore the third dimensions of age and gas phase metallicity. The disks in particular demonstrate strong co-dependency between mass--size--age in a well defined plane, where at a given disk stellar mass younger disks tend to be larger. These findings are in broad agreement with work at higher redshift suggesting disks that formed earlier are physically smaller.
\end{abstract}




\section{Introduction}

Given the future multi-band broad wavelength deep large surveys on the horizon and the huge quantities of money and person power being invested \citep{Weinberg:2013ue}, it is a reasonable endeavour to develop software that can maximally utilise this information for scientific insight. Domains that have long made use of such data on large scales, but have never been formally combined, are extended source decomposition \citep[e.g.][]{Peng:2002vw, de-Souza:2004tt, Erwin:2015ug, Robotham:2017tl} and multi-band spectral energy distribution (SED) fitting \citep[e.g.][]{da-Cunha:2008vn, Noll:2009ty, Robotham:2020to}. \profuse, presented for the first time here, is our effort to combine these domains to produce physical models of galaxies.

In \citet{Robotham:2017tl} we described the first public version of the galaxy structural decomposition software package \profit. The published version of \profit{} was released with only single band fitting as an option, where multi-band fits could be achieved using cascading priors (fit the deepest band first, then use that solution as an initial condition and/or a prior for fitting adjacent bands). Since then, general purpose multi-band fitting has been possible where images could either be a single wavelength and the end goal was a single (potentially multi-component) profile to describe the data (with different world coordinate system and depth etc), or images could represent different wavelengths and smoothly varying model parameters could be fitted to describe them. While these approaches achieved reasonable results qualitatively, they have the obvious weakness of not being constrained by physics- it is quite possible to create erroneously rapid changes in e.g. bulge colours that could never be created by modern spectral templates \citep[e.g.][]{Bruzual:2003wd, Eldridge:2009tn} with a plausible star formation history.

Before developing an SED fitting tool that could be combined with \profit, the general purpose source extraction and analysis package \profound{} was created \citep{Robotham:2018tx}. This aided the generation of suitable segmentation maps and initial conditions for running \profit{} in a fully automated manner. \profound{} has since been used more widely as a general purpose photometry tool across a range of wavelengths, and serves as the main photometry tool for the legacy Galaxy And Mass Assembly \cite[GAMA;][]{Driver:2011uq, Liske:2015ub, Bellstedt:2020wl} survey, ongoing Deep Extragalactic VIsible Legacy Survey \citep[DEVILS;][]{Davies:2018td, Davies:2021vc}, and the future Wide-Area VISTA Extragalactic Survey \citep[WAVES;][]{Driver:2019wr}. In the case of GAMA, its functionality (compared to the published version) has been expanded to include the point spread function (PSF) fitting of unresolved mid-infrared (MIR) and far-infrared (FIR) images, allowing us to produce high quality SEDs over 20 bands of UV-FIR data \citep{Bellstedt:2020wl}. It has also been applied to radio data \citep{Hale:2019tc}, proving its versatility across a vast range of wavelengths.

The quality (depth and resolution) and wavelength breadth of the full imaging data available in the final data release of GAMA (Driver et al. in press) and ongoing DEVILS surveys lends itself ideally to SED fitting, and with this goal in mind (and the longer term aim of producing physical galaxy decomposition) the Bayesian inference software tool \prospect{} was created \citep{Robotham:2020to}. This allows highly flexible star and metallicity formation histories to be extracted from broad band SEDs, and has been used to invert the physical growth of galaxies for hundreds of thousands of GAMA and DEVILS sources to date \citep[][respectively]{Bellstedt:2020vq, Thorne:2021un}. The design and modularity of this code was written with later integration into \profit{} in mind. This aspect is important since a lack of coherent structure would have required a complete refactoring of one or both code bases, which would be a significant task.

In this work we make the first recognised effort in fully combining the extended source decomposition and SED domains in a rigorously self-consistent generative Bayesian framework, combining the inference power of the software tools \profit{} (galaxy decomposition) and \prospect{} (SED generation and fitting) to produce the new \R{} software package \profuse\footnote{https://github.com/asgr/ProFuse}. To further aid and automate its application, \profuse{} also utilises the source extraction and analysis software \profound{} to automatically define structural initial conditions and to identify and parameterise stars to produce per band PSFs on-the-fly.

Comparable but ultimately dis-similar efforts to extract component wise spectral properties exist in the literature. In terms of the broadband moderately-resolved data it is designed for, there are clear similarities with {\sc MegaMorph} \citep{Haussler:2013tz, Vulcani:2014up}, which allows for smooth functional forms to link properties across adjacent bands. \citet{Johnston:2017wg} extends this capability to use these outputs (which in effect produce the relative bulge and disk image contribution as a function of spatial position and wavelength) to then fit SEDs in coarsely binned integral field spectrograph (IFS) spaxels using {\sc pPXF} \citep{Cappellari:2004us}, which then allows spatial inference of the star formation and metallicity history and other properties that can be extracted from such SEDs but spatially separated into bulges and disks. This combination of decoupled structural decomposition followed by stellar properties being extracted from the resultant SEDs is common to other recent IFS efforts, with the major differences being whether the focus is on bulge-disk kinematic separation or stellar formation history separation \citep{Coccato:2011uh, Scott:2017ud, Tabor:2017wc, Mendez-Abreu:2019wj, Oh:2020ta, Costantin:2021ui}. 

Predating IFS work, conceptually similar efforts using long-slit based spectroscopic techniques have also been carried out \citep{MacArthur:2009tw, Johnston:2012wi, Johnston:2014uj}. Simple splits between bulge dominated and disk dominated regions of galaxies allowed some insight into their dis-similar star formation histories and gas and/or stellar metallicities. Going back further, investigations into the stellar populations of bulges and disks used coarser photometric data and comparisons to relatively simple evolutionary models \citep[e.g.][]{Mollenhoff:2004vz}. A common feature of all the above literature efforts is that there is a distinct separation between the stellar modelling and structural decomposition steps (although the order and precedence of these steps varies by application). The nearest to a `fully' self consistent spectral decomposition model is arguably \citet{Taranu:2017tq}, which creates a 3D equilibrium model and projects these particles to compute likelihoods. However, little spectral information is incorporated beyond the computed kinematics and optical $g$ and $r$ images.

\profuse{} differs from the above approaches in that it is fully generative and Bayesian in terms of how it solves the problem of reconstructing physical structures (but in this paper and initial software release, more specifically bulges and disks). All structural components fitted are allowed to have their own unique star formation histories and structural properties, and these are forward generated onto model images that are directly compared to observed data to compute the model constraining likelihoods. The huge advantage of this approach is that physically implausible spectra for each component (which in broadband data would mean extreme colours) are naturally penalised in the fitting process. It also means that structural properties (e.g.\ disk and bulge sizes) receive information from all bands available. Combining these steps together therefore maximises the data available for modelling, minimises the number of free parameters to fit, hugely speeds up the modelling process, and allows us to infer complex relationships between the parameters via well explored posterior samples (e.g.\ does disk size correlate with the stellar age of the disk? etc).

Importantly, \profuse{} also opens up more physical avenues to structural model selection, e.g.\ is a given galaxy a true bulge-disk system, or a single smooth profile? It has proven to be extremely difficult to select preferred models on purely flux based structural analysis \citep{Allen:2006vs, Lange:2016wx, Cook:2019wa}, and better automating this process will become significant for ever larger surveys. The flexibility of coupling resolved structure directly with the stars that form the components will pay dividends in the near future of deep, well-resolved, multi-band facilities \cite[e.g.\ James Webb, Vera Rubin, Euclid, Nancy-Grace Roman; see][respectively]{Gardner:2006ul, Ivezic:2019ub, Amendola:2013vz, Spergel:2015wr}.

\profuse{} is designed as a full {\it image through to inference tool}, where all the routine aspects of source identification, segmentation, sky subtraction and point spread function modelling are handled automatically (or if desired, provided in a more manual manner). To aid adoption and comprehension, long form reproducible examples are included with the software natively and online\footnote{https://rpubs.com/asgr}.

In this paper we discuss the core methodology and implementation of \profuse{} (Section \ref{sec:method}), explore a detailed application to a test galaxy (Section \ref{sec:example}), and apply it to the low redshift sample of 6,664 GAMA galaxies that have already been analysed with \prospect{} in \citet{Bellstedt:2020vq} (Section \ref{sec:GAMA}). We finally investigate the most interesting results relating to the stellar and structural properties of these bulges and disks (Section \ref{sec:bulgedisk}).

Where relevant we use the same concordance $\Lambda$CDM \citet{Planck-Collaboration:2016tp} cosmology as used in \citet{Bellstedt:2020vq}, i.e.\ H$_0$ = 67.8 (kms/s)/Mpc, $\Omega_M$ = 0.308 and $\Omega_\Lambda = 0.692$. We note for comparison purposes that all stellar properties discussed are equivalent to being stellar mass weighted (as opposed to light weighted).

\section{Methods}
\label{sec:method}

To achieve the fusion of the \profit{} and \prospect{} software packages some limitations on source flexibility had to be made (at least in this first implementation). The pragmatic solution was to limit the profile options available on the \profit{} side to either single \sersic{} or double \sersic{} profiles \citep{Sersic:1963uw, Graham:2005uu}. In our experience to date (roughly five years of usage) these options represent nearly all popular use cases for fitting extended extra-galactic sources- the additional profiles available rarely get used (the exception being the Moffat profile for PSF fitting, usually to fit unresolved sources). Beyond this profile limitation there is a large amount of flexibility in terms of how \profuse{} can be used for inference, with effectively all the SED fitting power of \prospect{} available to use for the both the bulge and disk components (if two components are desired). Naturally there is a ramp up in complexity of use as users apply more sophisticated models, but as much as is possible \profuse{} simplifies the user interface and abstracts away much of the complicated model building.

The over-arching conceptual flow of how \profuse{} works is as follows:

\begin{enumerate}
 \item The user provides a list of all target images, the redshift ($z$), and $x$ and $y$ image pixel location of the target source in each image and the desired cutout size (default $600  \times 600$ pixels). Optionally the sky and sky-RMS images can also be provided (if these are not provided, they will be determined internally). A filter name (over 300 exist within \prospect) or response table/function must be provided for each target image. PSFs can also be optionally provided to accompany each image. Currently, \profuse{} only allows a single image to be provided per band (where longer term the option to have multiple target images in each band will be added) but within that limitation the different input images may have different world coordinate system (scale, size, translation and rotation) allowing native images to be provided rather than a common projection scheme. This has the effect of minimising data volumes (since in practice you would need to project to the smallest pixel scale image) and reducing pixel covariance (unavoidable as you up-sample an image). Magnitude zero points (to bring images into the AB system) must be provided, and per image gain can also be optionally provided (in terms of the usual electrons per ADU).
 \item The user specifies the type of model. On the \profit{} side, local and global options are allowed for 1 (single \sersic), 1.5 (disk + PSF bulge), 2 (double \sersic{} with fixed or free bulge and/or disk), or 2.5 (triple \sersic{} with fixed or free bulge and/or thin/thick disk) profiles, e.g.\ you maybe want to fit all images with a common \sersic{} profile but allow the size ($R_e$) to vary per band (either smoothly or discretely). On the \prospect{} side all the usual options available in the normal SED fitting mode are available, so completely separate models can be specified for the bulge and disk. Table \ref{tab:ProFuseOptions} gives an overview of the highest level options available in \profuse. More flexibility (e.g.\ profile types other than those listed here, but available in \profit) is available via manual editing of the fitting objects, but only expert users should attempt this.
 \item The user runs the resulting fitting object through an inference engine of their choosing. For the tests outlined in this paper we make use of the dual phase \highlander{} package developed for use with \profit{} and \prospect{} \citep{Thorne:2021un} (discussed in more detail later).
\end{enumerate}

\begin{table*}
\begin{tabular}{l|l|ll|ll|ll}
                &        & \multicolumn{2}{l}{Bulge} & \multicolumn{2}{l}{Disk 1} & \multicolumn{2}{l}{Disk 2} \\
Model Type      & N Comp & Str.        & SED         & Str.         & SED         & Str.          & SED        \\
 \hline
Single          & 1      & \sersic        & Free        & .            & .           & .             & .          \\
Bulge-Disk      & 1.5    & PSF         & Free        & \sersic         & Free        & .             & .          \\
Bulge-Disk      & 2      & \sersic        & Free        & \sersic         & Free        & .             & .          \\
Bulge-Disk-Disk & 2.5    & \sersic        & Free        & \sersic         & Free        & Coupled       & Free      
\end{tabular}
\caption{Overview of high level fitting flexibility. For the structural component (`Str.') the main options are for a free \sersic{} profile, a PSF profile, or a `coupled' component where the second disk must follow the geometry (axial ratio and ellipticity) of the first disk. For the `SED' component, all the options available to \prospect{} can be accessed when labelled `Free'.}
\label{tab:ProFuseOptions}
\end{table*}

Steps (i) and (ii) above are in practice a single (but lengthy) call to a setup function in \profuse. Internally this creates the complex fitting objects for the user to later run inference on, and these objects can be saved to aid later re-analysis (e.g.\ running the same model with more sample chains etc). In principle these fitting objects can be constructed entirely manually (allowing more granular control), but this is a complex task and should in general be avoided. If finer control is needed a better route is to run the setup routine as normal, and then make changes to the fitting object returned.

Inside the setup function different routines will run depending on the data provided. In a minimal run of the user only providing the images and basic fitting instructions the following happens internally:

\begin{figure}
 \includegraphics[width=\columnwidth]{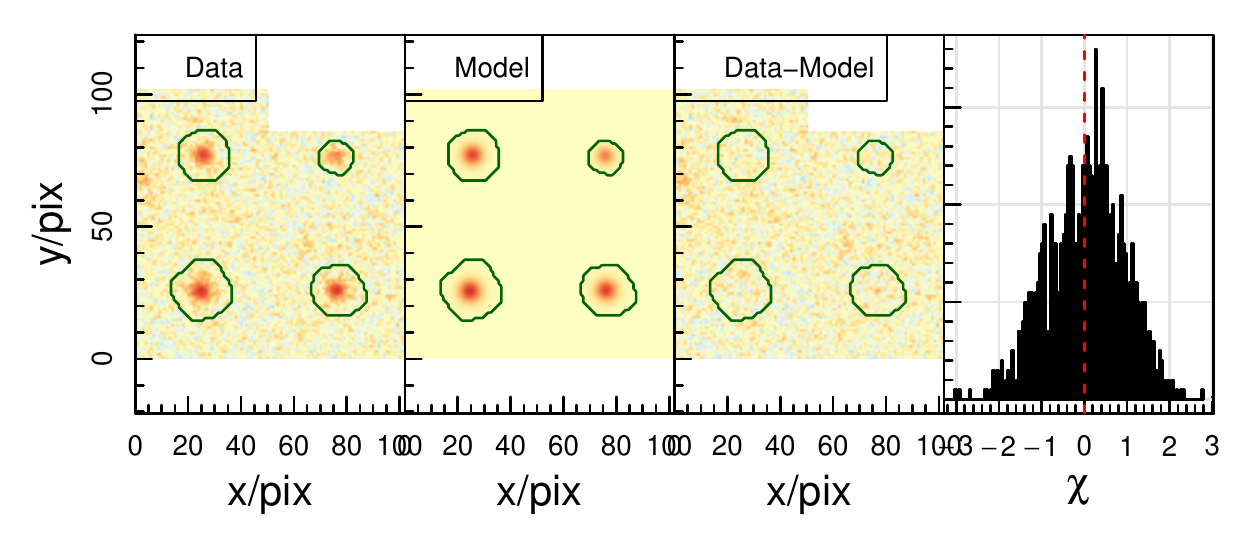}
 \caption{Example of the automatic PSF extraction and modelling that \profuse{} carries out if no PSF is provided. The $r$-band example here found the four most similar bright PSFs (left panel) and fitted them simultaneously with a global Moffat profile (second panel). The 2D residuals (third panel) and collapsed histogram (fourth panel) demonstrates the PSF model is highly accurate in this case. Any analytic model available to \profit{} can be used to model the PSFs, including ellipticity and boxiness parameters if deemed necessary. The results here are for example galaxy TGS426Z278 used in Section \ref{sec:example}.}
 \label{fig:allstar}
\end{figure}

\begin{enumerate}
 \item The images are cutout (by default $600 \times 600$ in size) and centred on the target extended source. This means very large mosaics of data can be provided as the input (pointers to on-disk images can be used), with the minimal in-memory cutouts made internally.
 \item \profound{} is run on each image using settings that can be provided by the user (otherwise it will use default parameters, which in practice work well on a range of image data). This creates segmentation maps per band, along with sky and sky-RMS images. \profound{} also generates reasonable initial starting points for the model requested, but in practice the \highlander{} optimisation engine suggested is designed to extract good solutions even from very poor initial conditions.
 \item If no PSF is provided in a particular band (users need not supply a PSF for every band) then candidate stars are identified in size magnitude space via identification of the stellar locus\footnote{This is done automatically by detecting the magnitude range where compact sources dominate the number counts, and sigma clipping the sizes to find the robust subset of PSF-like objects. Users can optionally specify the magnitude range to use if the automatic detection does not work well, but in our experience is this is very rarely necessary.} (note this only works well when there are a good number of stars (roughly a dozen) on an image, if this is not possible then a PSF should be provided manually). By default the four most PSF-like objects are extracted and simultaneously fitted with a Moffat profile using \profit{} and the \highlander{} inference engine. Figure \ref{fig:allstar} shows extracted and modelled stars using the example data included with the software, where we can see that the model is extremely good, with almost pure noise residuals in the third and fourth panels.
 \item An inverse variance weighted stack is made of all images.
 \item The stacked image is re-analysed with \profound{} to make the final segmentation map and to determine reasonable starting conditions for \profit{} for either 1, 1.5, 2 or 2.5 component profiles. This also determines reasonable fitting limits for all relevant free parameters automatically (something that had to be done by users manually in the published version of \profit). All images are now cut down to the minimal rectangle that contains all identified object pixels.
 \item If the gain is not provided for all images, then using Poisson statistics a reasonable approximate gain is determined for each required image using the sky pixels. The image, sky image, gain and segmentation map is combined with the sky-RMS image to make a final sigma image to be used for each image inside \profit.
 \item The data created or provided in the above step is passed into the usual \profit{} data setup function for each image, creating a meta list of the different data structures required per image.
 \item Internal methods are determined to reshape a set of target parameters and generate the correct models for the \profit{} and \prospect{} components of the multi-band model.
 \end{enumerate}

The output of the above is a fairly large fitting object (in terms of components), since each target image will contain cutout versions of the image, sky, sky-RMS, PSF, sigma and segmentation maps. In general the actual in-memory object size is quite small since everything is cut down to the minimum viable image size, e.g. the example galaxy provided in the package vignette has 9 bands of imaging and only takes on 40 MB in-memory. Given the ability to automatically extract the PSF by extracting stars and fitting a common Moffat profile, even the setup stage can take a few minutes for a common target of 9 bands of optical and near infrared (NIR) data. If PSFs are provided then the setup should only take a few seconds for anything but the largest target galaxies (the \profound{} stages are very fast in practice).

\subsection{Parameter Priors}

In the most automated mode of \profuse{} priors cannot be provided explicitly. This is largely because the user would need prior knowledge of the source extraction and initial conditions to be able to specify these in a meaningful manner. To combat this limitation \profuse{} was designed to have two modes of operation: in the first everything including the fitting is carried out in a single call; in the second (as intimated above) an intermediate fitting object is made that can be further edited (e.g.\ priors can be embedded), and fitting can then be carried out.

In practice, adding priors for the \prospect{} and \profit{} components of \profuse{} requires some knowledge of both of those codes, so the average user will probably not attempt such advanced execution. In which case the priors are automatically generated as improper Uniform priors between limits of physicality and computation for both \profit{} and \prospect{} (in logarithmic space for all scale parameters). E.g.\ star formation rates and sizes are not allowed to be negative, and the centre of the object has to exist within the bounds of the data.

Given people often apply a limit to the \sersic{} index ($n$), this parameter uniquely in \profuse{} has the option to define an upper limit when fitting. The default is 5.3, offering a $\sim30$\% margin above the $n=4$ de Vaucouleurs profile index \citep{de-Vaucouleurs:1948ty}. In practice this should usually be set to higher values, e.g.\ the upper limit of 8 used in \citet{Kelvin:2012un} and used later in this work for our analysis of GAMA. Since poor sky subtraction can cause very strong systematics in recovered \sersic{} index \citep{Kelvin:2012un} we use the lower default of value of 5.3 to reduce catastrophic failures.

\subsection{Parameter Inference}

Once this fitting object has been set up, users are able to carry out parameter inference with an engine of their choice (there are many hundreds available just within \R) for as long as they deem necessary. Optimal inference is a massive and unsolved problem in statistics in itself, so this paper will not aim to provide a comprehensive discussion of the options available. Pragmatically we can suggest the use of the \highlander{} package as offering a good mixture of speed and quality of fit \citep[as discussed and used in][]{Thorne:2021un}. \highlander{} works by alternating phases of $N$-dimensional genetic algorithm based optimisation (Covariance Matrix Adapting Evolutionary Strategy; CMA-ES) followed by phases of Component-wise Hit-And-Run Metropolis Markov Chain Monte Carlo (CHARM-MCMC). This can be done as many times as desired, but the default is two phases of each, with a final longer CHARM phase recommended if posterior parameter exploration is desired. The default for each phase is 1,000 iterations (where a single iteration means sampling over all $N$ parameters, so this naturally scales with the number of parameters being fitted).

In very complex fits using more components, we have found better global solutions from increasing the number of total fitting phases whilst also reducing the iterations per phase. This encourages a more aggressive hyper-dimensional search for good solutions, where the down side is less time is spent exploring the optimal solution region. To combat this it is possible to separately specify the number of posterior samples during the final CHARM exploration phase, where this should usually be in the range of 1,000--10,000 for a well explored posterior.

In all cases the likelihood being optimised by any inference engine will be the sum of \profit{} $data - model$ log-likelihoods as outlined in \citet{Robotham:2017tl}. The result of this is the model will naturally attempt to better explain the pixel data with smaller uncertainty (usually corresponding to deeper images) but will be penalised for poorly fitting any band present. Unresolved images (usually the MIR and FIR) can (and ideally, should) also be provided, because whilst they provide little constraint on any \profit{} structural parameters, they offer useful information regarding the star formation history (particularly recent star formation in the case of the FIR) for \prospect. We discuss the incorporation of unresolved data in more detail in Section \ref{sec:unres}.

\subsection{AGN Modelling}

As well as allowing for the above 1, 1.5, 2, and 2.5 component extended models using \prospect{} to construct the stellar history of a galaxy, it is possible to also generate an AGN component and spatially associate it with the bulge component of the galaxy. Physically this likely makes most sense when fitting a 1.5 component model (with a PSF bulge) since you are already anticipating that the bulge will be entirely unresolved in the regime where the AGN dominates the light enough to make fitting for it viable (as will be the case for a true AGN). A simple version of AGN fitting with \prospect{} was discussed in the original paper, but it has since been over-hauled to offer the full flexibility needed based on the models provided by \citet{Fritz:2006tz} as discussed in detail in \citet{Thorne:2022uy}.

\subsection{Unresolved Data}
\label{sec:unres}

Another aspect of \profuse{} that should make it powerful in general application is its natural ability to predictively assign unresolved flux (often in the MIR and FIR) to the correct component given the \prospect{} model uncovered. E.g. in the case of extreme star formation in the disk and an old red-and-dead stellar population representing the bulge, it is clear that almost all of the unresolved FIR flux identified in the image correctly belongs to the disk. This need not always be the case though, \profuse{} is able to equally identify situations where we observe significant recent star formation in the bulge (e.g. a star burst event) and allocate an appropriate (non-zero) fraction of the FIR flux to the bulge in this case. This is practically impossible to do under the normal circumstances of running \profit{} on each band separately, since there would be no reasonable route to inform the FIR priors in such a manner that both of the above scenarios are allowed.

\profuse{} further offers a fitting mode where marginally- or un-resolved data can be flagged as such (by setting the {\sc doprofit} option to FALSE), and the structural decomposition is not applied to that band. Instead only the total flux within the specified (or automatically detected) fitting region is used to compute a likelihood. This means that the sum of the various components in these bands should ideally match the integrated flux, with the usual penalty of the flux uncertainty giving more or less weight to this tension. These additional likelihood terms increase the computation time (given the addition of extra \prospect{} band passes to compute) and requires more interim data storage for the fitting model. Since GAMA is somewhat unusual in having access to 20 bands of UV-FIR data, we have focussed on resolved fitting of the optical and NIR bands in the rest of this paper since this will be more representative of the typical use case for non-GAMA surveys in the future.

\section{Application To An Example Galaxy}
\label{sec:example}

\begin{figure}
 \includegraphics[width=\columnwidth]{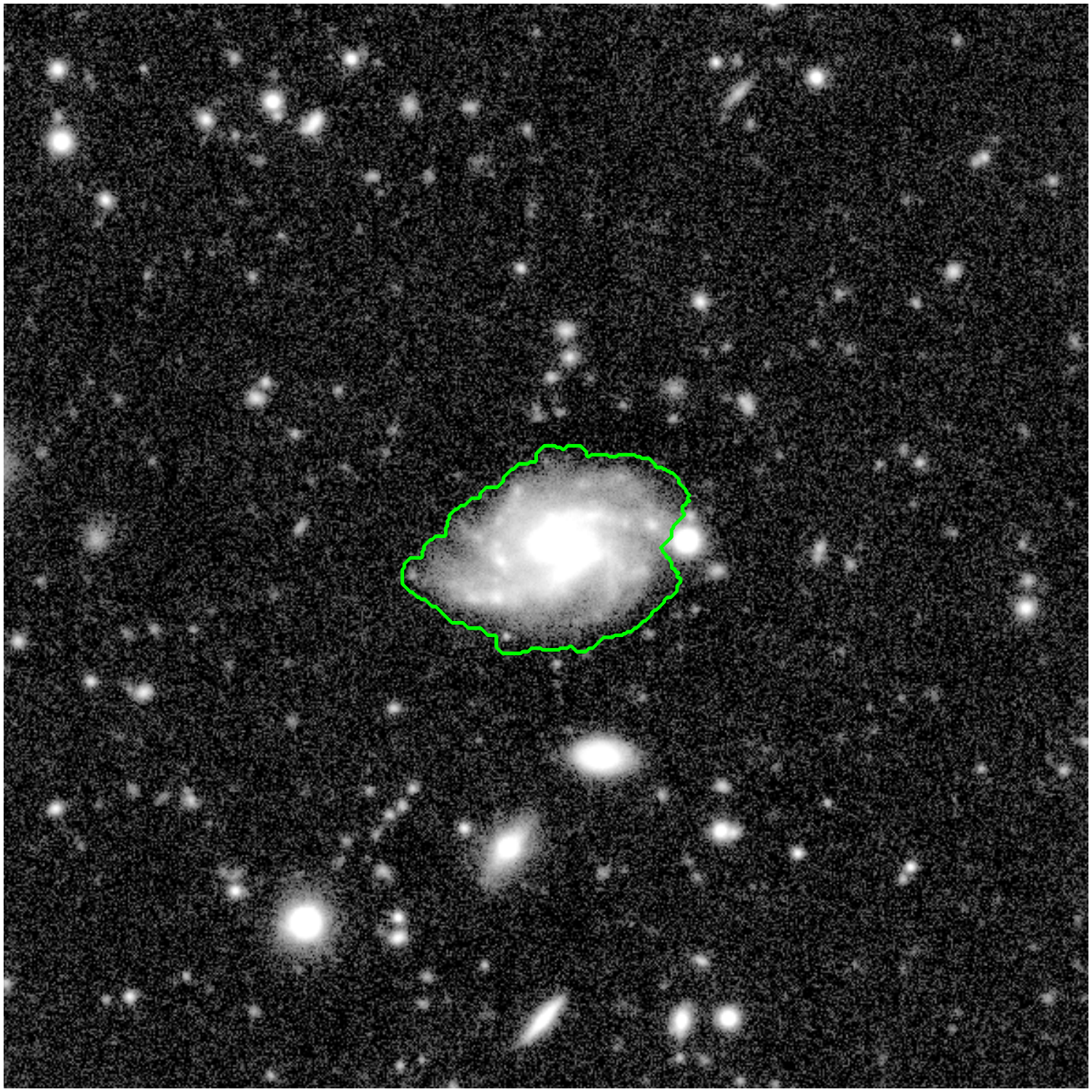}
 \includegraphics[width=\columnwidth]{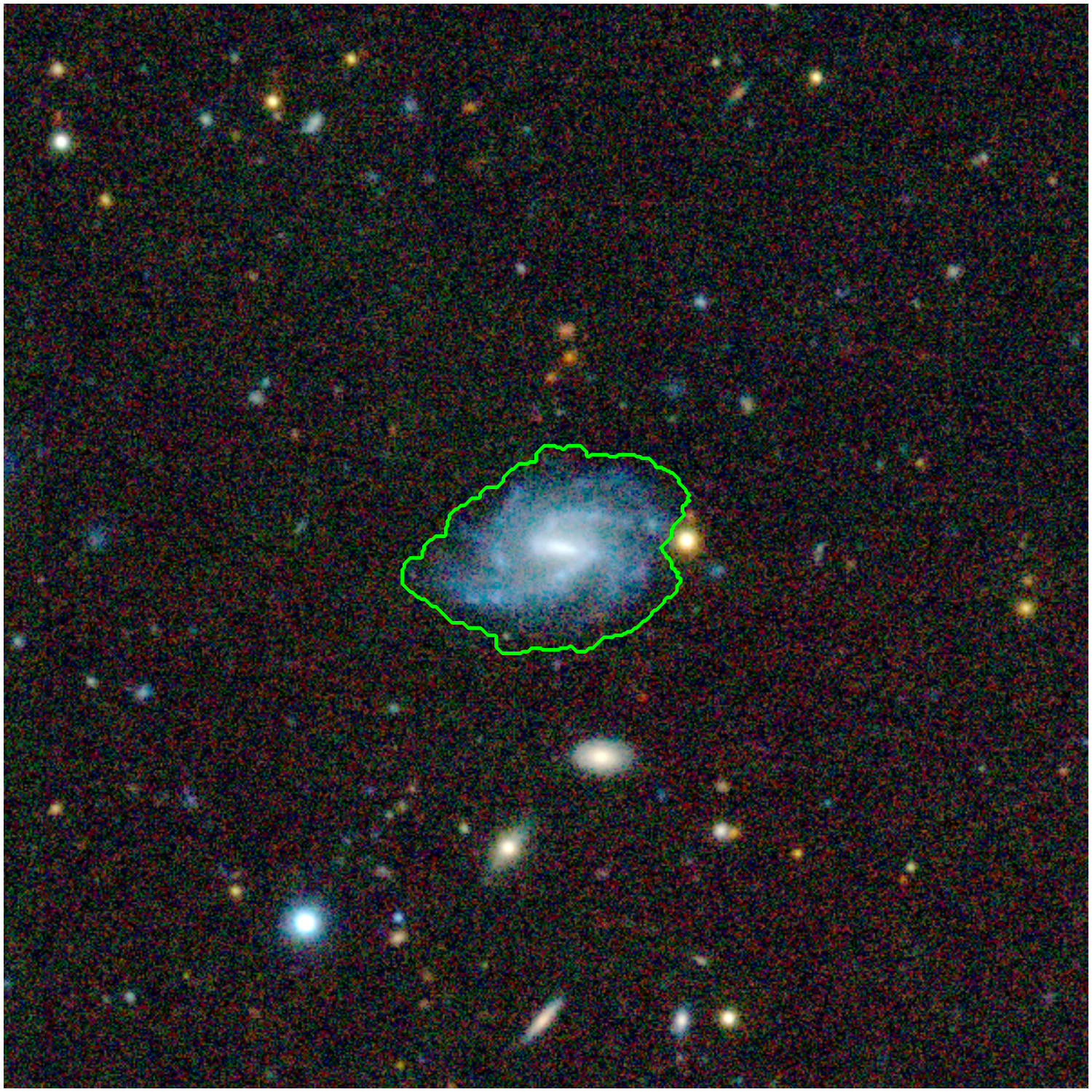}
 \caption{Example galaxy TGS426Z278. Top panel is the full 9-band $u$--Ks inverse variance weighted stack; bottom panel is the RGB image mapped from asinh scaled H$ig$ band images. The example was cut out large enough such that at least 4 stars could be identified in all band $u$--Ks. The green region shows the pixels selected for fitting based on the \profound{} segmentation image extracted from the 9-band stacked image (top panel).}
 \label{fig:Ex_RGB}
\end{figure}

To aid users, a comprehensive example is provided with the package. This makes use of pixel-matched 9-band ($ugri$ZYJHKs) imaging taken from the GAMA survey \citep{Bellstedt:2020wl} that comes included as part of the \profound{} package. The example multi-band data that comes with \profound{} has a few extended sources available to test fitting with \profuse, but here we focus on one of the more extended examples in the field as shown in Figure \ref{fig:Ex_RGB}. The example galaxy (unique name TGS426Z278) has been targeted by 2dFGRS \citep{Colless:2001tb} and has a redshift $z=0.0447$. Visually it is clearly a spiral galaxy, with some evidence of a redder central region (visually white in the image due to the scaling) which suggests it might be reasonable to fit a bulge and disk model. 

Running the 9-band automatic setup takes about two minutes, which includes identifying, extracting and fitting the PSFs in all bands. With the fitting object assembled, we use \highlander{} to fit a simple circular de Vaucouleurs profile \citep{de-Vaucouleurs:1948ty} (analogous to a \sersic{} index $n=4$ profile) bulge and exponential (analogous to a \sersic{} index $n=1$) elliptical disk model with a skewed Normal star formation history (SFH) and linearly mapped metallicity history (ZH) with free final $Z$ \citep[see][for details of the SFH and ZH implementation]{Robotham:2020to}. For convenience, the skewed Normal SFH used has the form:

\begin{eqnarray}
SFR(\rm{age}) &=& m_{\rm{SFR}} \frac{e^{X({\rm age})^2}}{2}, \\
X(\rm{age}) &=& \left(\left[\frac{{\rm age} - m_{\rm{peak}}}{m_{\rm{period}}}\right] e^{m_{\rm{skew}}}\right)^{{\rm{asinh}}\left(\frac{{\rm age} - m_{\rm{peak}}}{m_{\rm{period}}}\right)},
\end{eqnarray}

where $m_{\rm{SFR}}$ is the peak star formation rate, $m_{\rm{peak}}$ is the age of the peak in star formation, $m_{\rm{period}}$ is the standard deviation of the star formation period, and $m_{\rm{skew}}$ is the skew of the Normal (where 0 is perfectly Normal, +ve is skewed to younger ages and -ve is skewed to older ages). The linear ZH model used forms metals in linear proportion to the fraction of stars formed, e.g.\ when half the stars in a given galaxy have been formed the metallicity will, by construction, be half that of the free final $Z$.

This is probably the simplest plausible bulge-disk system that can be fitted to such data, and the resultant model (with all initial parameters determined automatically) requires 16 free parameters.

We use \highlander{} for parameter inference with two phases of CMA and CHARM with 1,000 iterations during each phase (where an `iteration' means a sequential modification of all parameters, so this naturally scales with model complexity). The example requires roughly one hour of CPU time to achieve a reasonable solution across all 9 images on a single modern laptop core. This is slower than running \profit{} separately on a single image (which might take 10--20 minutes with similar \highlander{} settings), but since the number of parameters is effectively less and we fit all 9-bands simultaneously with a \profuse{} model the total time is 2--4 times less. This is ignoring the \prospect{} side, which on its own typically takes 10 minutes for a single version of the SFH and ZH model (which we are fitting separately for the bulge and disk in \profuse).

\begin{figure}
 \includegraphics[width=\columnwidth]{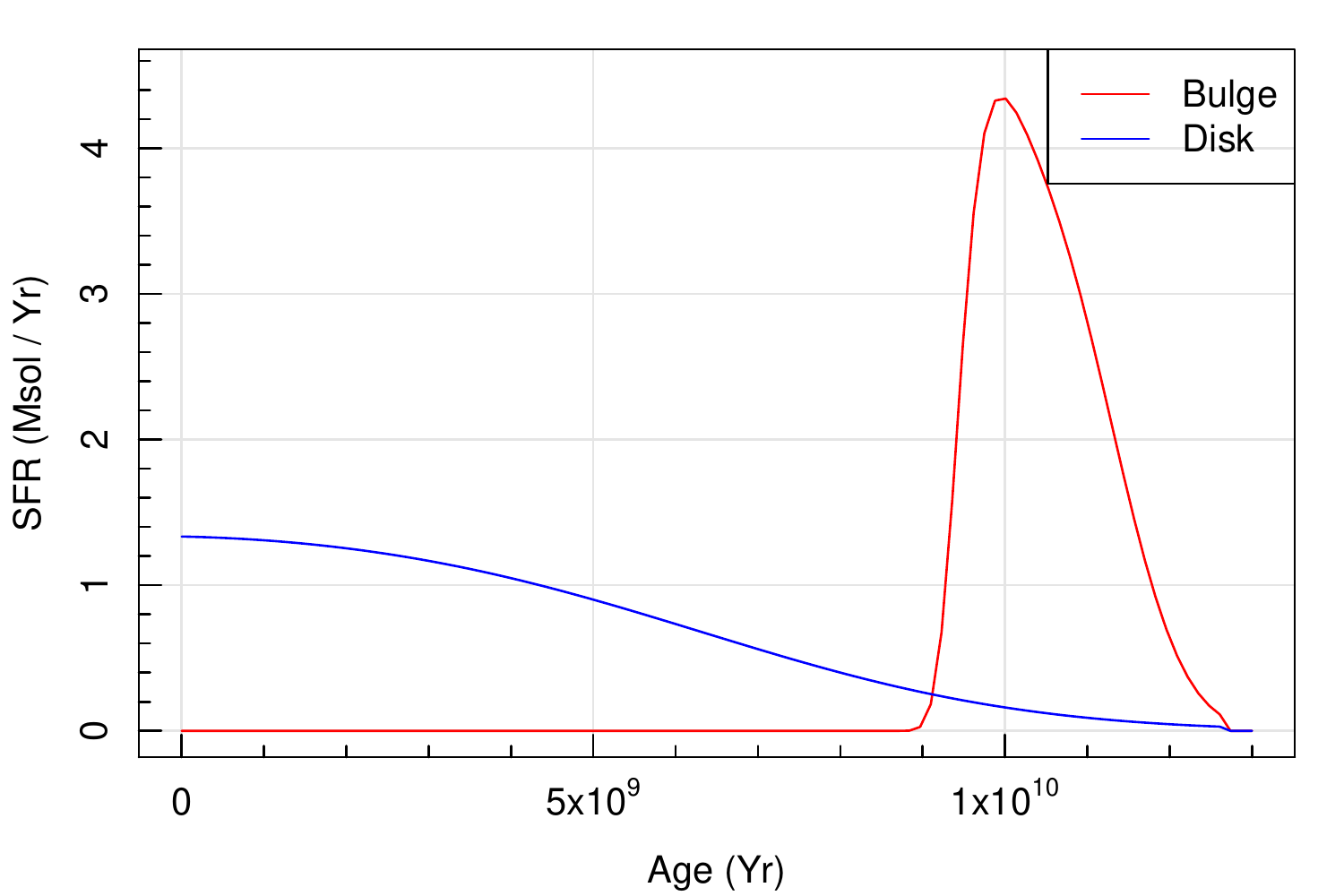}
 \caption{Best-fitting SFHs for the bulge and disk using \profuse{} on example galaxy TGS426Z278 (as seen in Figure \ref{fig:Ex_RGB}) using 9-bands of imaging data.}
 \label{fig:SFHs}
\end{figure}

The end result is a physical model with SFHs, ZHs, stellar masses and star formation rates (and all other \prospect{} properties) for the bulge and disk that runs factors faster than the laborious approach of attempting to fit all bands separately with \profit{} and then using \prospect{} to fit the resultant bulge and disk properties. The latter approach is also prone to catastrophic failure unless some artificial (and unphysical) coupling between bands it attempted \citep{Haussler:2013tz, Vulcani:2014up}.

\begin{figure}
 \includegraphics[width=\columnwidth]{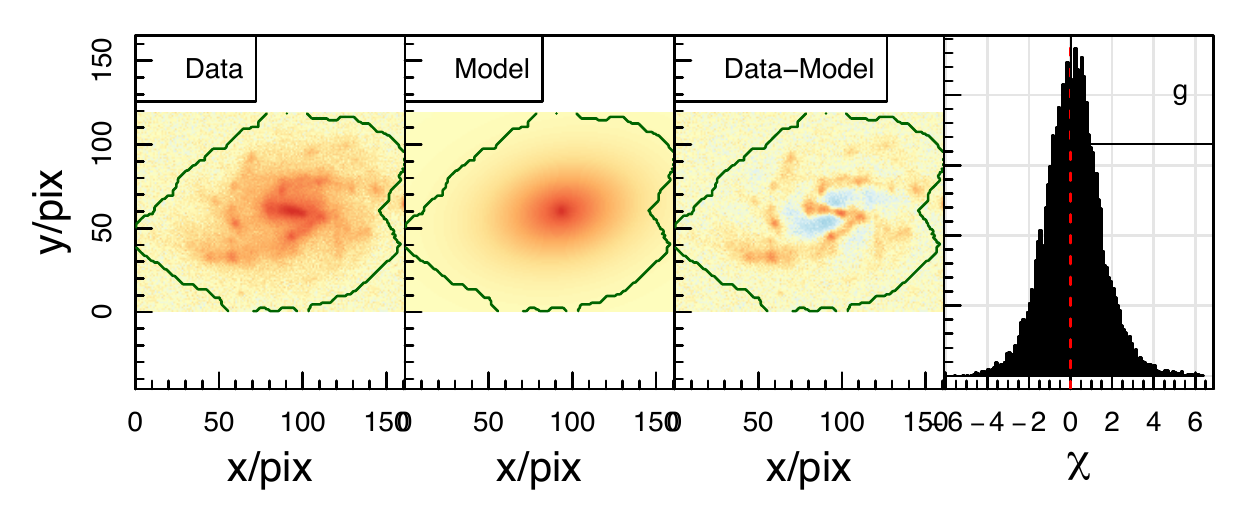}
 \includegraphics[width=\columnwidth]{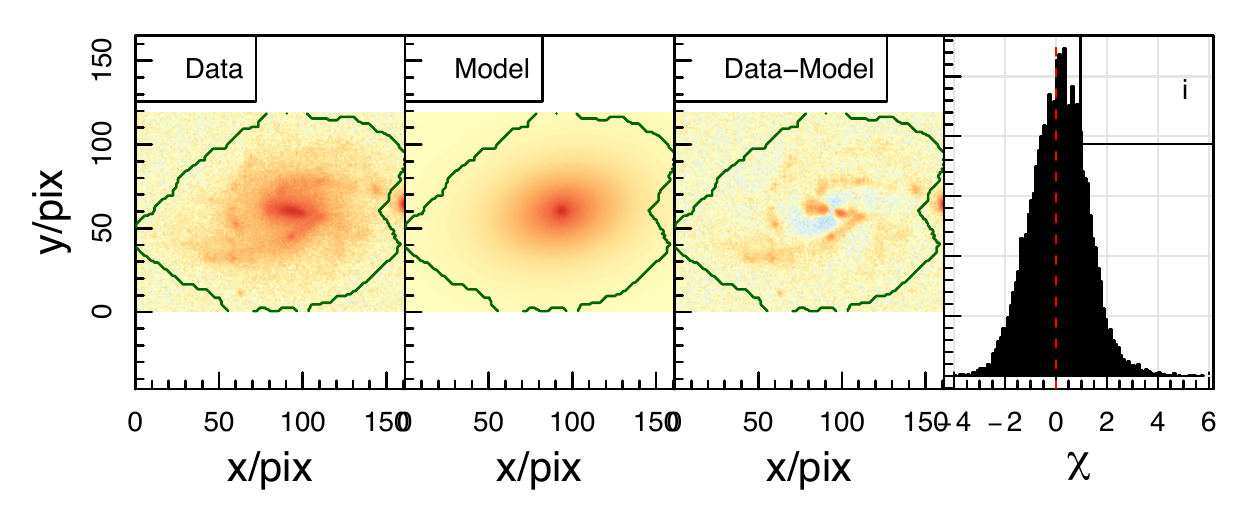}
 \includegraphics[width=\columnwidth]{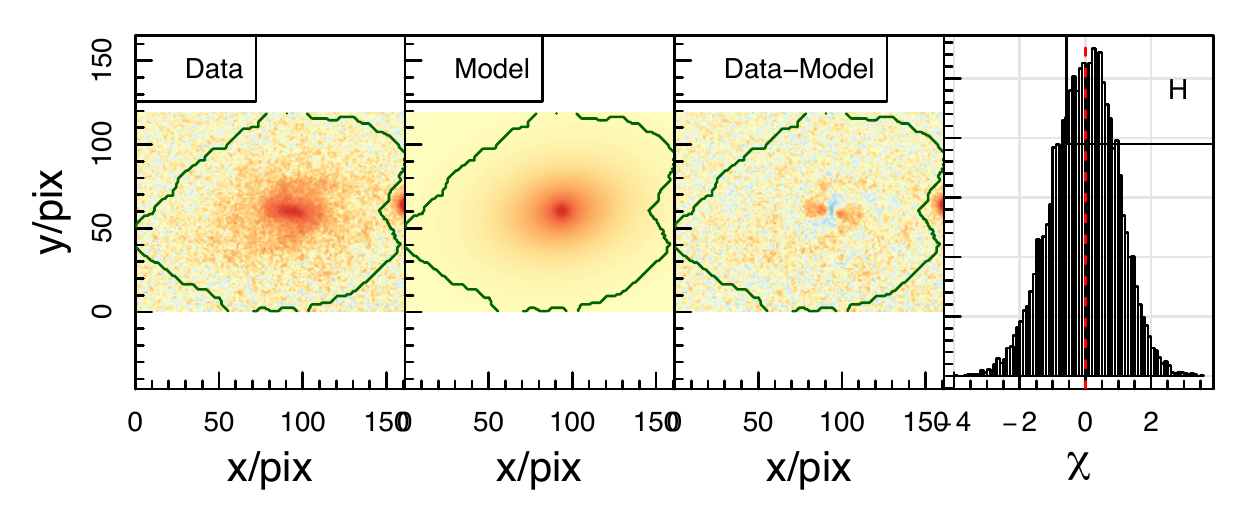}
 \caption{Example Data-Model results for 3 bands (giH) of the 9 bands used to constrain the target model of example galaxy TGS426Z278 (as seen in Figure \ref{fig:Ex_RGB}). In each row of panels we see the imaging data (left panel), model (second panel), spatial residual (third panel) and error normalised histogram (right panel). These are very similar to the standard analysis figures generated by \profit{} and presented in \citet{Robotham:2017tl}. The desired outcome in all cases is for the third panel (the residual `Data - Model') to be close to yellow, with no dramatic over subtraction (blue) or under subtraction (red) by our model. Unsurprisingly, given the smoothness of the model, the spiral arms tend to remain red in the bluer $u$ and $i$ bands, but there is little spiral structure in the H band and residuals are close to featureless.}
 \label{fig:DataModel}
\end{figure}

\begin{figure}
Bulge SED:\\
 \includegraphics[width=\columnwidth]{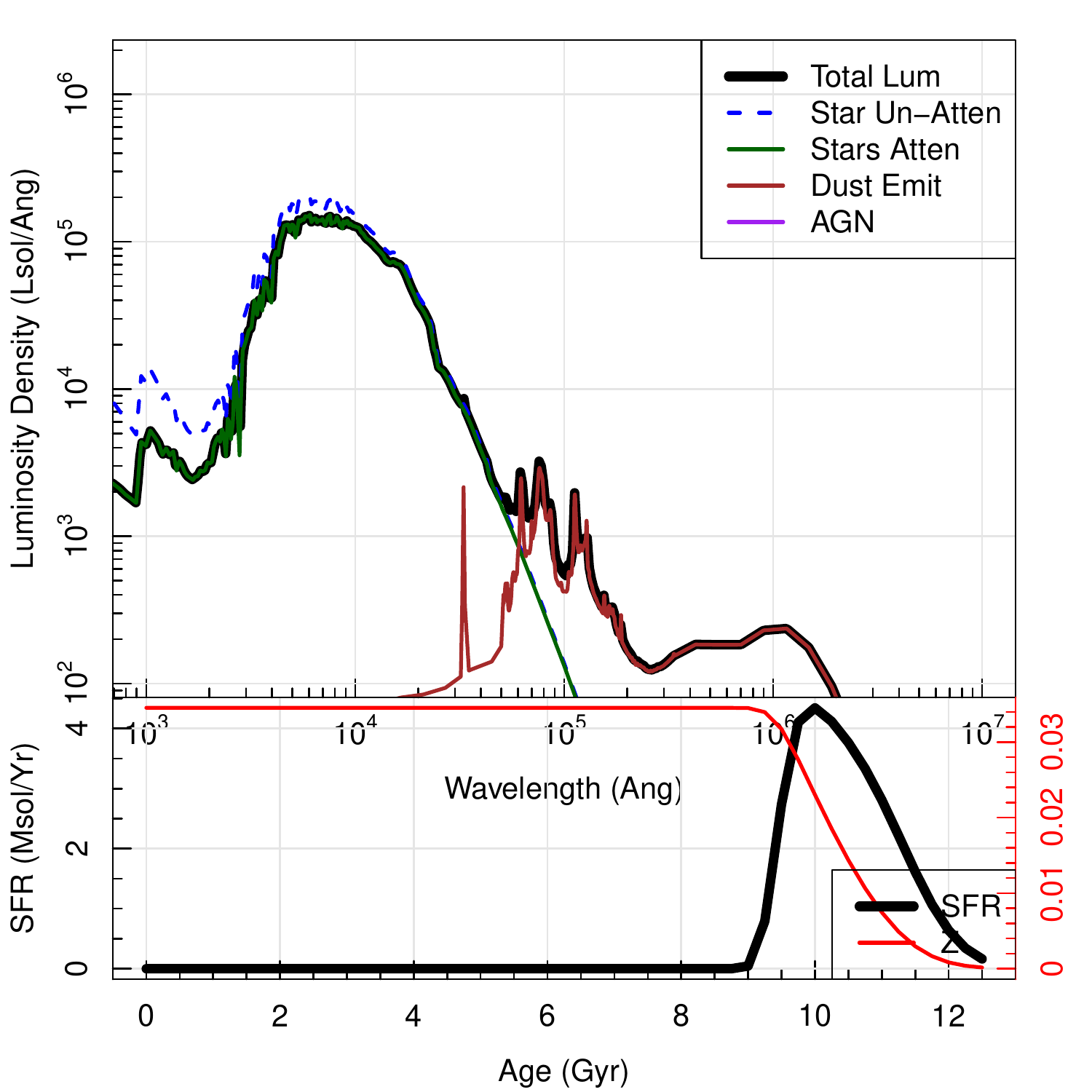}
 Disk SED:\\
 \includegraphics[width=\columnwidth]{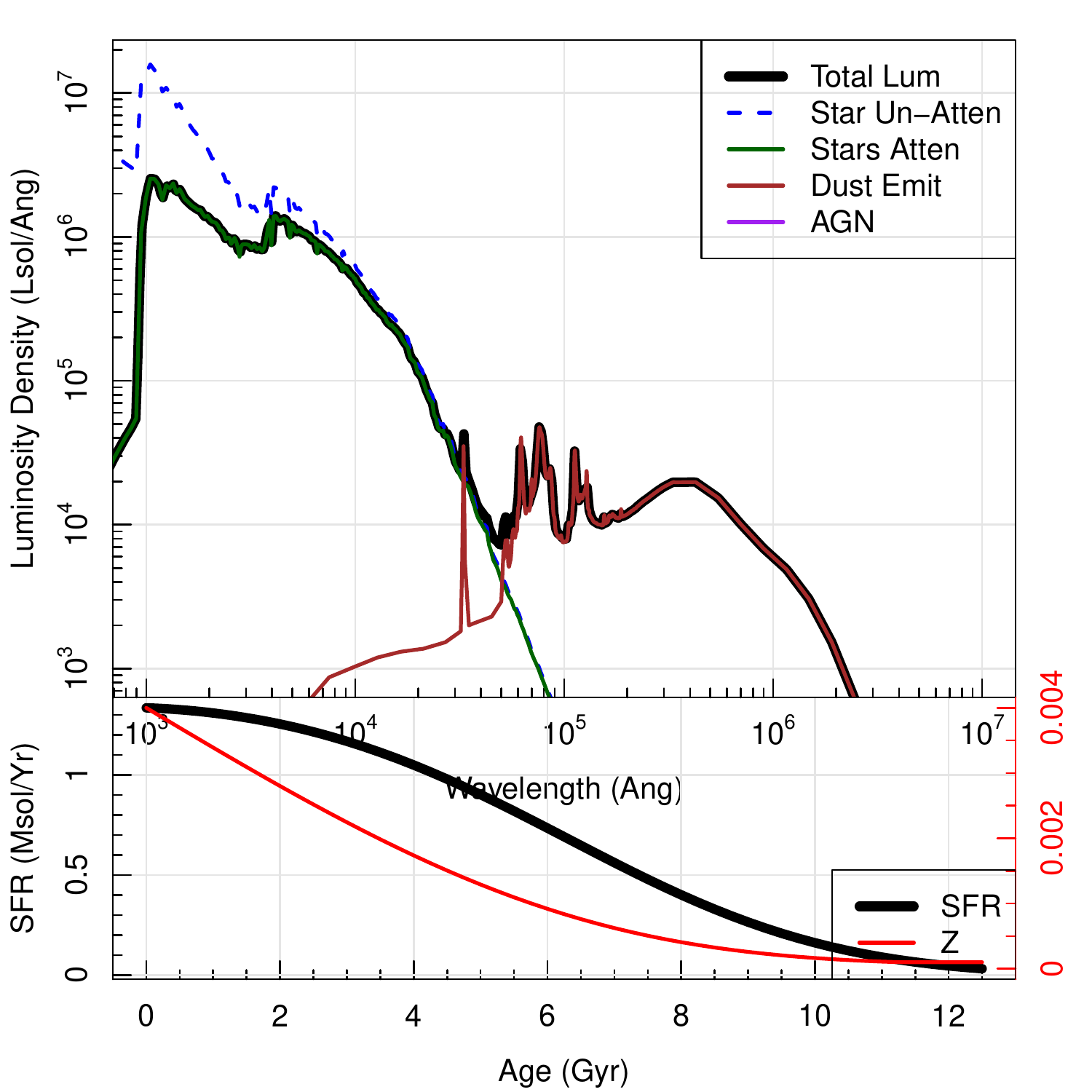}
 \caption{Extracted bulge (top panels) and disk (bottom panels) SEDs from \profuse{} for the example galaxy TGS426Z278. Within the grid of panels the top shows the \prospect{} intrinsic luminosity spectrum (pure light blue dashed, attenuated in dark green, re-emitted by dust in brown, final observable in black). The lower panels show the star formation and metallicity enrichment histories (SFH left axis, ZH right axis). In this case the AGN component was marginalised out.}
 \label{fig:SEDs}
\end{figure}

\begin{figure}
 \includegraphics[width=\columnwidth]{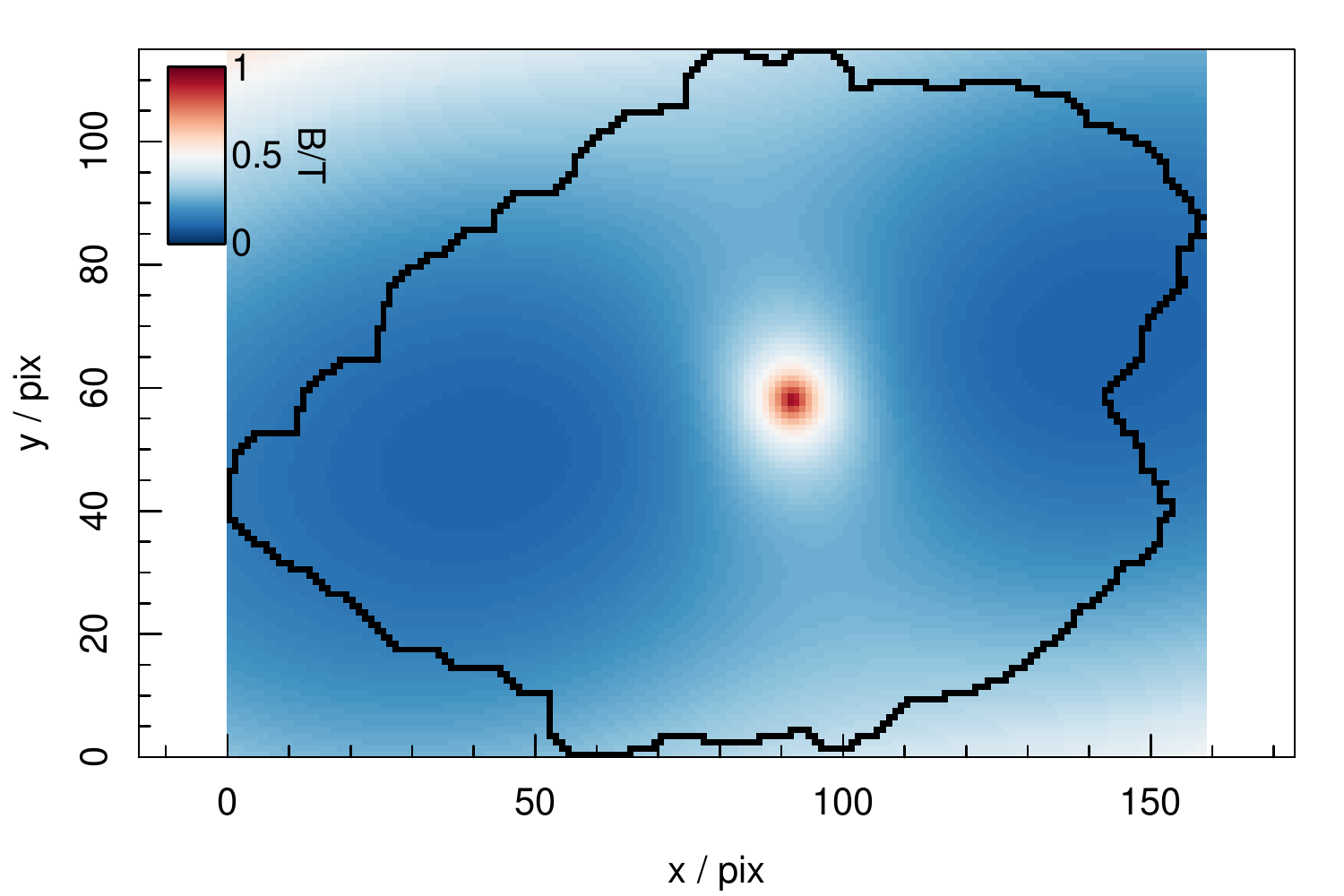}
 \caption{The model predicted intrinsic total mass formed $B/T$ ratio per pixel (without PSF convolution). The mass in the disk dominates almost everywhere, but as should be expected the bulge contribution increases towards the centre.}
 \label{fig:model_B_T}
\end{figure}

The result of the fit is a full SFH for both components, as seen in Figure \ref{fig:SFHs}. The model also produces per-band predictions of the bulge and disk model, where we show data and model residuals for the $g$, $i$ and H bands in Figure \ref{fig:DataModel}. As should be expected given the elliptical smoothness of the model, we have stronger residuals in the bluer bands where strong spiral arm features are present, and little structural residual in the NIR. The third useful output that can be assembled is the observed SED for the separated bulge and disk in Figure \ref{fig:SEDs}. The clear interpretation here is that the disk is significantly brighter than the bulge at all wavelengths (brighter luminosities in this Figure) and the bulge is redder in colour (steeper drop off in luminosity for bluer wavelengths). The end result of this is we have extracted a classical bulge and exponential disk galaxy. A simple reconstruction of the per pixel bulge-to-disk mass ratio is shown in Figure \ref{fig:model_B_T}, revealing a particularly compact bulge.

A natural consequence of using \profuse{} is we get all the component-wise properties from \prospect{} automatically, e.g. we recover a total stellar mass formed (assuming a Chabrier Initial Mass Function \citep{Chabrier:2003ta}, as used in \prospect) of $1.2\times10^{10}$\msol for the disk and $6.9\times10^{9}$\msol for the bulge, i.e. a disk that is 1.7 times more massive than the bulge. The total stellar mass formed by the bulge and disk is therefore $1.9\times10^{10}$\msol, i.e. this is a very typical slightly sub-$M^*$ galaxy. In comparison the light models generated vary between the disk being a factor 25 times brighter ($u$ band) and 2.9 times brighter (Ks band), reflecting the significantly more ancient star formation history of the bulge and the subsequent dimming of its stellar populations over time. We see a much bigger difference in the current star formation rates, defined in \prospect{} as the average SFR over the last 100 Myrs. These are 1.35 \msol/yr for the disk, and statistically 0 \msol/yr for the bulge, i.e.\ 100\% of the current star formation is occurring in the disk (not the 96\% that would be implied from pure $u$ band flux scaling).

Spatially we also have access to the normal outputs of \profit{} (depending on which fitting options were selected). In this case the main spatial parameters of interest are the bulge and disk $R_e$, which are 2.2' and 10.9'' respectively. Given the redshift of the example galaxy ($z=0.0447$) this corresponds to 2.0 kpc (bulge) and 9.8 kpc (disk), which is reasonable for a slightly sub-$M^*$ galaxy.

The interpretative possibilities from the full \profuse{} fits are extremely rich, the above only touching upon some of the simpler analyses that can be extracted. A caveat to these fits, and results in following sections, is that we are assuming a particular model for the SFH and ZH (and also a certain IMF and dust template etc), and other models (e.g.\ closed box metallicity evolution) might individually produce equally reasonable results. However, this combination of models have been shown to reproduce the cosmic star formation history of the Universe \citep{Bellstedt:2020vq, Thorne:2021un}, the variation of evolution observed in simulated semi-analytic galaxies \citep{Robotham:2020wo}, and even the evolution of the mass metallicity relationship \citep{Bellstedt:2021vx}. Investigating the full impact of relaxing these assumptions is beyond the scope of this initial work, but we have at least demonstrated it is a meaningful foundation to build from.

\section{Application to the Low-Redshift GAMA Sample}
\label{sec:GAMA}

\profuse{} was designed with the upcoming era of rich multi-wavelength surveys in mind. Like the foundational SED code \prospect{} used to generate the light output, \profuse{} works best when we have access to a broad wavelength range since this can constrain the SFH and ZH of the bulge and disk components. It is also important that the galaxies in question are reasonably well-resolved, where the expected $R_e$ of the components are certainly no smaller than the average PSF across all bands. At low redshift the ideal data set to test \profuse{} on is therefore the GAMA survey since it offers well-resolved and reasonably deep imaging data covering the $u$--Ks bands \citep{Driver:2016vh}. It also has shallower less well-resolved data that covers the UV \citep[GALEX;][]{Morrissey:2007vl}, MIR \citep[WISE;][]{Wright:2010tg} and FIR \citep[Herschel;][]{Pilbratt:2010wp}, but very few objects are both detected and resolved in GAMA in these wavelengths so we have dropped them from this proof-of-concept analysis. As above, we use \highlander{} to carry out our inference, where we increased the number of iterations to 1,000 (CMA) and 1,000 (CHARM) run twice back-to-back (where again an iteration in this context means all available parameters are modified once).

\begin{figure}
 \includegraphics[width=\columnwidth]{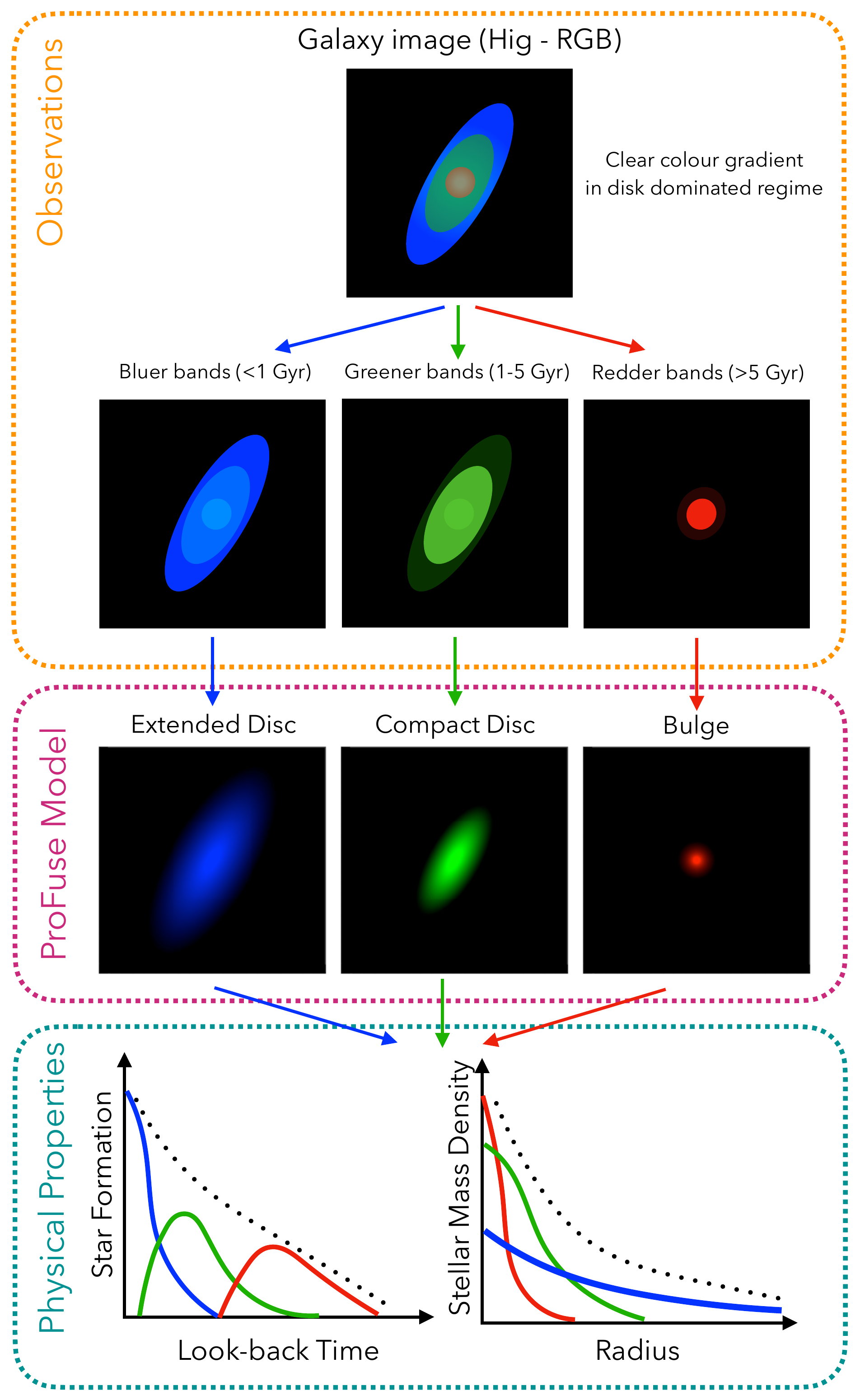}
 \caption{A schematic view of how the presence of separated SFHs in bulge and disk (two in this case) can produce quite complex colour gradients. Spectral structural decomposition (SSD) with \profuse{} offers a route to untangle such complex observations. The caveat to applying such a complex model is that the data needs to be deep enough and resolved enough to justify it.}
 \label{fig:BDDcartoon}
\end{figure}

The ideal sample to test \profuse{} with has already been well justified and studied in previous GAMA papers: the $z<0.06$ sample that has proved itself well suited to single band bulge-disk decomposition \citet{Lange:2016wx}, and also \prospect{} SED fitting \citet{Bellstedt:2020vq}. This sample contains 6,664 galaxies with complete coverage in the optical $ugri$ \citep[VST/KiDS;][]{Shanks:2015va, Kuijken:2019vi} and NIR ZYJHKs \citep[VISTA/VIKING;][]{Edge:2013uv} bands. In this initial work we will explore the three simplest models that are likely to explain our data in a reasonable manner: a single component elliptical free \sersic{} index model [FS from here]; the same bulge-disk model as discussed in Section \ref{sec:example} (with the same limitations on profile and geometry) [BD from here]; and a bulge disk model where the bulge is limited to be just a PSF [PD from here]. We expect one of the two bulge-disk models (BD or PD) to work better for galaxies that have a canonical classical bulge and star forming disk, and the FS model to work better for early-type galaxies with a more homogenous stellar population but non-circular geometry. For all three models we achieved a better than 99\% fitting success rate (i.e.\ less than 1\% of galaxies were missing one or more models), which is exceptionally high compared to published automated galaxy profiling work \citep{Kelvin:2012un, Cook:2019wa}. Much of this success is due to the presence of multiple bands compensating for data quality issues that often just affect a single image.

We also processed all of this low redshift GAMA sample using the three component bulge and double disk mode [BDD from here]. For this run the interest was whether we needed this added complexity due to regularly finding strong colour gradients along the GAMA galaxy disks as found in other work \citep[e.g.][]{MacArthur:2004ta, Mollenhoff:2006ws, Bakos:2008ts}. In principle linear or complex colour gradients can be formed in BDD mode since both the disk components have entirely decoupled SFHs and ZHs. A simple cartoon of this paradigm is shown in Figure \ref{fig:BDDcartoon}.

Whether these disk components can be thought of as physically separate is debatable of course, since a kinematically simple disk might possess colour gradients due to its formation history, as can kinematically distinct thick and thin disks. In practice, with our GAMA quality data, very few galaxies require the additional complexity of BDD versus BD, and the double disks serve to degenerately recreate the properties of the single disk in the BD mode. Some global comparisons are made to demonstrate this fact in Appendix \ref{app:BDD}, but we will ignore the BDD mode for the main analysis in this paper since the simpler BD mode mimics its general behaviour.

\begin{figure*}
 \includegraphics[width=18cm]{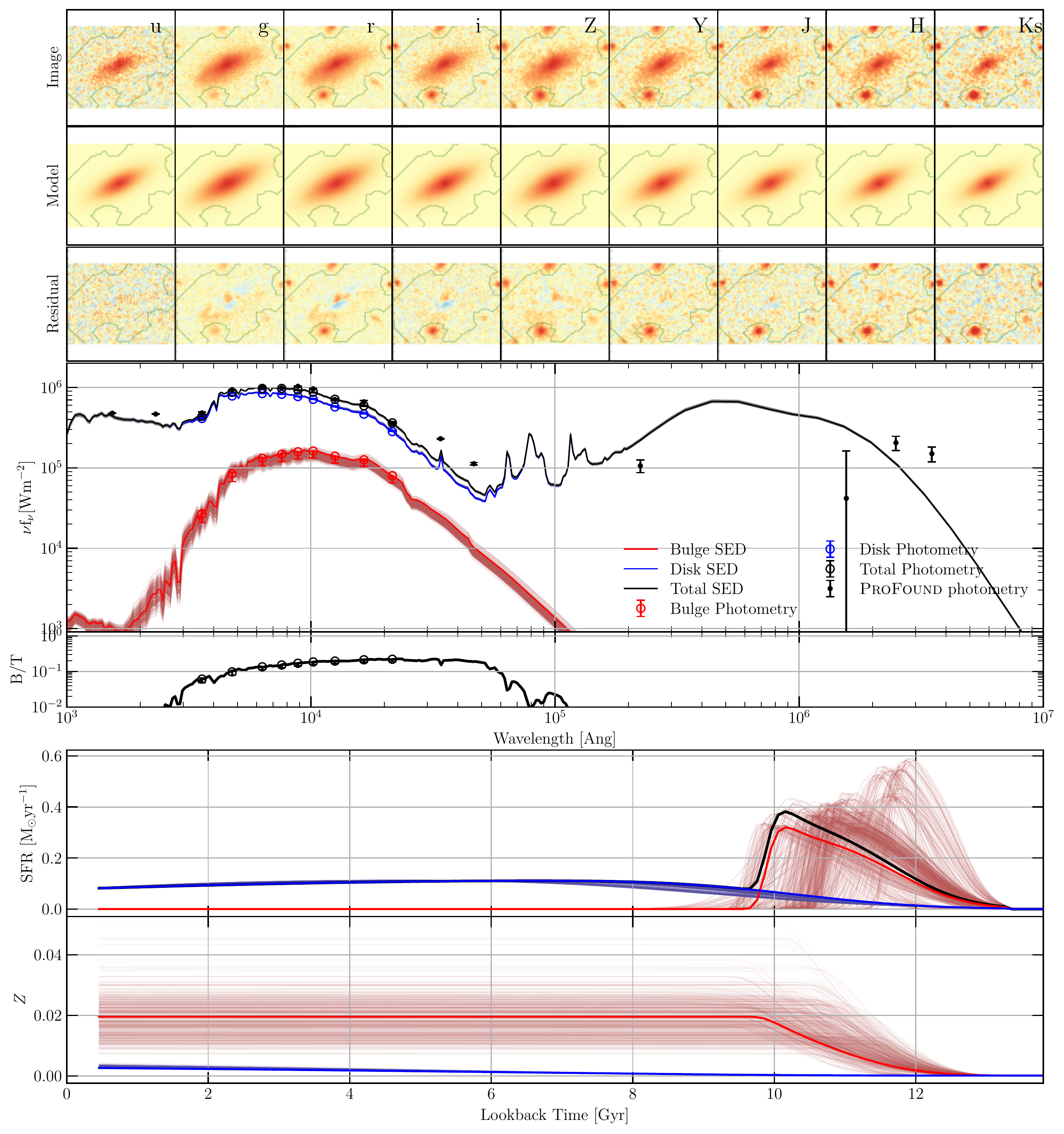}
 \caption{Full set of visual diagnostics for GAMA 7688 (an example of a moderately resolved object, approximately at the $50^{th}$ percentile of our sample). Top panels show the 9-band data, model and residuals. The second panel from top is the SED output of the bulge/disk/total model compared to the 20 band GAMA data including the implied spectral bulge-to-total flux ratio (where we can see the predicted \profuse{} model recreates the unseen UV/MIR/FIR data very well). The fourth panel from the top shows the SFH of the bulge/disk/total. The bottom panel shows the ZH for the bulge/disk/total. This particular galaxy prefers a small old metal rich bulge and a younger more metal poor disk. In all panels light coloured lines are random posterior samples showing the spread in uncertainty in the model convergence. This means, for instance, that we constrain the SFH and ZH of the disk remarkably well, but the bulge has a large variety of ancient SFH allowed (since 10-12 Gyr stellar populations are extremely similar in SED appearance). Also, the bulge ZH shows a similarly large spread, reflecting the classic degeneracy between age and metallicity.}
 \label{fig:ProFuse_7688_Summary_BD}
\end{figure*}

For the \prospect{} component of our \profuse{} GAMA modelling, we closely followed the approaches of \citet{Bellstedt:2020vq} and \citet{Thorne:2021un}. To better compare to this previous work we use the optional \citet{Bruzual:2003wd} stellar template library in \prospect, which implicitly assumes a \citet{Chabrier:2003ta} initial mass function. Each of our components is allowed to have a skewed Normal SFH and linearly mapped metallicity evolution with free final $Z$. We set the initial \citet{Charlot:2000tk} dust parameters to $\tau_{\text{screen}}=0.16$ and $\tau_{\text{birth}}=0.63$ for our components, which is informed by the distribution modes recovered in \citet{Thorne:2021un}.

For all 6,664 GAMA galaxies we generate a range of outputs and diagnostics. \highlander{} outputs include the last iterations of the CMA-ES and CHARM MCMC (including all 1,000 posterior chains). Using these outputs we can reconstruct the SFH and ZH with uncertainties from the chains, and compute total photometry to compare to GAMA including UV/MIR/FIR bands that are available in GAMA but not used as part of the \profuse{} process due to the lack of spatial information present.

Figure \ref{fig:ProFuse_7688_Summary_BD} is an example of the kind of summary plot that we have generated for every galaxy for the three main modes explored in detail below (FS/BD/PD). This example is the BD output for GAMA 7688, a moderately well-resolved source approximately at the $50^{th}$ percentile of our sample in terms of the intrinsic $R_e$ of our FS model. Here we see a classic old red bulge and young blue disk formation scenario, where we have significant stellar mass in both the bulge and disk components ($B/T\sim 0.3$). For reference Appendix \ref{app:resolve} presents examples of a marginally- and well-resolved GAMA galaxy in the same plot summary format. In general \profuse{} returns successfully converged model solutions across a broad range of resolutions.

\subsection{High Level Run Comparisons}

Two obvious quantities of interest when deciding what mode to run in is how often the fit is preferred and how long the fitting process takes. There are many ways to specify the fit preference when carrying out model selection, and in general this is considered to be an unsolved problem in Bayesian analysis. For simplicity we just compare the reduced Chi-Squared ($\chi^2 / \nu$) for the various runs on the same objects, where in general values close to 1 will represent better fits (much below 1 tends to indicate over-fitting, and much above 1 under-fitting).

\begin{figure}
 \includegraphics[width=\columnwidth]{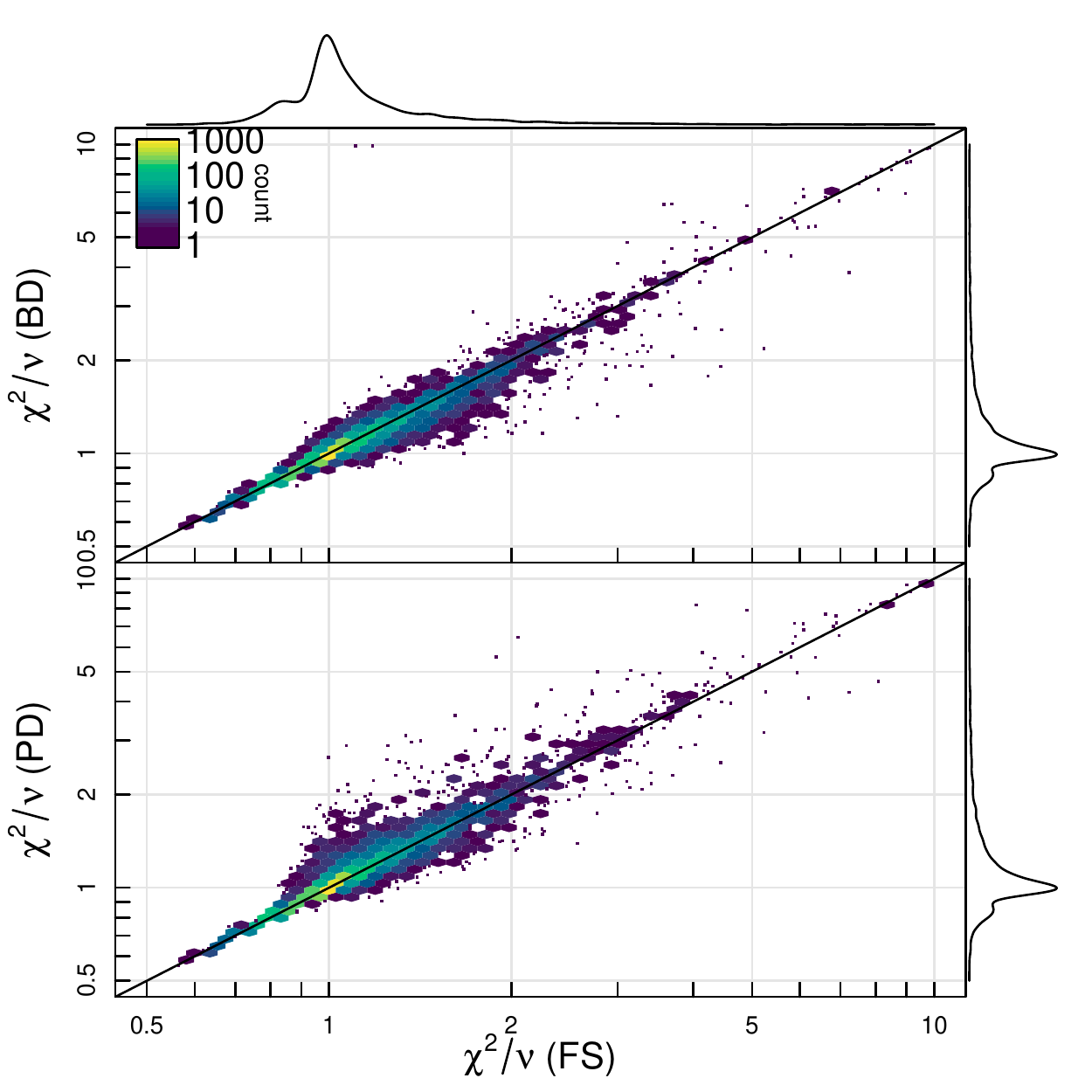}
 \caption{Comparison of FS, BD and PD \profuse{} reduced Chi-Squared ($\chi^2 / \nu$). In general multiple models can often do a similarly reasonable job of fitting the data, but the flexibility of the BD model is clearly better for a certain subset of galaxies (presumably those comprised of true extended bulges and disks).}
 \label{fig:RedChi_FSvBDvPD}
\end{figure}

Figure \ref{fig:RedChi_FSvBDvPD} presents the comparison of $\chi^2 / \nu$ for the FS, BD and PD runs. FS and BD models are of particular interest since they are at the opposite extremes in terms of fit complexity (BD being technically more complex with 16 parameters compared to 13 for FS). In general the fitting is highly correlated, i.e.\ if we are able to achieve a good BD fit then we can usually also find a reasonable FS fit. However, the outliers to this statement tend to favour BD fits, hence the large cloud of fits in the bottom-right of the Figure. Given the added flexibility available to the BD model, this outcome is expected. It is notable that the PD models have bigger extremes of poor fits. This appears to be due to FS models being able to reasonably represent true bulge-disk systems (but not as well as the BD model), and BD models can approximately describe the light profile of an elliptical galaxy (but not as well as the FS model). PD models however cannot properly capture elliptical galaxies in particular, causing a wing of poorer fits.

\begin{figure}
 \includegraphics[width=\columnwidth]{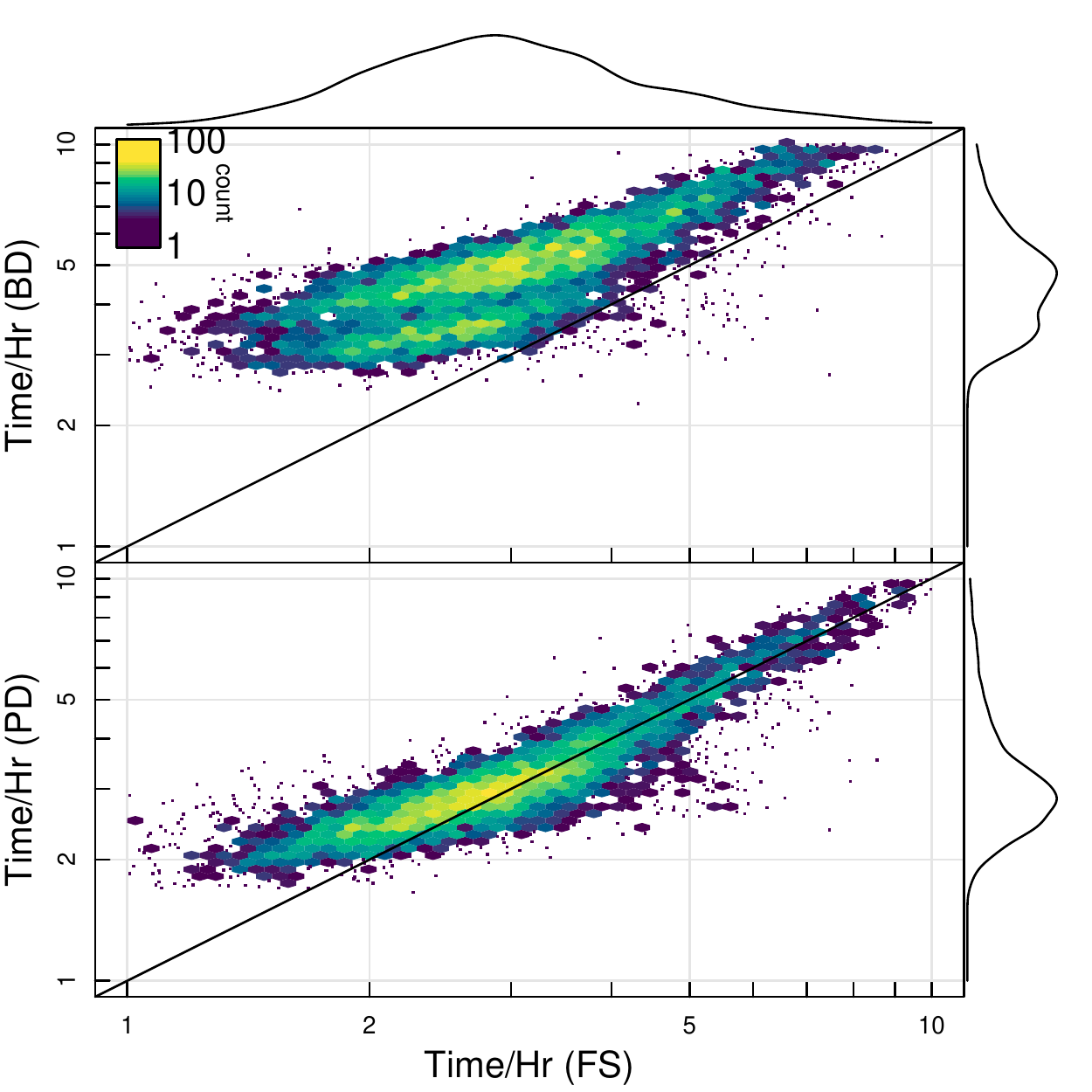}
 \caption{Comparison of FS, BD and PD \profuse{} fitting time.}
 \label{fig:Time_FSvBDvPD}
\end{figure}

The added flexibility of the BD model does come with a significant penalty in terms of computation cost. Figure \ref{fig:Time_FSvBDvPD} compares the fitting computation time in hours for both types of fits, where FS fits average near to 2.5 hours per fit, whilst BD fits are closer to 3.5 hours with more extreme outliers in general. In this case both runs were using the same number of iterations in total. For a given image with a fixed number of pixels to be modelled we expect the BD fit to take about 70\% longer to fit on average (taking the mean difference, 40\% taking the median difference). The extended fitting time is longer than the 20\% addition you would expect simply due to the increased complexity of the model (moving from 13 to 16 parameters). This is because the lower level \profit{} software has to generate two galaxy images in the BD mode, but only one for FS. This process is not twice as long however, since there is no additional overhead in empty image creation (which can be a significant portion of creation time for some models) and the free \sersic{} fitting available to the FS model will sometimes create computational expensive profiles (generally those with higher \sersic{} indices). The latter is why some of the fits can very occasionally take longer for the FS model. The bimodality present in the top panel of Figure \ref{fig:Time_FSvBDvPD} is due to compact bulges that are PSF-like being evaluated rapidly, whilst more extended sources take longer to evaluate. PD sit between FS and BS models in terms of typical fitting time, which is consistent with their relative complexity.

\subsection{Comparison Between \prospect{} and \profuse}

We already have SED fitting based SFH and ZH results from \prospect{} for previous runs using the \profound{} extracted 20-band GAMA photometry \citep{Driver:2016vh, Bellstedt:2020wl, Bellstedt:2021vx}. This serves as a useful comparison data set for our new \profuse{} analysis of the same low redshift galaxies. The FS model in particular should be very similar since it was run with the same free \prospect{} parameters and only a single SED component. We should not expect the results to be identical since the \profound{} photometry used for the earlier \prospect{} analysis was not restricted by geometry, instead \profound{} uses summed flux within matched apertures that attempt to approximate total photometry via curve of growth convergence in resolved bands and PSF photometry otherwise. The other major difference is for \profuse{} we are only using 9-band optical-NIR data (at least in this initial work), whilst \prospect{} used the full UV-FIR available to GAMA.

\begin{figure}
 \includegraphics[width=\columnwidth]{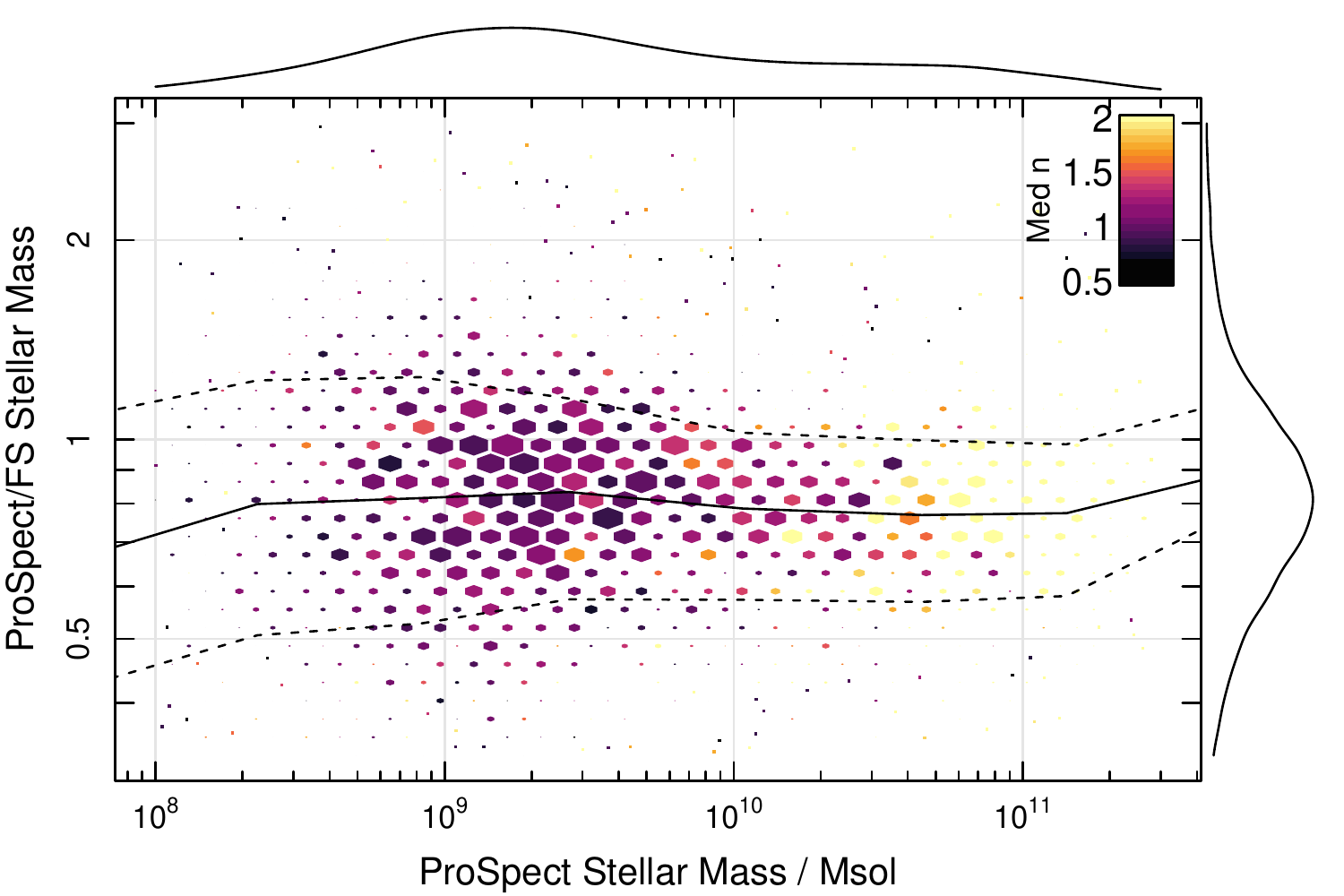}
 \caption{Comparison of \prospect{} and FS \profuse{} stellar mass formed as a function of \prospect{} stellar mass. Colouring is by \sersic{} index, where black represents disk-like profiles and yellow spheroidal-like profiles.}
 \label{fig:prospectVprofuseFS_SM}
\end{figure}

One of the more fundamental properties returned from SED fitting is the stellar mass (either formed or remaining). Figure \ref{fig:prospectVprofuseFS_SM} demonstrates the similarity of the formed stellar mass between \prospect{} and the FS \profuse{} model. They are consistent at the 0.1 dex level over the whole sample analysed here, with the FS \profuse{} model tending to find more formed stellar mass on average. Given \profuse{} is able to integrate its light profiles to arbitrarily large radii (infinity by default) it should be expected that more mass can be found. On average such profiles are able to capture more flux beyond the limited apertures extracted with \profound{}, and this is reflected in the bias we see here for lower stellar masses and is consistent with the predicted flux difference computed in the Ks band \citep[which is a reasonable single band tracer of stellar mass][]{Robotham:2020to}. There is no strong trend with \sersic{} index $n$ though (shown in colour here), which suggests we are not suffering run-away integration issues for large $n$ \citep{Kelvin:2012un}. We note that much larger discrepancies than this can be generated when switching between stellar population models \citep[e.g.][]{Bruzual:2003wd, Eldridge:2009tn} using the same SFH and ZH (see Thorne et al in prep.).			

\begin{figure}
 \includegraphics[width=\columnwidth]{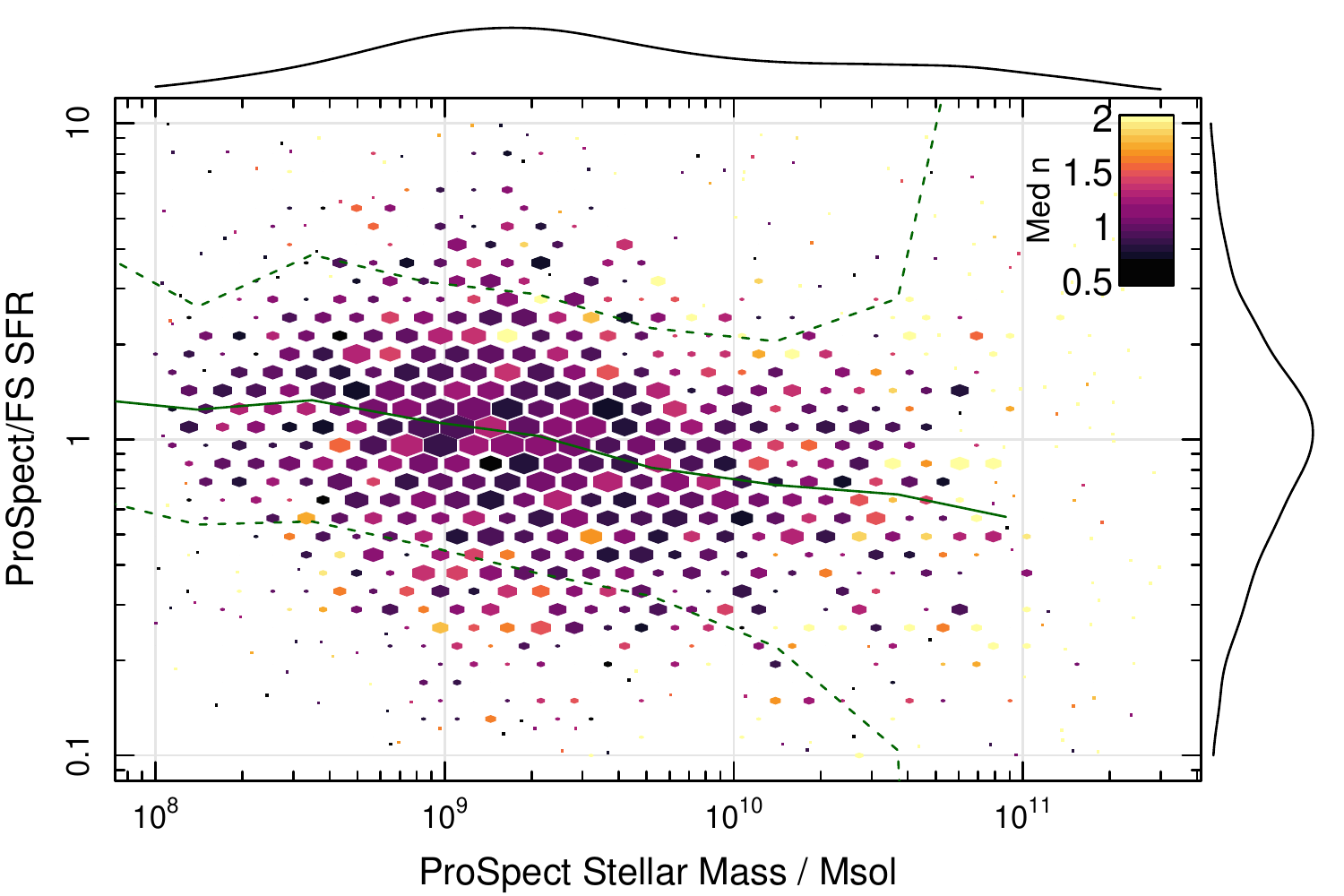}
 \caption{Comparison of \prospect{} and FS \profuse{} SFR as a function of \prospect{} stellar mass. Colouring is by \sersic{} index, where black represents disk-like profiles and yellow spheroidal-like profiles.}
 \label{fig:prospectVprofuseFS_SFR}
\end{figure}

\begin{figure}
 \includegraphics[width=\columnwidth]{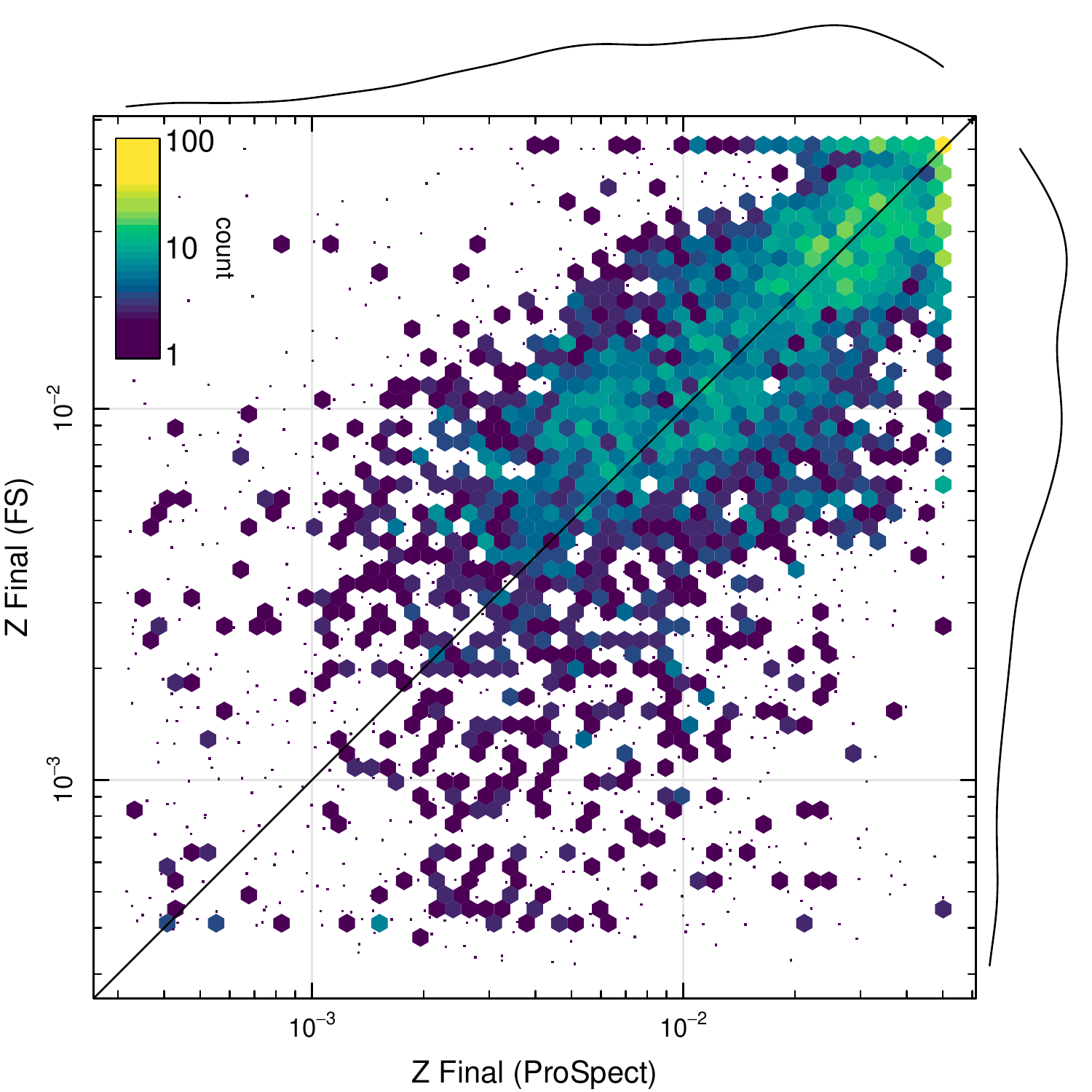}
 \caption{Comparison of \prospect{} and FS \profuse{} final gas phase metallicity ($Z_{\text{final}}$). The density of points is reflected by the Viridis colour scale shown.}
 \label{fig:prospectVprofuseFS_Z}
\end{figure}

Two other key outputs of interest from \prospect{} (and therefore also available in \profuse) are the recent SFR (the star formation rate over the last 100 Myrs) and the final gas phase metallicity ($Z_{\text{final}}$). Figure \ref{fig:prospectVprofuseFS_SFR} presents the comparison of star formation rates, whilst Figure \ref{fig:prospectVprofuseFS_Z} presents the comparison of final gas phase metallicities.

The star formation rate displays more bias and dependence on the stellar mass than the stellar mass differences presented above. Given the \profuse{} model does not use the UV and MIR/FIR as observational constraints this is not entirely surprising- without these data the recent star formation history has much more flexibility. The spread of the data noticeably increases for the most massive galaxies which are also the ones that have relatively little star formation (on average) hence small differences between \prospect{} and \profuse{} becoming {\it relatively} large in this regime.

The $Z_{\text{final}}$ metallicity shows the least consistency between \prospect{} and \profuse, but given the difficultly in inferring gas phase metallicity from broad band data the agreement is subjectively reasonable.

\subsection{Comparison Between \profuse{} Models}

\begin{figure}
 \includegraphics[width=\columnwidth]{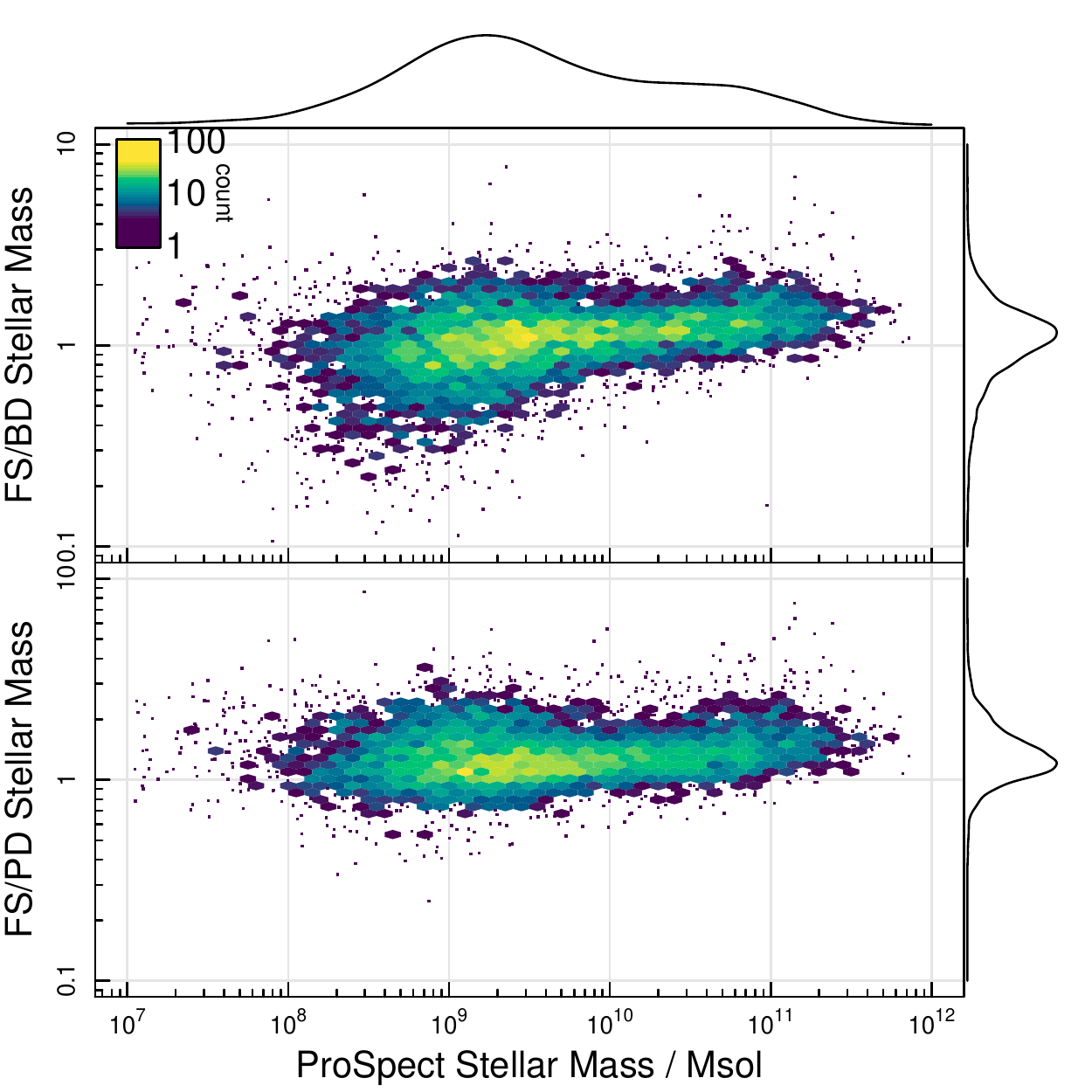}
 \caption{Comparison of stellar masses for our FS, BD and PD models compared as a function of the \prospect{} stellar mass recovered in \citet{Bellstedt:2020vq}.}
 \label{fig:SM_FSvBDvPD}
\end{figure}

\begin{figure}
 \includegraphics[width=\columnwidth]{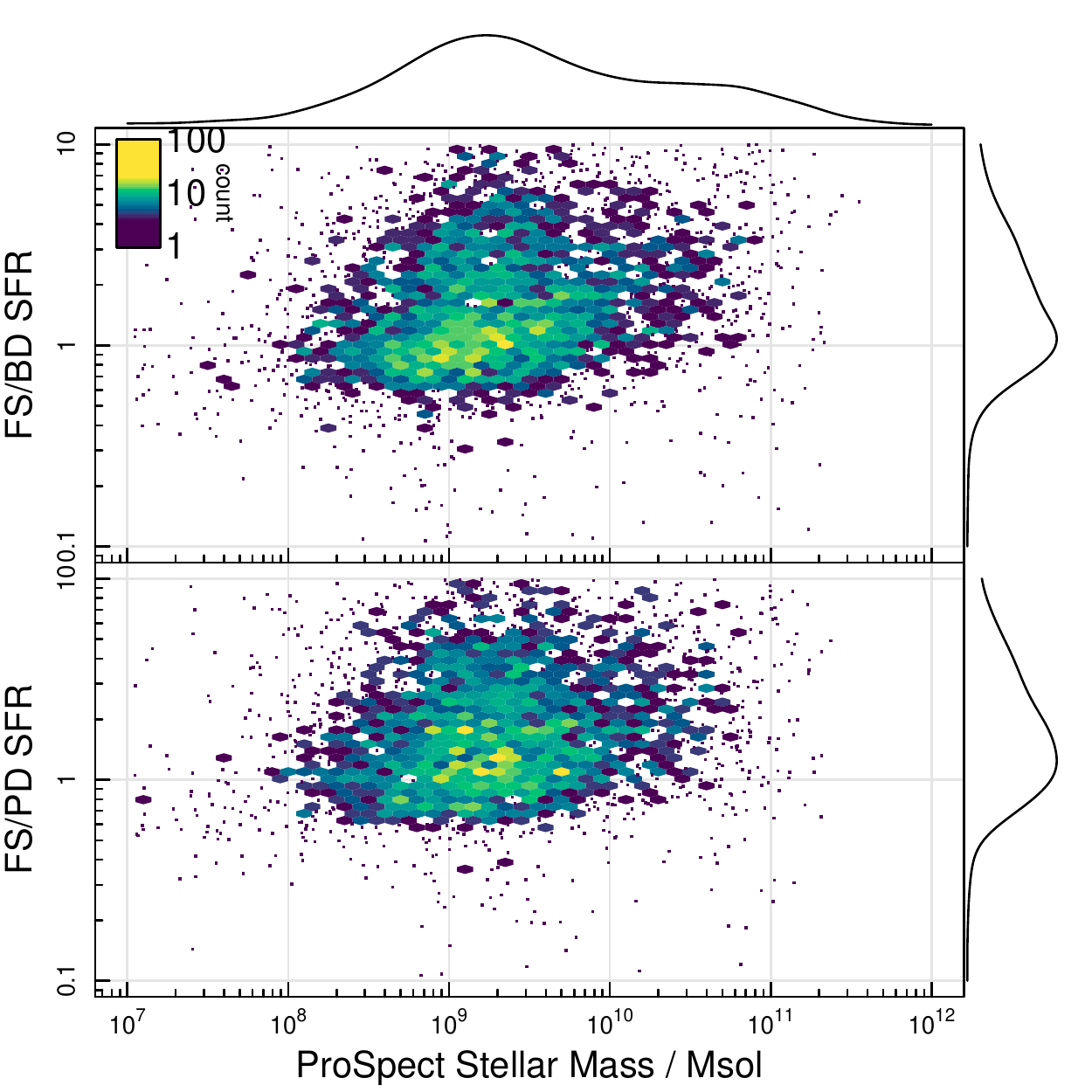}
 \caption{Comparison of star formation rates for our FS, BD and PD models compared as a function of the \prospect{} stellar mass recovered in \citet{Bellstedt:2020vq}.}
 \label{fig:SFR_FSvBDvPD}
\end{figure}

With the three different \profuse{} models we naturally expect to see differences in the inferred parameters, but in general we should hope we find consistency in the physical parameters we extract and compute. Two of the most fundamental parameters we extract from the \prospect{} side of \profuse{} are the stellar mass and star formation rate. Figure \ref{fig:SM_FSvBDvPD} demonstrates the consistency we see in the inferred stellar mass formed as a function of the \prospect{} stellar mass formed (so this is consistent for each panel), and Figure \ref{fig:SFR_FSvBDvPD} demonstrates the SFR agreement for the different models.

In general the stellar mass formed agrees well (consistent with the above results comparing the FS model to \prospect). The main trend we see is that the BD model recovers more stellar mass for lower stellar mass galaxies, and less for more massive galaxies. The PD model recovers the least stellar mass, which should not be surprising given the strong geometric constraints imposed by the model (a purely PSF bulge and exponential disk). We see larger dispersion with the SFR results and more complex behaviour. Whilst the FS model has larger outliers in SFR compared to the PD/BD models, the latter tend to systematically find larger star formation rates thanks to the separate disk model.

It is a bit more complex to directly compare the gas-phase metallicity of a two component system to a single component one since we have to make some assumptions about the gas enrichment model, but a reasonable approximation can be achieved as follows. For any system the gas-phase metallicity can be defined as mass in metals divided by the mass in gas

$$
Z_{\text{gas}} = \frac{M_{\text{metal}}}{M_{\text{gas}}},
$$

where for the linear metallicity mapping used in \prospect{} we will form a constant amount of metals per unit stellar mass formed (this is implicit in the model). This means

$$
M_{\text{metal}} = \alpha M_{\text{star}}
$$

where $\alpha$ is an arbitrary scaling of the stars formed into metals (the actual value of this will not matter, as we see in later working). It is therefore clear that

$$
M_{\text{gas}} = \frac{\alpha M_{\text{star}}}{Z_{\text{gas}}}.
$$

When summing components to find the average gas-phase metallicity we compute

\begin{eqnarray*}
\bar{Z}_{\text{gas}} 	= \frac{M_{\text{metal}_1} + M_{\text{metal}_2}}{M_{\text{gas}_1} + M_{\text{gas}_2}} \\
			= \frac{\alpha M_{\text{star}_1} + \alpha M_{\text{star}_1}}{\frac{\alpha M_{\text{star}_1}}{Z_{\text{gas}_1}} + \frac{\alpha M_{\text{star}_2}}{Z_{\text{gas}_2}}}.
\end{eqnarray*}

The $\alpha$ divides out, leading us to

$$
\bar{Z}_{\text{gas}} = \frac{M_{\text{star}_1} + M_{\text{star}_1}}{\frac{M_{\text{star}_1}}{Z_{\text{gas}_1}} + \frac{M_{\text{star}_2}}{Z_{\text{gas}_2}}}.
$$

\begin{figure}
 \includegraphics[width=\columnwidth]{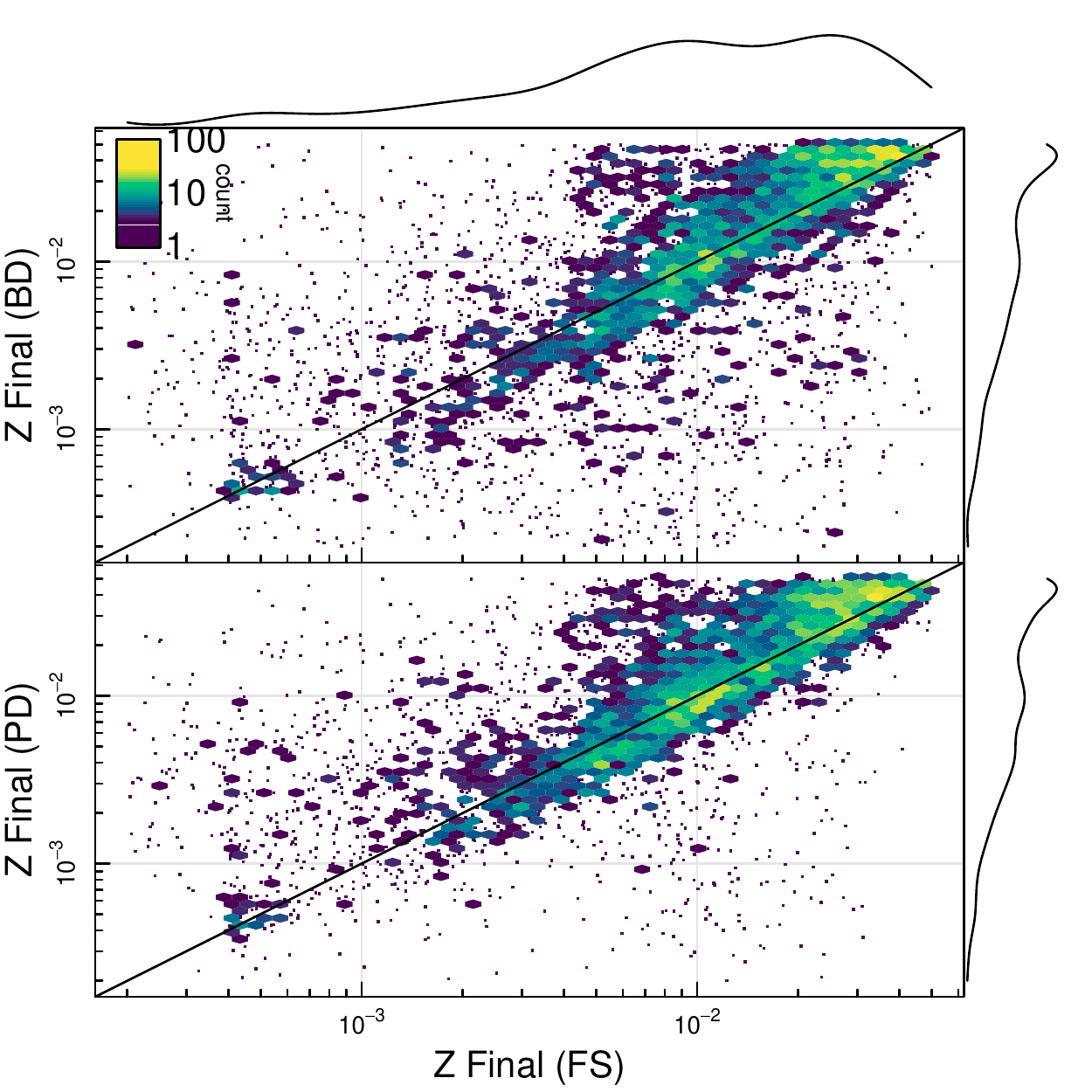}
 \caption{Comparison of metallicity for our FS, BD and PD models.}
 \label{fig:Z_FSvBDvPD}
\end{figure}

$M_{\text{star}}$ and $Z_{\text{gas}}$ are naturally computed by \prospect, making a global $\bar{Z}_{\text{gas}}$ derivable for our two component BD and PD \profuse{} models. Figure \ref{fig:Z_FSvBDvPD} shows the agreement between $Z_{\text{final}}$ for the three \profuse{} models, where $Z_{\text{final}}$ is simply $Z_{\text{gas}}$ at the epoch of observations. Given the difficulty in computing $Z_{\text{final}}$ the level of agreement is encouraging. The BD model appears to show the most disagreement compared to the FS model

\subsection{Selecting the Best Model}

Model selection is a complex problem, and is formally speaking unsolved except in the most trivial cases in Bayesian analysis. The reasoning is simple: except in the regime where one of your models is the correct interpretation of reality, classic goodness-of-fit metrics cannot be blindly used to select preferred models. Since almost all astronomy models are in detail `wrong' (which is not to say they are not useful) this means in practical terms human insight is usually required to define preferred models \citep[e.g.][]{Allen:2006vs, Cook:2019wa}.

\begin{figure}
 \includegraphics[width=\columnwidth]{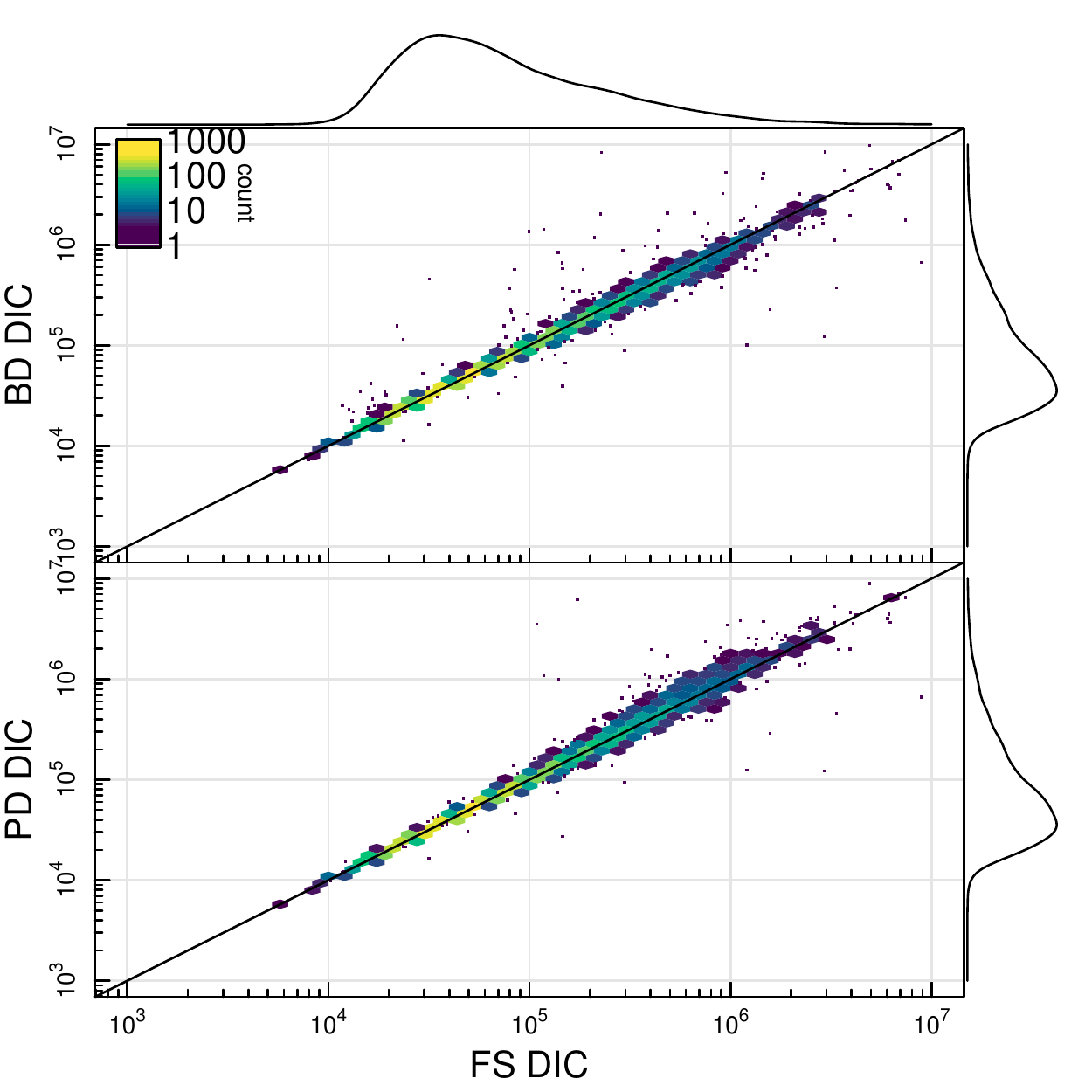}
 \caption{Comparison of the Deviance Information Criterion (DIC) for our FS, BD and PD models.}
 \label{fig:DIC_FSvBDvPD}
\end{figure}

It is still instructive to compare some traditional goodness-of-fits metrics to investigate the similarity in fitting quality we tend to find with \profuse. Figures \ref{fig:RedChi_FSvBDvPD} and \ref{fig:DIC_FSvBDvPD} show the reduced $\chi^2$ ($\chi^2/\nu$, where $\nu$ is the degrees-of-freedom) and the Deviance Information Criteria (DIC, the log-likelihood of the fit modified by the effective degrees-of-freedom) respectively. In both of these figures the lower value is generally considered to be the preferred one, with the caveat we should not prefer $\chi^2/\nu$ much less than 1 since this would imply over-fitting of the data. The heaviest concentration of $\chi^2/\nu$ is seen at 1 for all three models, with the longer tail stretching to larger values that implies under-fitting (i.e.\ the model needs to be more complex to achieve a reasonable fit to the data). The $z$ scaling is logarithmic, so the vast majority of all 3 models are clustered near 1 (as can be seen in the PDF projections surrounding the main panels), which is interesting since it suggests there are many situations where all 3 models (despite being quite dissimilar in design) do a reasonable job of representing the data.

In general we see a marginal preference on average for the BD model followed by the FS and lastly the PD. This broadly follows the known flexibility of the model, but given the DIC in particular penalises the model complexity it suggests the BD model often offers a notable improvement in the fitting even accounting for the increased flexibility.

To determine the preferred model for a particular galaxy we use a combination of the above fitting preferences using the DIC for the different components in combination with the removal of models with extremely faint bulges (where they provide less than 1\% of the stellar mass) or stellar mass dominant bulges (more than 50\% of the stellar mass). On this GAMA sample we find 68.3\% of galaxies prefer a FS model, 15.3\% BD and 16.4\% PD.

\begin{figure}
 \includegraphics[width=\columnwidth]{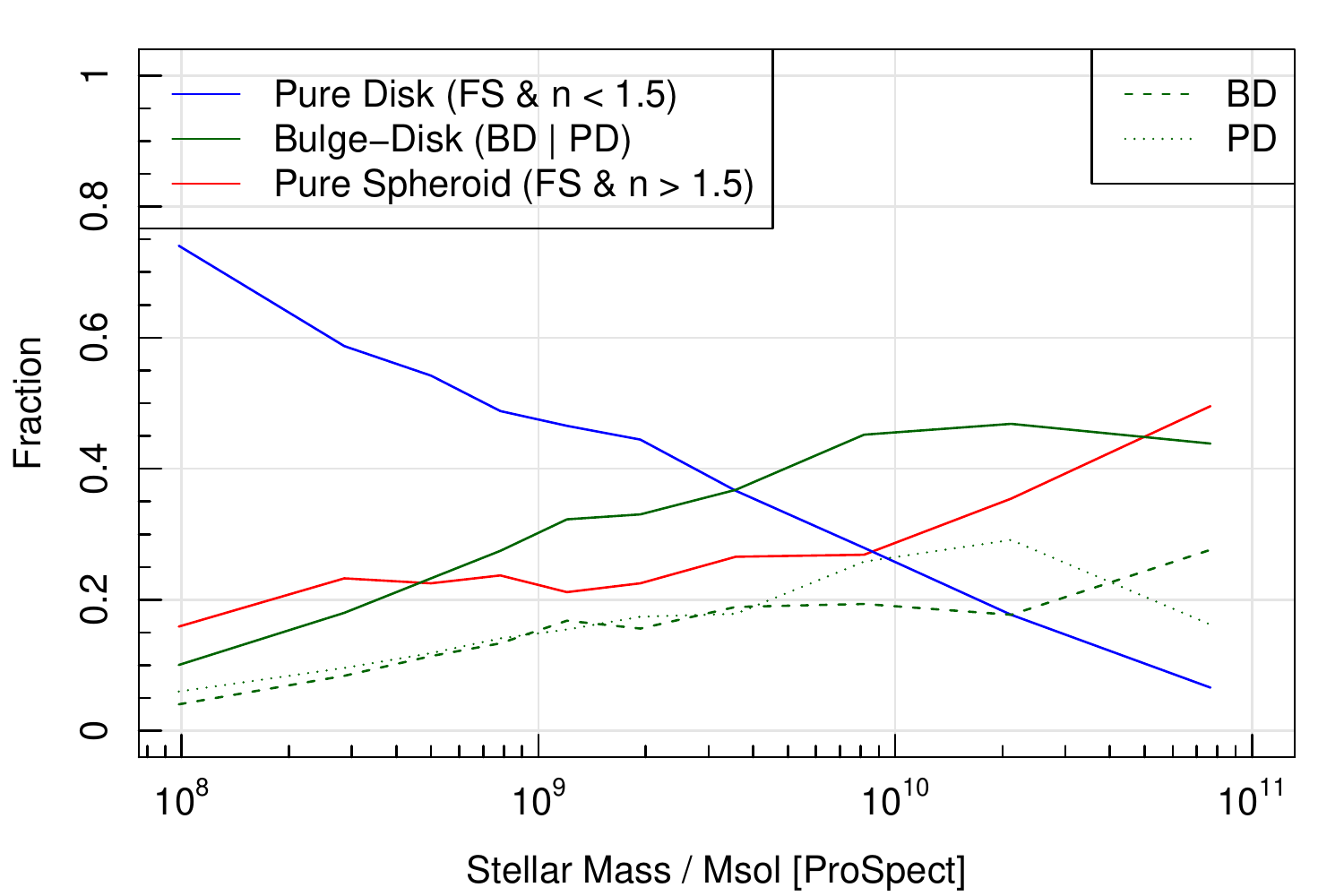}
 \caption{Comparison of model selection preferences as a function of stellar mass. Pure disks dominate for lower stellar mass galaxies, and pure spheroid galaxies (ellipticals) dominate at the massive end.}
 \label{fig:ModelFractions}
\end{figure}

Figure \ref{fig:ModelFractions} shows the fractions for the models, with the FS model split further by \sersic{} index (where $n<1.5$ is interpreted as a pure disk and $n>1.5$ is a form of spheroidal or elliptical profile, as per \citealt{Trujillo:2006uj}). As might be expected, the pure disk fraction drops steadily with stellar mass and galaxies requiring bulge-disk models peaks at around $M^*$. In comparison, spheroidal-like galaxies become more common for more massive galaxies, representing $\sim30$\% of galaxies at $10^{11}$\msol. Note adjusting the pure disk $n<1.5$ threshold does adjust these fractions, but the distribution is quite smooth and the general trends seen in Figure \ref{fig:ModelFractions} are preserved.

\subsubsection{Comparison to Previous GAMA Structural Fits}

GAMA has produced a number of structural fitting outputs over more than a decade. Given differences in setups, not all can be easily compared to the new \profuse{} fits, in particular we are directly creating stellar mass $R_e$ rather than band specific light $R_e$. To date two free \sersic{} (FS) models have already been published and released publicly: \citet{Kelvin:2012un} and Casura et al. (in prep.). The former of these was run on a mixture of Sloan Digitial Sky Survey optical images \citep[SDSS;][]{Abazajian:2009tr} and VISTA/VIKING NIR images using {\sc GALFIT} (specifically a pipeline named {\sc SIGMA}), and the latter KiDS $gri$ bands using \profit. Because we fit in stellar mass space directly, in general you might expect the redder bands available to be the most comparable to \profuse{} \citep{Robotham:2020to}, however there is also a tradeoff in depth and PSF quality \citep[see][]{Driver:2016vh} meaning we need to balance the comparison. To compare to \citet{Kelvin:2012un} we selected the Z-band, which is the deepest of the VISTA/VIKING data. To compare to Casura et al. (in prep.) we selected the $i$-band, which is the reddest of the three bands analysed, and quite deep with a good PSF in general.

\begin{figure}
 \includegraphics[width=\columnwidth]{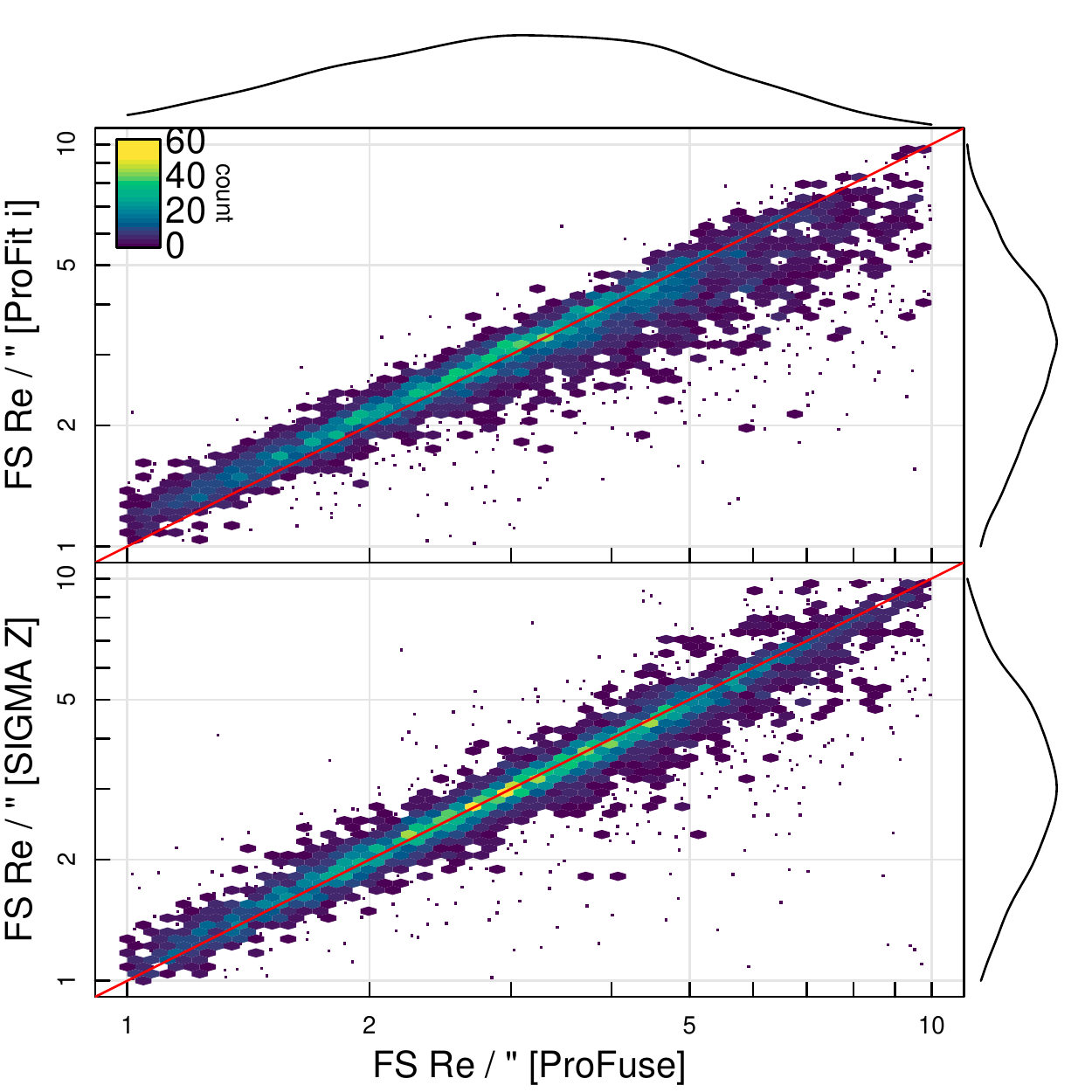}
 \caption{Comparison of $Re$ between \profuse{}, {\sc GALFIT} (aka {\sc SIGMA} in this work) and \profit. In general all three agree quite well, with {\sc SIGMA} agreeing better with \profuse{} for the most compact sources.}
 \label{fig:CompareRe}
\end{figure}

Figure \ref{fig:CompareRe} compares the FS model $R_e$ for the three models discussed above. In general all are quite consistent, with perhaps the most compact sources being more similar between \profuse{} and {\sc SIGMA} (\profit{} tends to find slightly larger sources in this regime). The reason for this shift could be the PSF quality (including the ability to model it), and perhaps a real physical shift due to galaxies appearing larger in bluer bands \citep{Kelvin:2012un}. The treatment of the background is also different between all three efforts, and this can have a marked impact on the fitted parameters, in particular the size and \sersic{} index \citep{Kelvin:2012un}. In general the level of agreement here suggests no pathologically worrying behaviour in any of the three analyses.

\begin{figure}
 \includegraphics[width=\columnwidth]{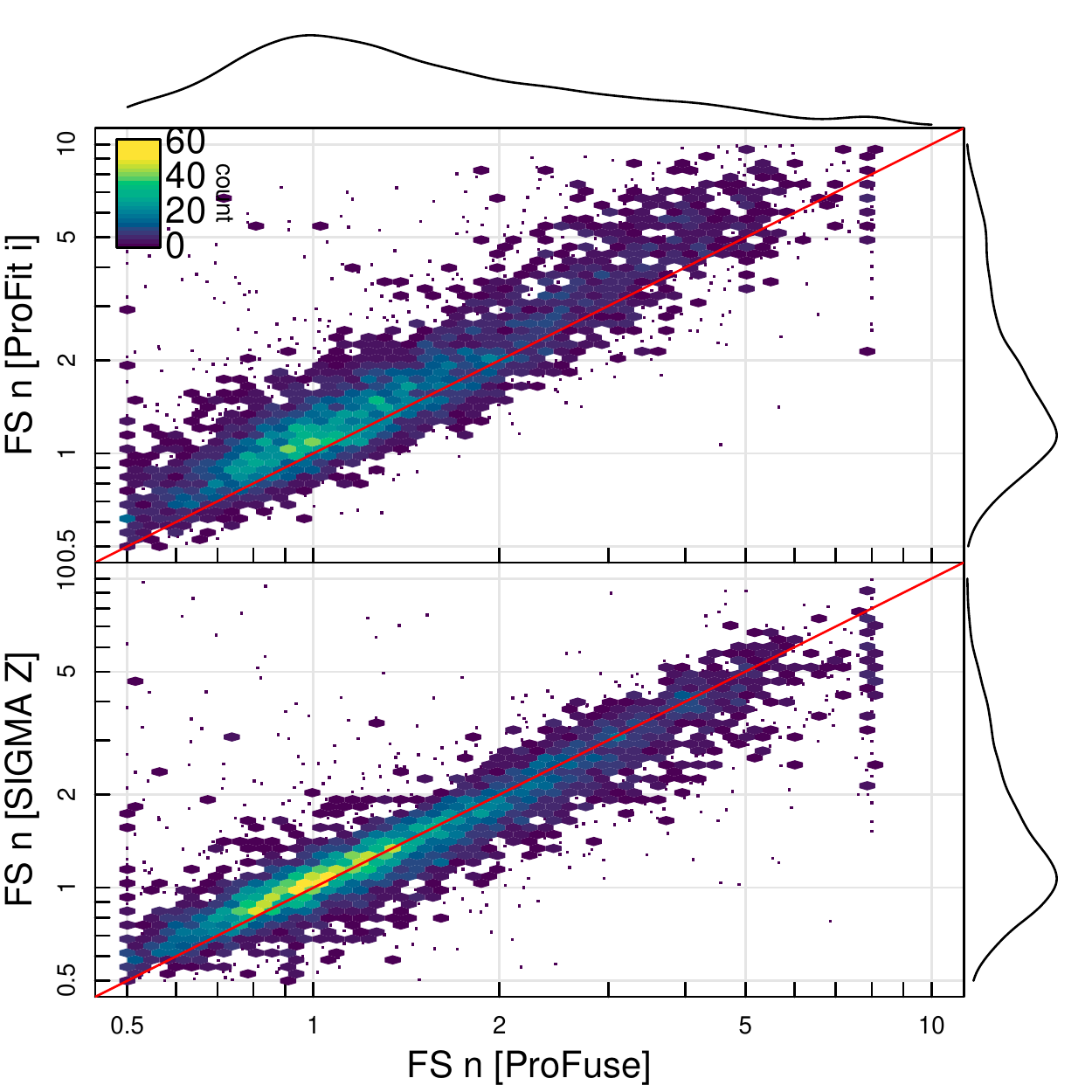}
 \caption{Comparison of \sersic{} index ($n$) between \profuse{}, {\sc GALFIT} (aka {\sc SIGMA} in this work) and \profit. {\sc SIGMA} agrees better with \profuse{} in general, with the \profit{} $i$-band fits tending to have systematically higher \sersic{} index ($n$).}
\label{fig:CompareNser}
\end{figure}

Figure \ref{fig:CompareNser} compares the FS model \sersic{} index ($n$) for the three models discussed above. The \sersic{} index is one of the harder parameters to infer consistently because of its dependency on both the PSF modelling in the core and the sky subtraction in the wings. For this reason we might expect to see poorer agreement between all three models. That caveat aside, we clearly see highly correlated behaviour between all three analyses. The \profit{} fits do show more spread and a systematic trend towards larger values of $n$, which could also be related to the slightly larger sizes seen in Figure \ref{fig:CompareRe}. The agreement with {\sc SIGMA} is very encouraging, and suggests the Z-band is a reasonable predictor of stellar mass structural parameters in the absence of the full generative modelling executed with \profuse.

\subsubsection{Comparison to GAMA Morphological Classifications}

GAMA has its own human-constructed catalogue of morphological classifications as described in GAMA Date Release 4 (Driver et al., in press), referred to as {\sc MorphGAMA} from here. The most meaningful cross comparison that can be made unambiguously from that earlier work is to compare our preferred FS models with the `elliptical' classes therein. Table \ref{tab:compare} presents the main results of the two way classifications between these approaches.

\begin{table}
\begin{center}
\begin{tabular}{@{}|c|c|c|c}
 & \profuse{} FS & \profuse{} BD | PD & Total \\
 \hline
 Elliptical & 387 (\textcolor{red}{81.6\%}) [8.5\%] & 87 (18.4\%) [4.1\%] & (474)\\
 Not Elliptical & 4,167 (67.3\%) [91.5\%] & 2,023 (32.7\%) [\textcolor{red}{95.9\%}] & (6,190)\\
 Total & [4,554] & [2,110] & \\
\end{tabular}
\end{center}
\caption{Two way comparison table of our preferred \profuse{} models and the `elliptical' morphological classifications. The percentages in parentheses correspond to the different normalising totals. The two percentages that we would want to be as close to 100\% as possible (since the cross associations are unambiguous) are highlighted in red.}
\label{tab:compare}
\end{table}%

Notably the {\sc MorphGAMA} `ellipticals' are selected as a \profuse{} FS model with high fidelity (81.6\% of `ellipticals' are flagged as having a preferred FS model, coloured red for clarity). The BD and PD preferred models hugely prefer the `not elliptical' class in {\sc MorphGAMA} (95.9\%, coloured red for clarity). The most complex overlap is where the `not ellipticals' are assigned with \profuse{}. Most reside in the FS model class (67.3\%). This might seem surprising, but this class also contains all our pure disk (or at least, extremely disk dominated) models. Of these 4,167 galaxies only 13.8\% possess \sersic{} index $n$ steeper than 2.5, which would be the regime of canonical elliptical galaxies. This means 86.2\% are quite disk-like in appearance, explaining why they might have been classed as `not elliptical' whilst also being in our FS population as disk-like models.

Overall the agreement between our automatic \profuse{} model selection and the {\sc MorphGAMA} classifications is excellent in the regime where they can be directly compared. To create internal consistency, and reduce dependence on human visual classification, we will focus on the \profuse{} model selections when creating sub classes for analysis in the remainder of this paper.

\section{Bulge and Disk Properties}
\label{sec:bulgedisk}

Investigating all aspects of the SFH and ZH decomposition of our preferred bulge-disk models is deferred to later work (Bellstedt et al.\ in prep.) but some of the higher level outcomes will be presented briefly.

Where relevant we present the bulges from our PD/BD fits, the FS fits with \sersic{} index $n>1.5$ (the latter we generically refer to as `spheroids' in Figures), and combined PD/BD disks with FS fits with \sersic{} index $n<1.5$. For completeness the BD/PD/FS separated versions of the disks can be found in Appendix \ref{app:Disk_SFR} (but we note that our disks appear to share universal properties once controlling for stellar mass, regardless of being a PD/BD or FS model).

\subsection{Star Formation Properties}

\begin{figure}
 \includegraphics[width=\columnwidth]{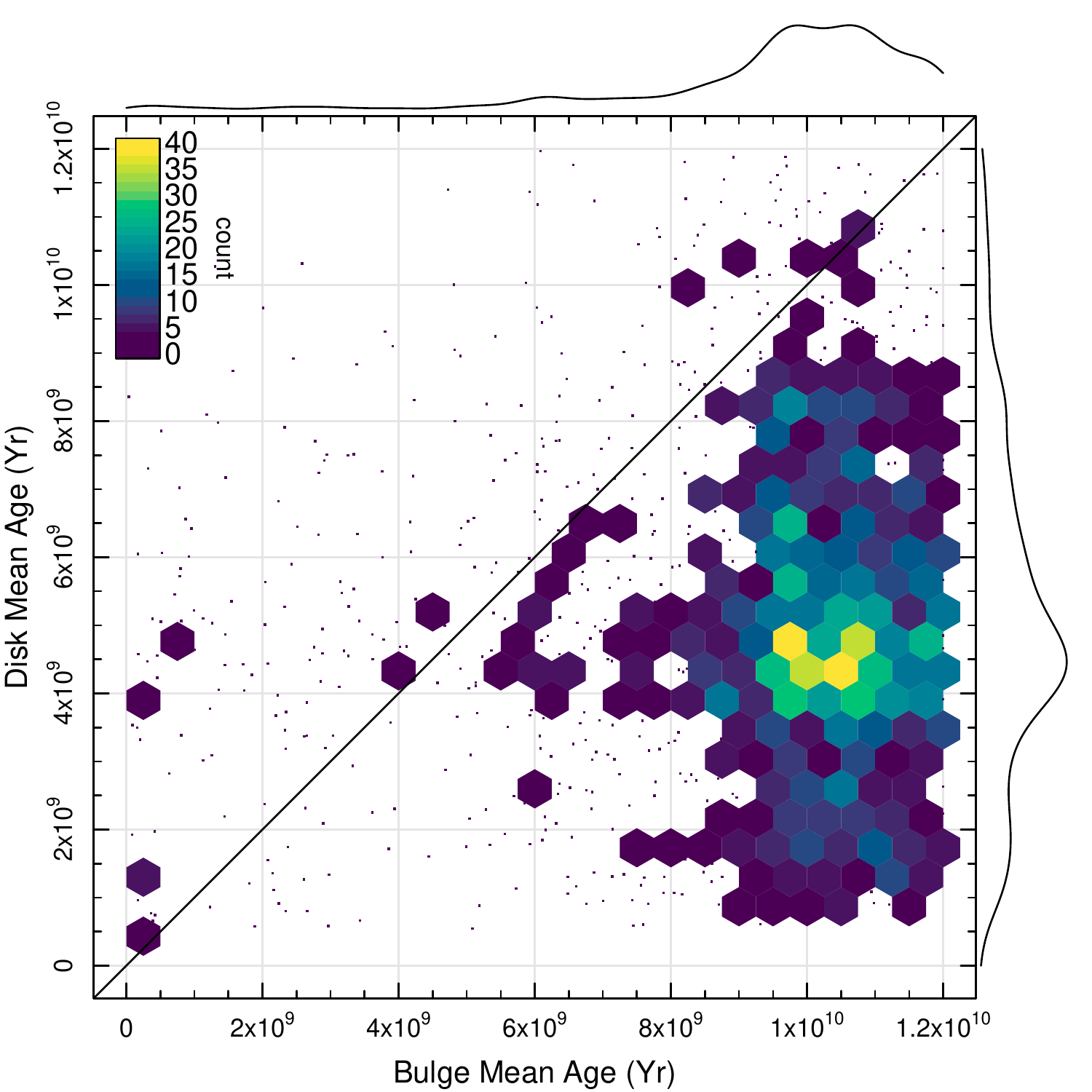}
 \caption{Mean age of bulges and disks for galaxies with BD/PD preferred models.}
 \label{fig:MeanAges}
\end{figure}

An interesting result from this initial application of \profuse{} is to compare the mean stellar ages of our preferred BD/PD model bulges and disks (where we have high confidence that the real galaxy contains both bulge and disk components). Figure \ref{fig:MeanAges} presents the direct comparisons for all 2,110 galaxies which prefer a BD/PD model, and it is immediately clear our \profuse{} disks are significantly younger in most cases (87.4\% of the examples here). This preference of disks being systematically younger than bulges agrees well with work from \citet{MacArthur:2009tw} (for a small well-resolved sample) and \citet{Fraser-McKelvie:2018wq} (for a larger but less well-resolved study). However, there have been counter examples to this general trend \citep{Johnston:2014uj, Barsanti:2021wu}, but these papers used cluster galaxy selections compared to the environmentally diverse and almost volume complete sample used here.

\begin{figure}
 \includegraphics[width=\columnwidth]{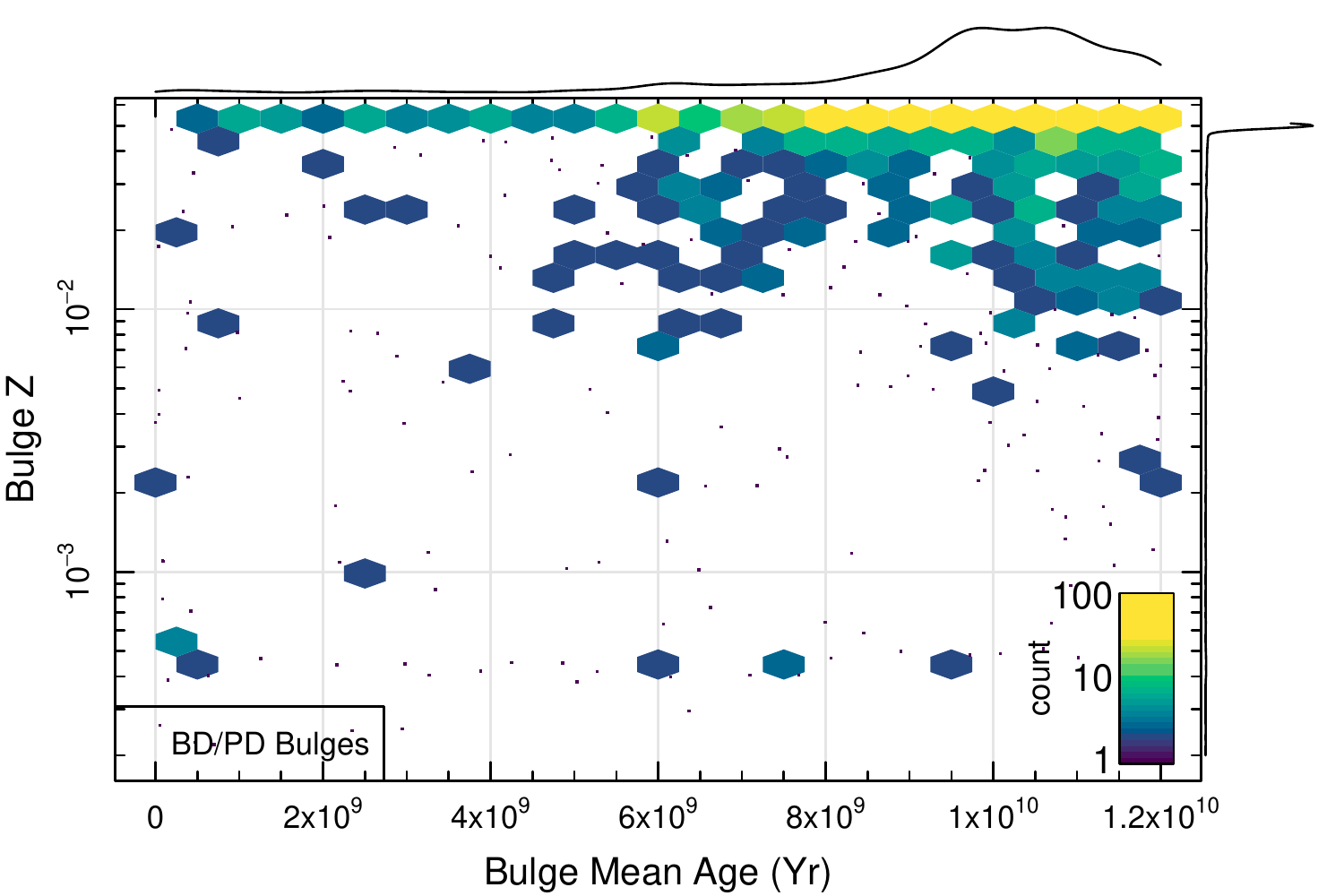} \
 \includegraphics[width=\columnwidth]{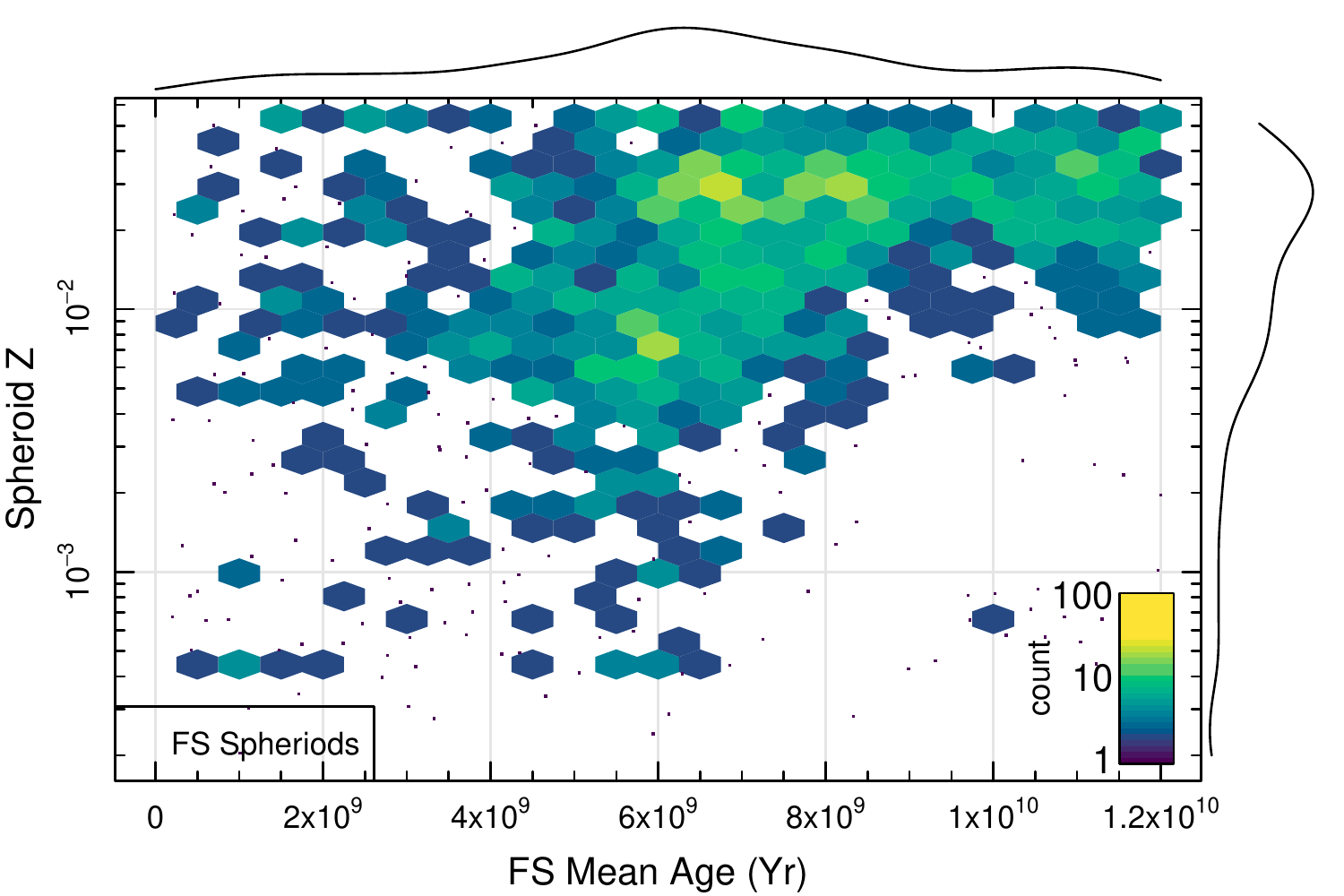} \
 \includegraphics[width=\columnwidth]{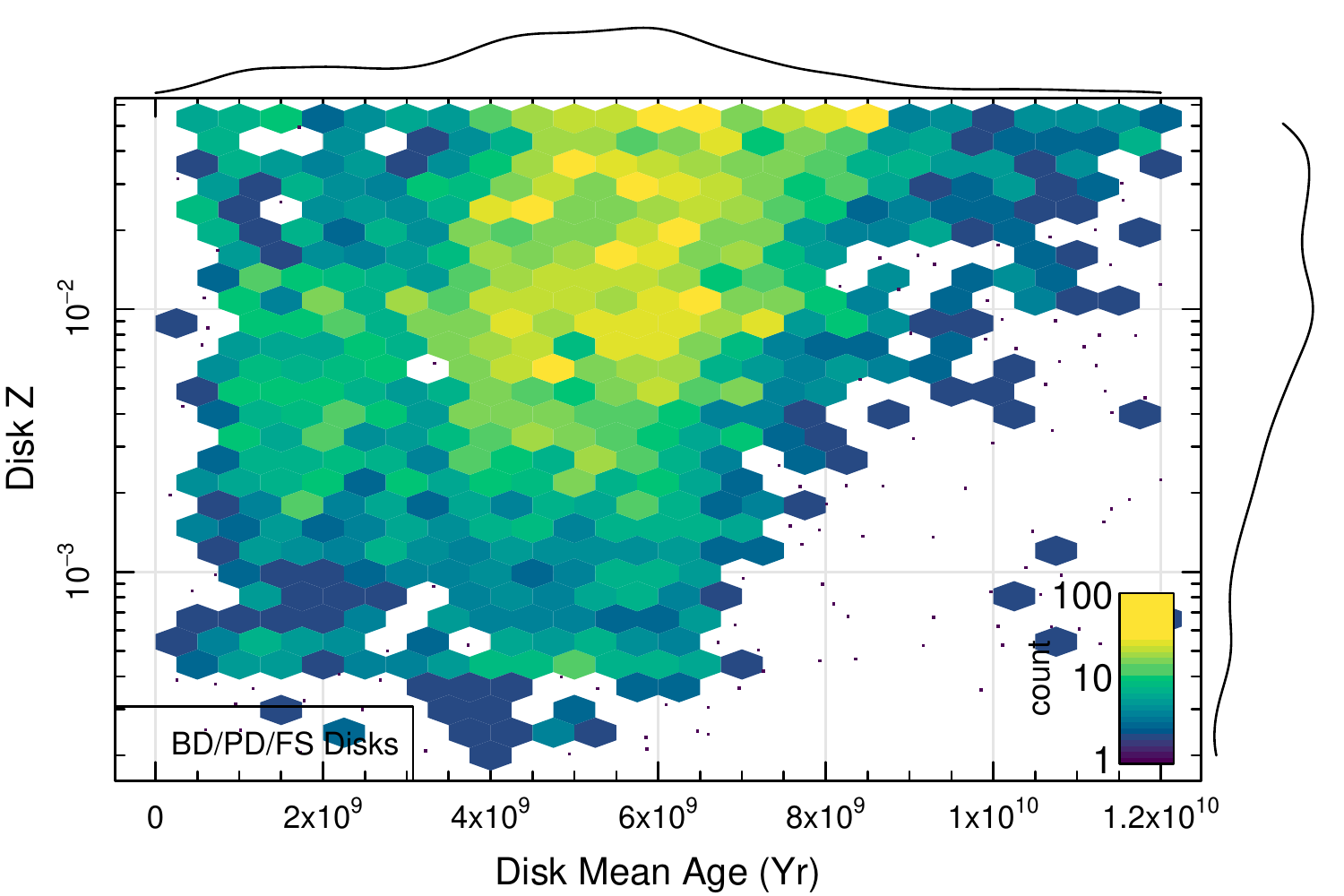}
 \caption{Component age versus metallicity. Top panel shows BD/PD bulges and middle panel FS spheroids. Our older bulges tend to also be the most metal enriched. Bottom panel shows disks. We see a much broader range of metallicities in disks compared to bulges.}
 \label{fig:AgevZ_all}
\end{figure}

Extending the comparison of ages, Figure \ref{fig:AgevZ_all} demonstrates the age versus metallicity behaviour of our BD/PD/FS combined samples. It is notable that the FS spheroids (including elliptical galaxies) have a different distribution in terms of mean age compared to bulges, showing evidence for both a young ($\sim$6Gyr) and old ($\sim$10Gyr) population. The more massive ellipticals possess broadly similar properties to the BD/PD bulges. We do see a tail of younger BD/PD bulges in less massive systems that overlaps with the younger low-mass FS spheroid population, but the FS spheroids are entirely missing ancient low-mass high-metallicity systems that exist in large number in the BD/PD bulges. Beyond these differences in stellar age, we find a strong preference for our BD/PD bulges and FS spheroids to be metal rich. Our disks inhabit a much broader distribution of metallicities, spanning down to quite low values. The few old ($\sim10$Gyr) disks that do exist possess high metallicities.

\begin{figure}
 \includegraphics[width=\columnwidth]{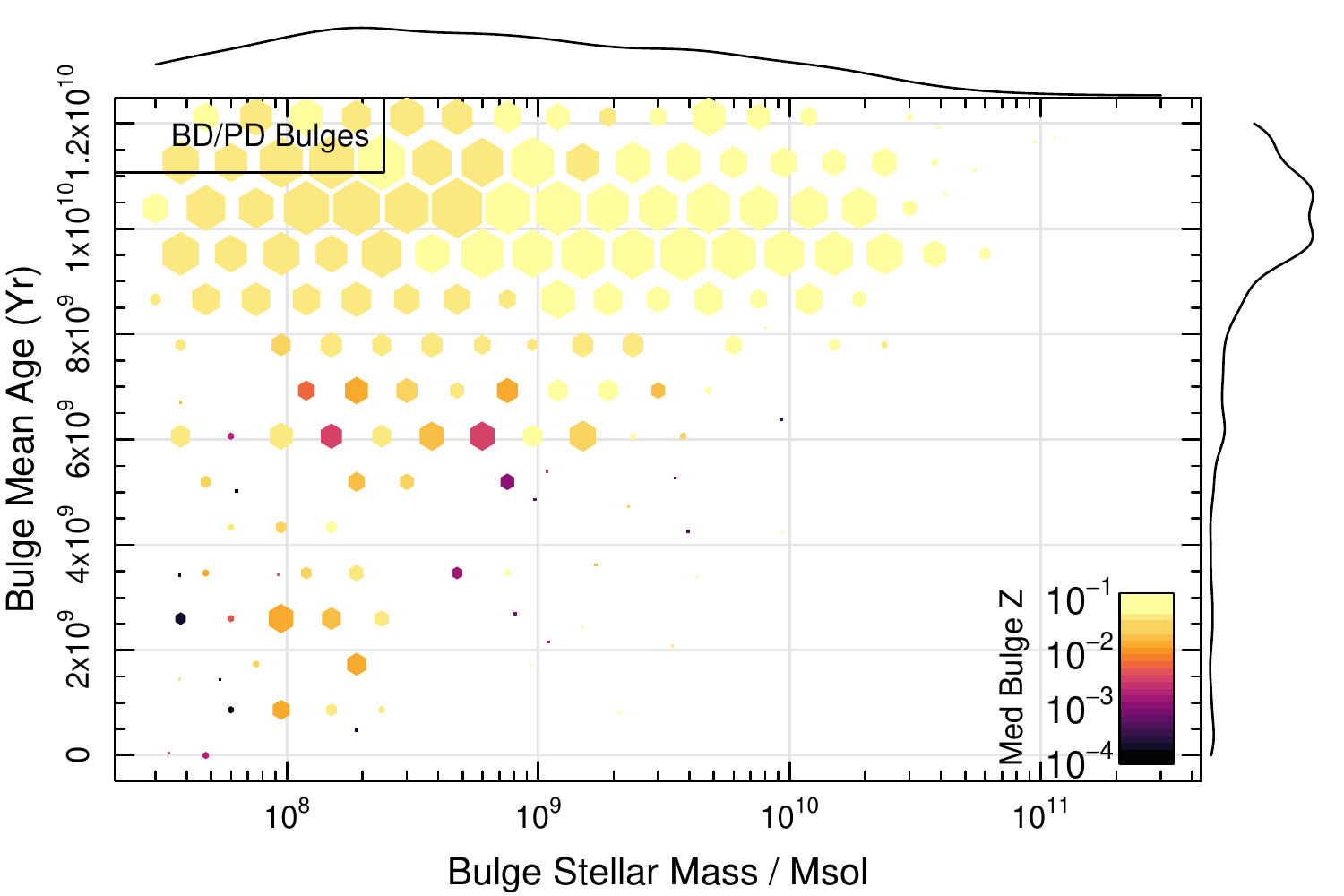} \
 \includegraphics[width=\columnwidth]{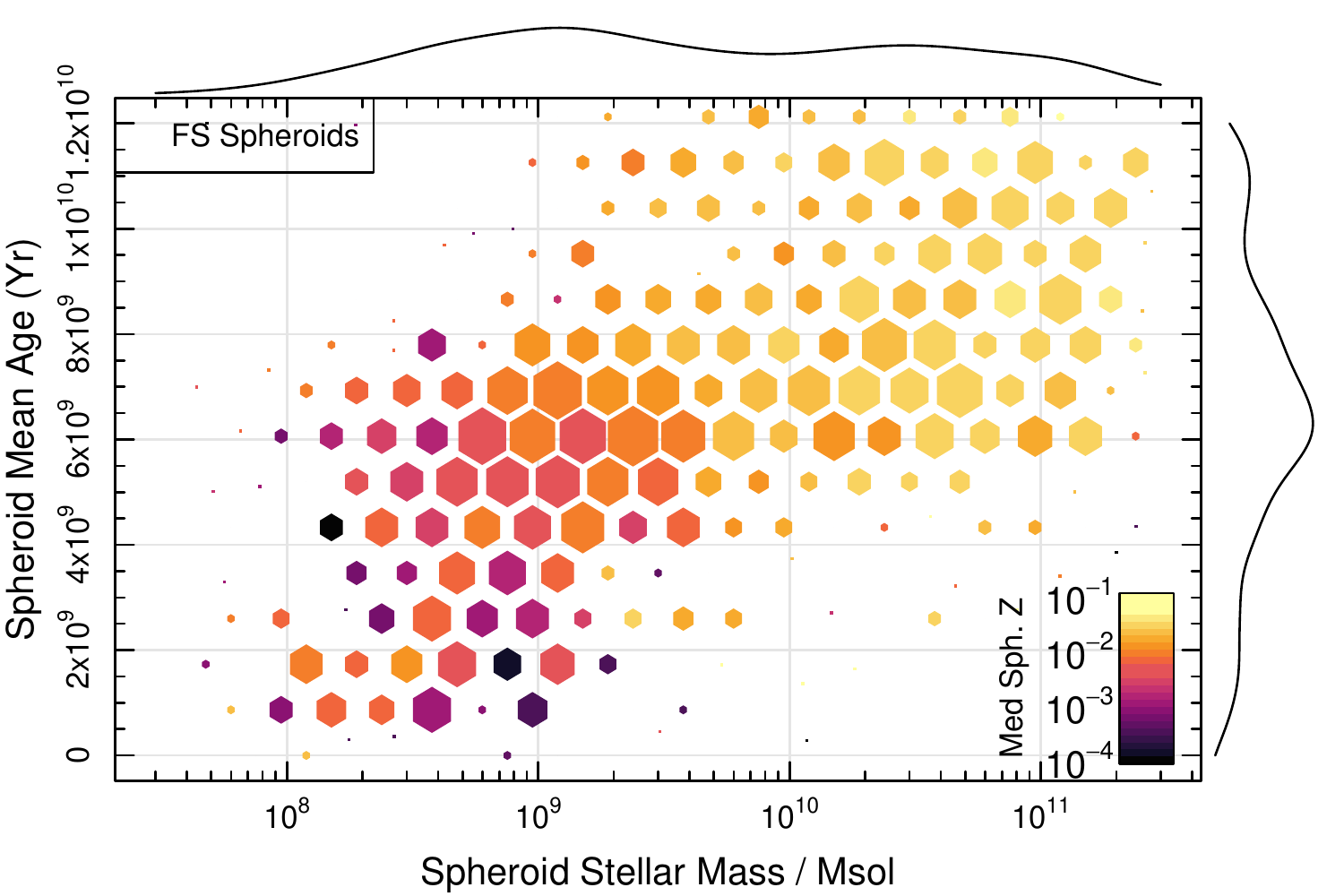} \
 \includegraphics[width=\columnwidth]{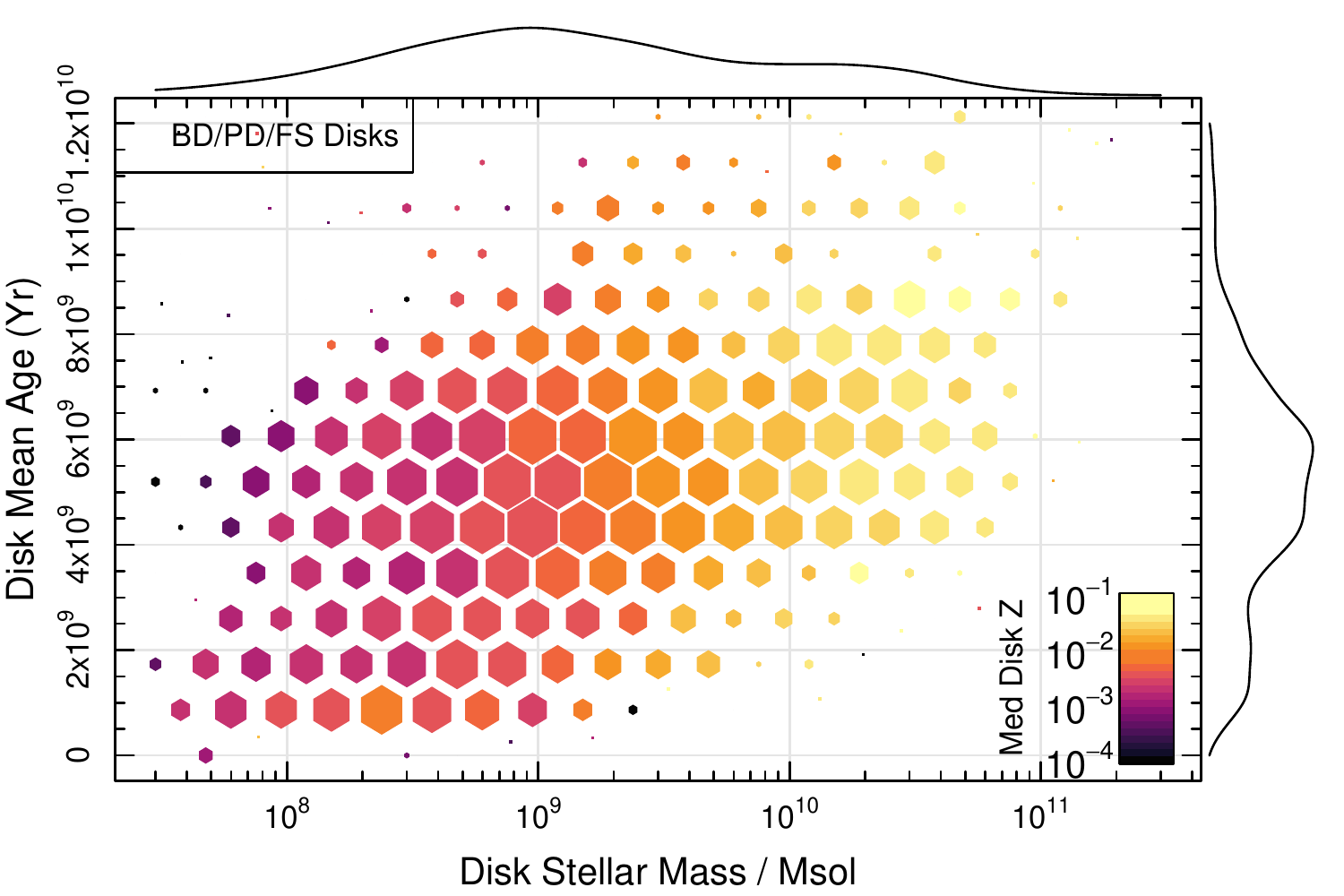}
 \caption{Component stellar mass versus age versus metallicity. Top panel shows BD/PD bulges and middle panel FS spheroids. Our bulges appear quite uni-modal in this parameter space, whilst we see a bimodal population of old metal-rich spheroids and young metal poor spheroids. Bottom panel shows disks, which show a strong mass-metallicity relationship. Size scaling reflects the logarithmic counts in cells.}
 \label{fig:SMvAgevZ_all}
\end{figure}

Delving further, Figure \ref{fig:SMvAgevZ_all} presents the component stellar mass versus mean age, coloured by the median metallicity of objects within the hexagonal cells. Our bulges are predominantly unimodal, being almost entirely old ($\sim$10 Gyr) and quite metal rich. Our FS spheroids are bimodal, having an older moderately enriched population above $10^{10}$\msol, and a distinctly younger and metal poorer population below this mass. This suggests there are strong caveats in attempting to combine bulge and spheroid components in general, and suggests quite different formation pathways for bulge-like components in the low mass regime in particular. BD/PD bulges above $10^{10}$\msol become rarer, this population being dominated by FS spheroids (canonical elliptical galaxies) that are older and more metal rich, but notably not as old and metal rich as our BD/PD bulges in general. At a given stellar mass our bulges appear to be older on average than our spheroids. The span of ages and the differences between the bulge and spheroid populations is consistent with figure 25 of \citet{Costantin:2021ui}, which spans a narrower mass range around $M^*$. The bulge population in \citet{Costantin:2021ui} has a distinct clustering between 8-10 Gyrs, as we see here.

The disks in Figure \ref{fig:SMvAgevZ_all} display quite different behaviour to the bulges and spheroids. They are younger than the bulge/spheroids (as we expect) but possess an extremely strong positive metallicity trend as a function of the component stellar mass, but as a smooth continuous function rather than a bimodality. Our most massive disks are approaching solar metallicity, whilst at $10^{8}$\msol our disks are extremely metal poor (in fact hitting our lower template limit). It is notable from Appendix \ref{app:Disk_SFR} that our BD/PD and FS disks display similar behaviour in the mass regimes where they overlap, suggesting these components have similar formation pathways regardless of the presence of a significant bulge.

\subsection{Structural Properties}

\begin{figure}
 \includegraphics[width=\columnwidth]{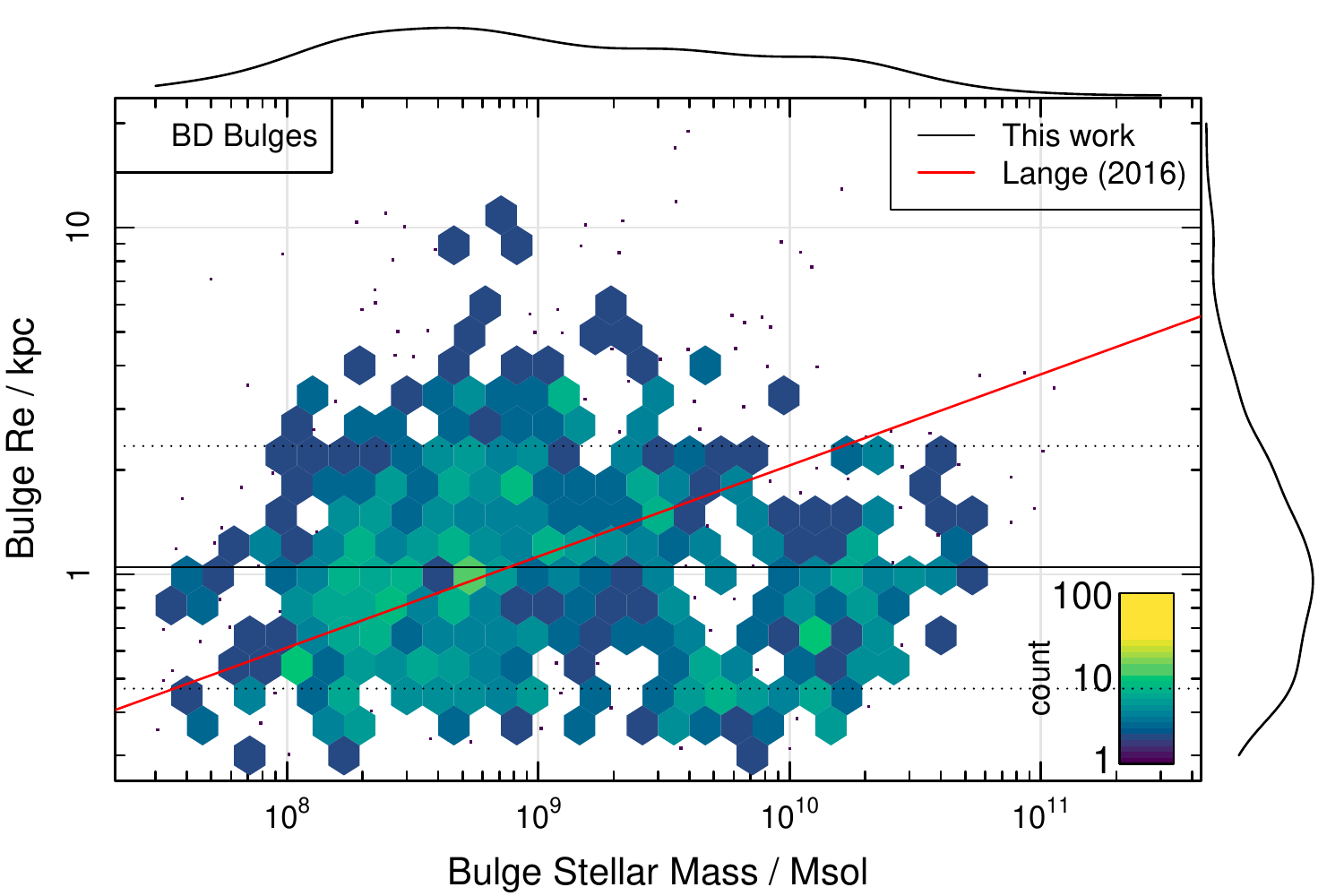} \
 \includegraphics[width=\columnwidth]{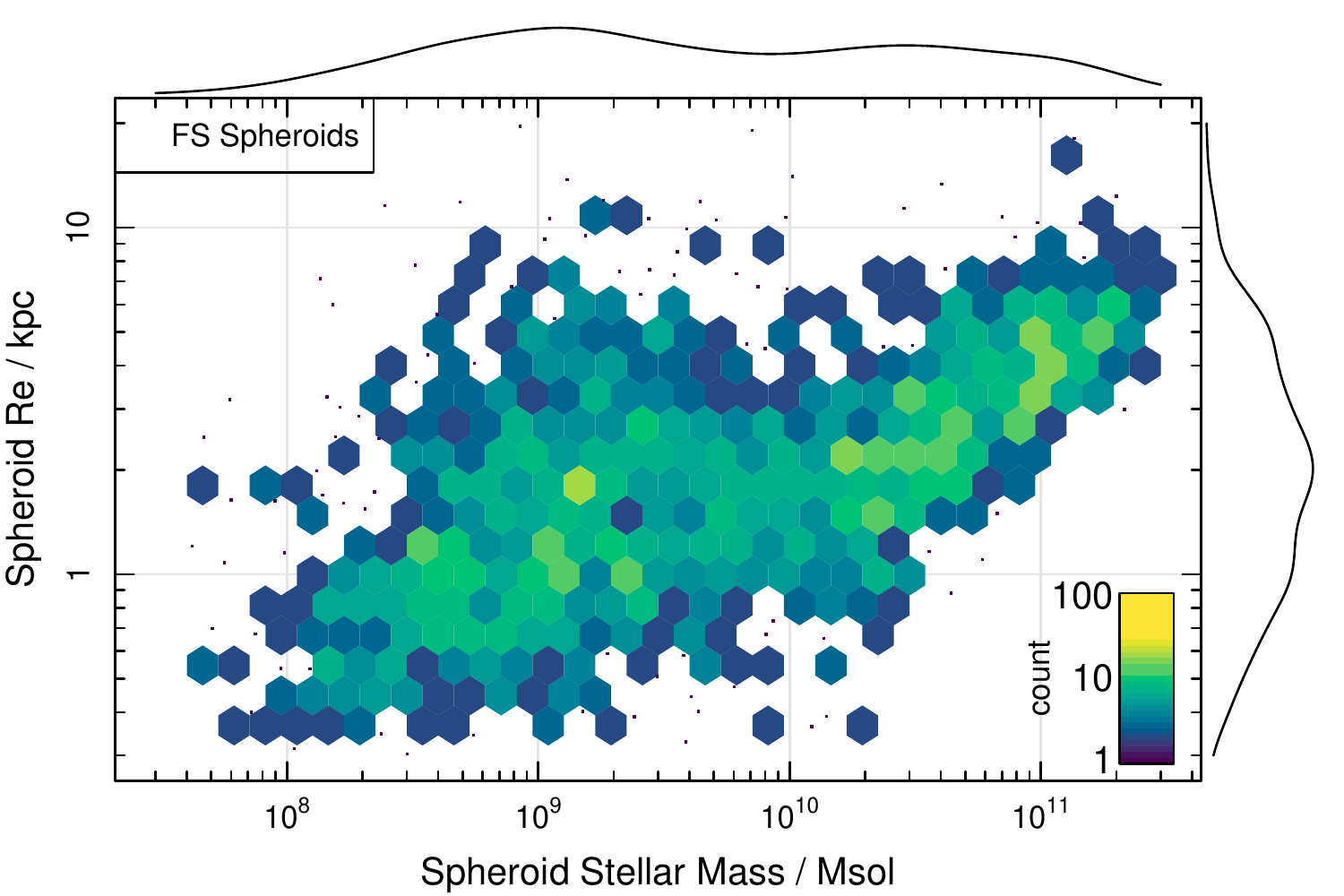} \
 \includegraphics[width=\columnwidth]{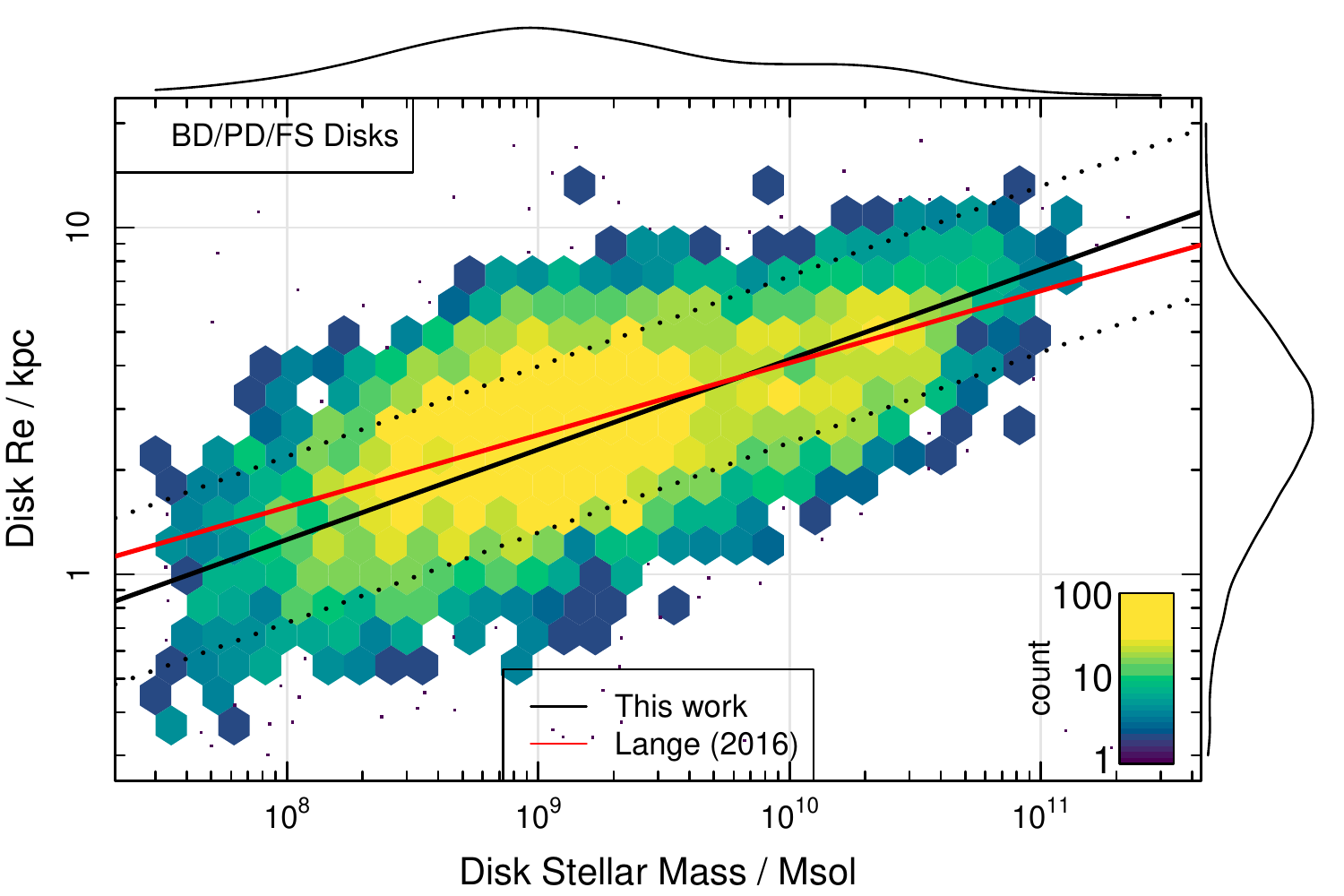}
 \caption{Component stellar mass versus size. Top panel shows BD/PD bulges and middle panel FS spheroids. We can see a distinct population of similarly sized bulges on the LHS and elliptical galaxies on the RHS of this figure. Bottom panel shows disks. We see a strong relationship between stellar mass and size across the full range of this Figure. Where relevant the size-mass relationship found in \ref{sec:SMfit} are shown as black lines (dotted indicates the intrinsic scatter), and the previous work of \citet{Lange:2016wx} is shown as a red line.}
 \label{fig:SMvRe_all}
\end{figure}

Figure \ref{fig:SMvRe_all} presents the main results for our bulge-like components (top and middle panels) and disks (bottom panel). The bulge-like components have two distinct features: at lower mass near constant size ($\sim 1$kpc) below $10^{10}$\msol, and a distinct turn up in size for the spheroids above $10^{10}$\msol. The lower mass components are dominated by true bulges and compact single component spheroids, while the more massive bulge-like components are almost entirely massive elliptical galaxies. These two populations are also evident in \citet{Lange:2016wx}, where in figure 9 sub-panel $d$ the `elliptical' class of galaxies shows a distinct upturn at the high mass extreme. Arguably the constant-size compact bulges are also discernible in figure 9 sub-panel $i$ (although in this work it was still fit with a positive slope regression, as discussed later). 

The disk components in Figure \ref{fig:SMvRe_all} are extremely consistent with the various disk populations identified in \citet{Lange:2016wx} figure 9, spanning sizes of $\sim 1$kpc at $10^{8}$\msol, and $\sim 5$kpc at $10^{11}$\msol. In our \profuse{} decompositions the apparent spread in the disk sizes reduces considerably between our smallest ($10^{8}$\msol) and largest ($3\times 10^{11}$\msol) disks: 0.38 dex and 0.15 dex respectively. Appendix \ref{app:Disk_Size} shows the PD/BD and FS disk populations significantly overlap, and predominantly describe the same locus of mass--size.

In terms of the structural decomposition, \profuse{} points towards a single consistent disk component (regardless of whether it possesses a significant bulge or not) and two classes of bulge-like populations: compact low mass systems (either embedded in a disk, or as an isolated component) and more extended massive elliptical objects. From our earlier discussion regarding the star formation properties of \profuse{} components it seems further clear that compact bulges in disks and compact isolated spheroids have distinctly different formation histories: the former being extremely old and the latter much younger (possessing similar mean ages to our disk components). The low-mass compact young spheroids are canonically the same population as the `little blue spheroids' presented in \citet{Moffett:2019tw} it would seem.

\subsubsection{\hyperfit{} of the Mass--Size Relationship}
\label{sec:SMfit}

The true bulge components presented in Figure \ref{fig:SMvRe_all} are consistent with having no size dependency on component stellar mass. The distribution in $R_e$ size is approximately Normal, and can be expressed as:

\begin{equation}
\log_{10} \left( \frac{R_e}{\text{kpc}} \right) = 0.02^{\pm 0.02} \pm 0.35^{\pm 0.04},
\end{equation}

where the first term represents the population mean and the second term is the standard deviation. The new size distribution for bulges and the older \citet{Lange:2016wx} work are shown in Figure \ref{fig:SMvRe_all}. Whilst we find no size dependency with bulge mass, the combined bulges in \citet{Lange:2016wx} were fitted with a positive slope.

The mass--size relationship of disks shows a clear component stellar mass dependency in Figure \ref{fig:SMvRe_all}. Similar to the work of \citet{Lange:2015vk} and \citet{Lange:2016wx} we use \hyperfit{} to define this distribution as a log-linear relationship incorporating intrinsic scatter as part of the fitting model \citep{Robotham:2015ws}. \hyperfit{} minimises the perpendicular residual plane, using simultaneous errors in the mass and size dimensions to correctly weight the data. This gives the following best fit relationship (the upper numbers represent the errors on the fitting terms):

\begin{equation}
\log_{10} \left( \frac{R_e}{\text{kpc}} \right) =  0.26^{\pm 0.04} \log_{10} \left( \frac{\text{M}^*}{10^{10}\text{M}_\odot} \right) + 0.62^{\pm 0.04},
\end{equation}

where $R_e$ is the half-mass size along the major axis and $\text{M}^*$ is the stellar mass of the disk component. This new mass--size relationship for disks and the older \citet{Lange:2016wx} work are shown in Figure \ref{fig:SMvRe_all} as black and red lines respectively.

The above regression has a Normal orthogonal intrinsic scatter of $0.23^{\pm 0.03}$. In the \citet{Lange:2016wx} the slope term for all late-type galaxies (which is the closest analogue to the above sample) is $0.208^{\pm 0.004}$ and the offset term is $0.61^{\pm 0.04}$. Within the stated errors of both fits these results are consistent, but our slope term is marginally steeper (a bit more than 1$\sigma$ difference).

\subsection{Combing Star Formation and Structural Properties}

Building on the above results, we can now explore stellar properties that appear to be associated with the mass--size relationship for bulges/spheroids and disks. Here we focus on mass--size--age and mass--size--$Z$ as having the most compelling results, but the more direct $g-i$ colour and current specific SFR (sSFR) of the disk are presented in Appendix \ref{app:Diskother} for completeness.

\subsubsection{Mass--Size--Age Relationship}

\begin{figure}
 \includegraphics[width=\columnwidth]{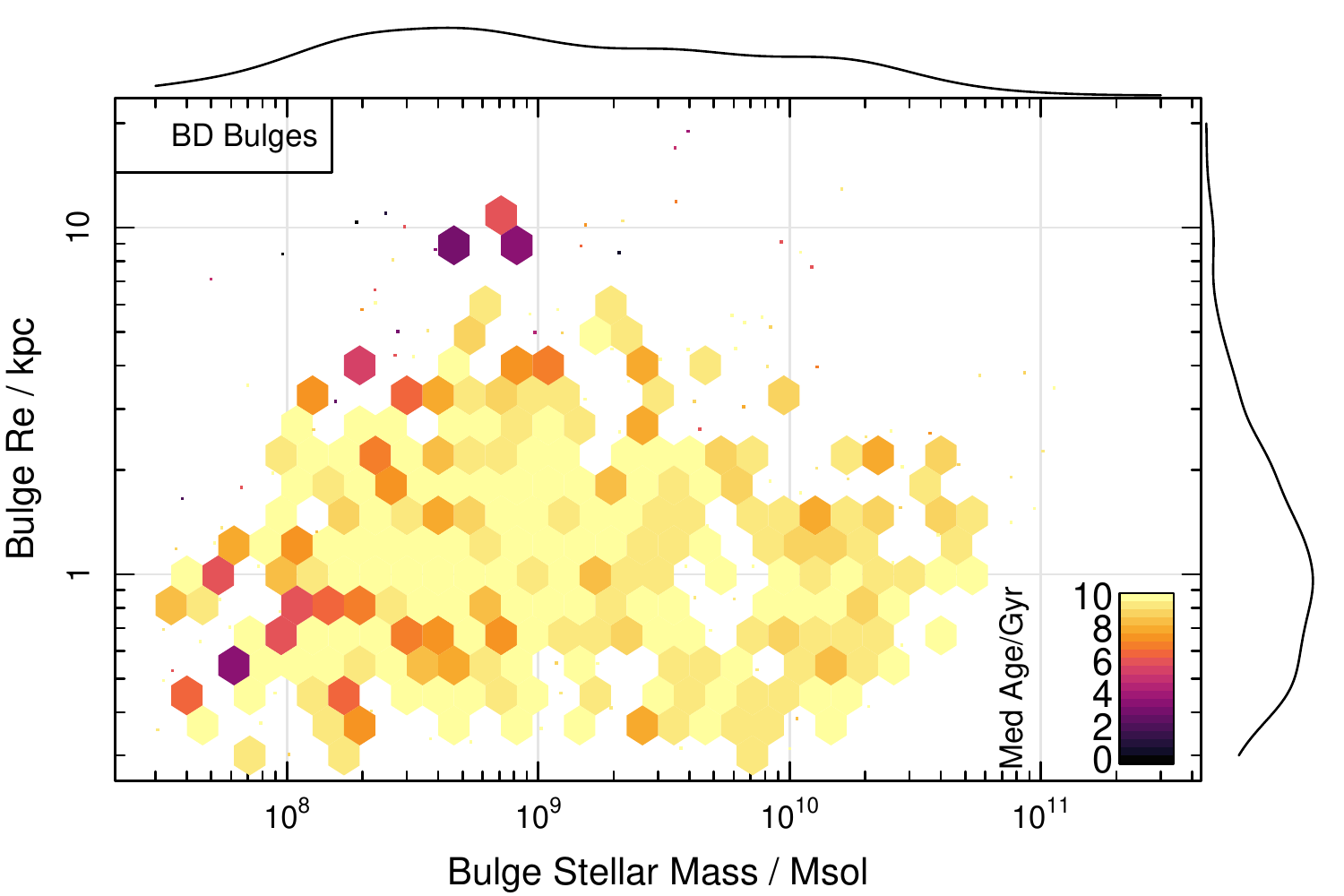} \
 \includegraphics[width=\columnwidth]{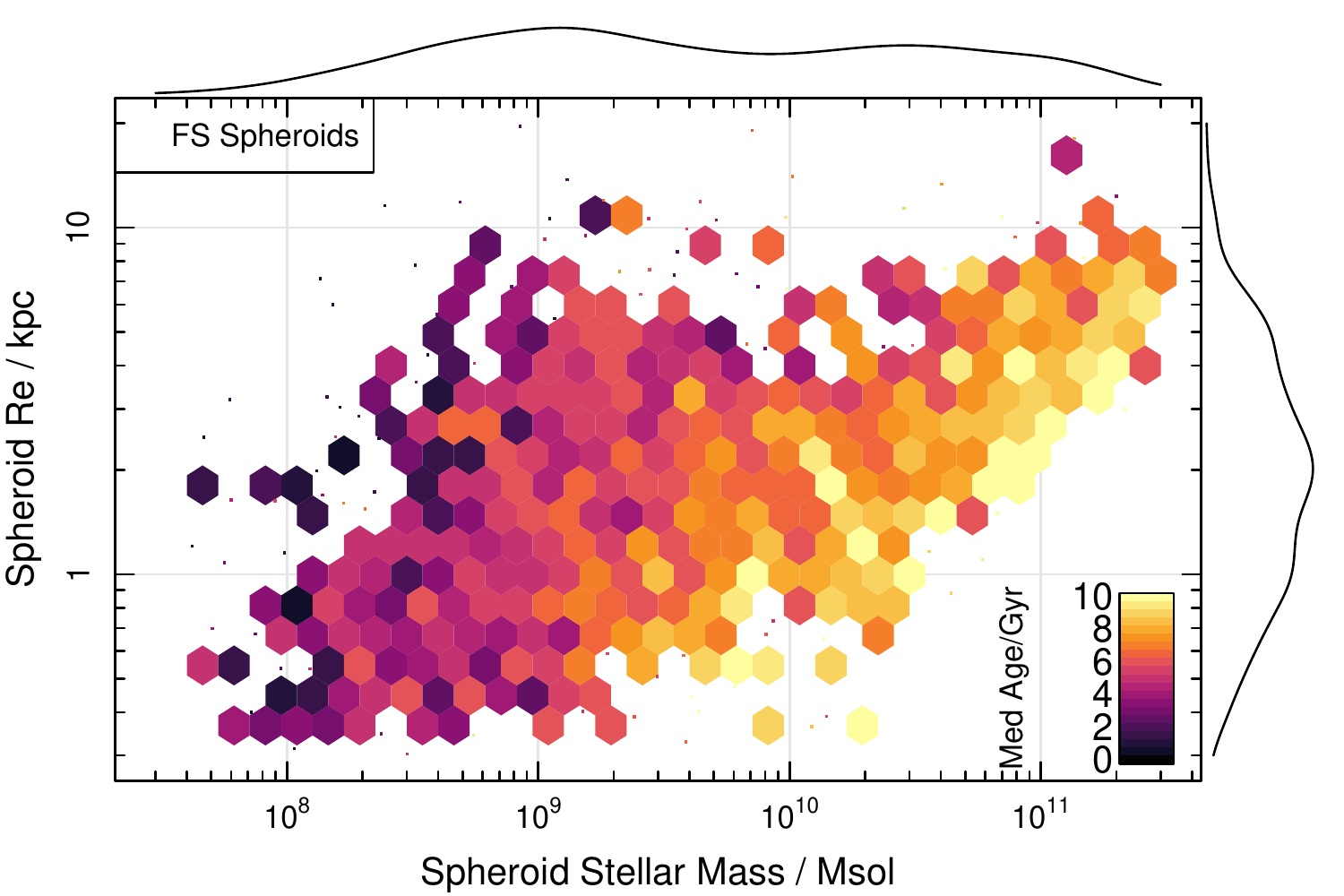} \
 \includegraphics[width=\columnwidth]{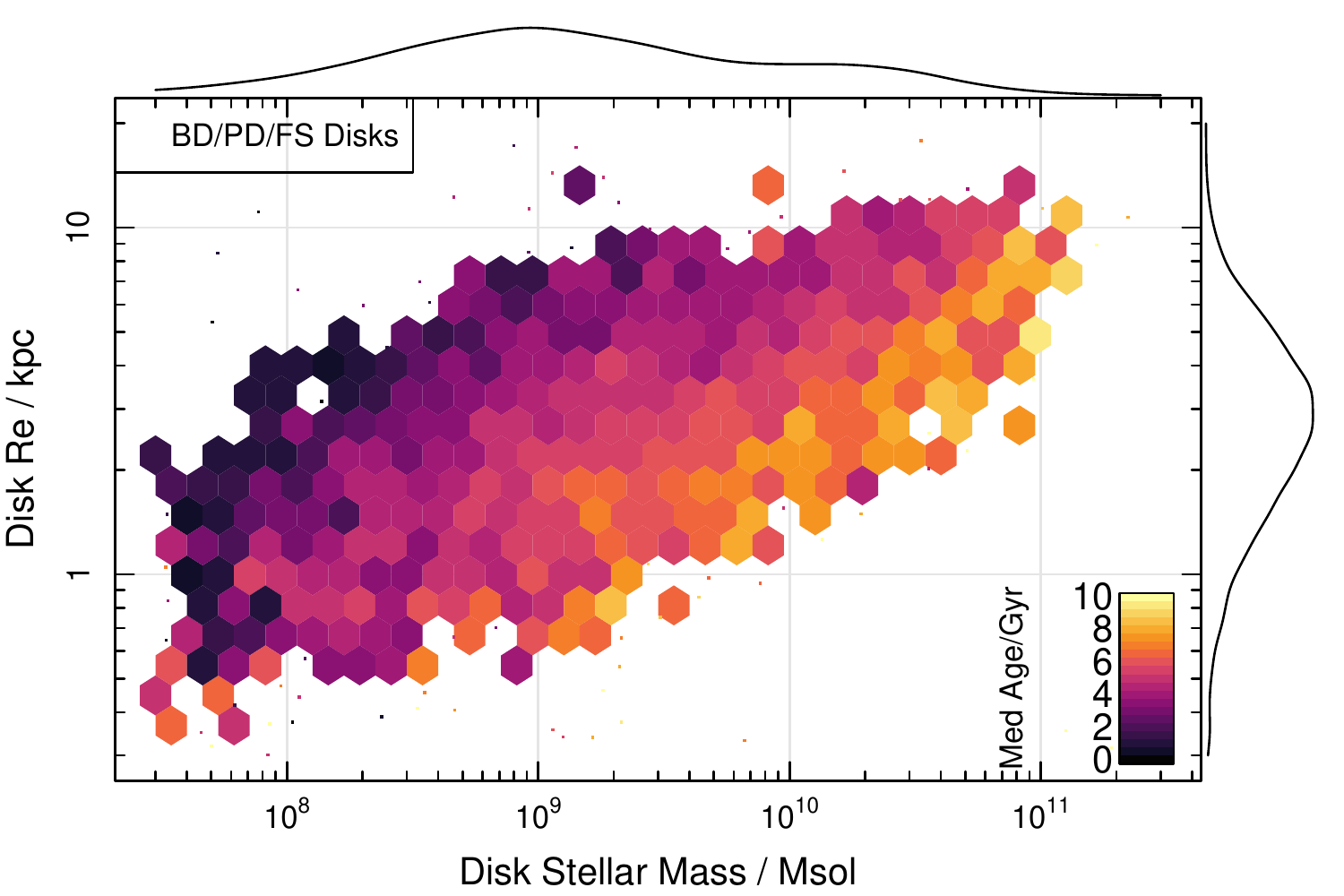}
 \caption{Component stellar mass versus size versus age. Top panel shows BD/PD bulges and middle panel FS spheroids. We can see the distinctly older elliptical galaxies on the RHS of this Figure. Bottom panel shows disks. We see a strong co-dependency between stellar mass, size and age that is well represented by a plane.}
 \label{fig:SMvRevAge_all}
\end{figure}

Figure \ref{fig:SMvRevAge_all} presents the mass--size relationship coloured by the median age within a hexagonal cell. There are no strong co-dependent trends seen for our bulge and spheroid populations, but the disks show strong co-dependence between stellar mass, size and age. The major trend is for more massive disks to also be older (as seen above), but we also see strong trends in age when looking vertically along a stellar mass window, where the larger disks for a given stellar mass are notably younger.

The presence of such a strong age dependence at a fixed mass for our disk populations suggests a plane that describes mass, size and age. We again fit this using \hyperfit{} (minimising the perpendicular plane residuals, as above), giving the following best fit relationship (the upper numbers represent the errors on the fitting terms):

\begin{eqnarray}
\log_{10} \left( \frac{R_e}{\text{kpc}} \right) = 0.47^{\pm 0.02} \log_{10} \left( \frac{\text{M}^*}{10^{10}\text{M}_\odot} \right) \\
-0.16^{\pm 0.01} \left( \frac{\text{Age}}{10^{9}\text{yr}} \right) + 1.56^{\pm 0.06},
\end{eqnarray}

where $R_e$ is the half-mass size along the major axis, $\text{M}^*$ is the stellar mass of the disk component and `Age' is the mean component stellar age. The above plane has a Normal orthogonal intrinsic scatter of $\pm 0.28^{\pm 0.01}$. Our mass dependence term has become notably steeper when adding the age dimension, which reflects the presence of visual diagonal striping in the bottom panel of Figure \ref{fig:SMvRevAge_all}. As should be expected from our visual analysis, the `Age' term has a negative slope, i.e.\ younger disks tend to be larger. The result is that a disk that is 2 Gyrs older (in terms of mean stellar age) might be expected to be a factor $\sim2$ more compact for the same component stellar mass.

\begin{figure*}
\includegraphics[width=\textwidth]{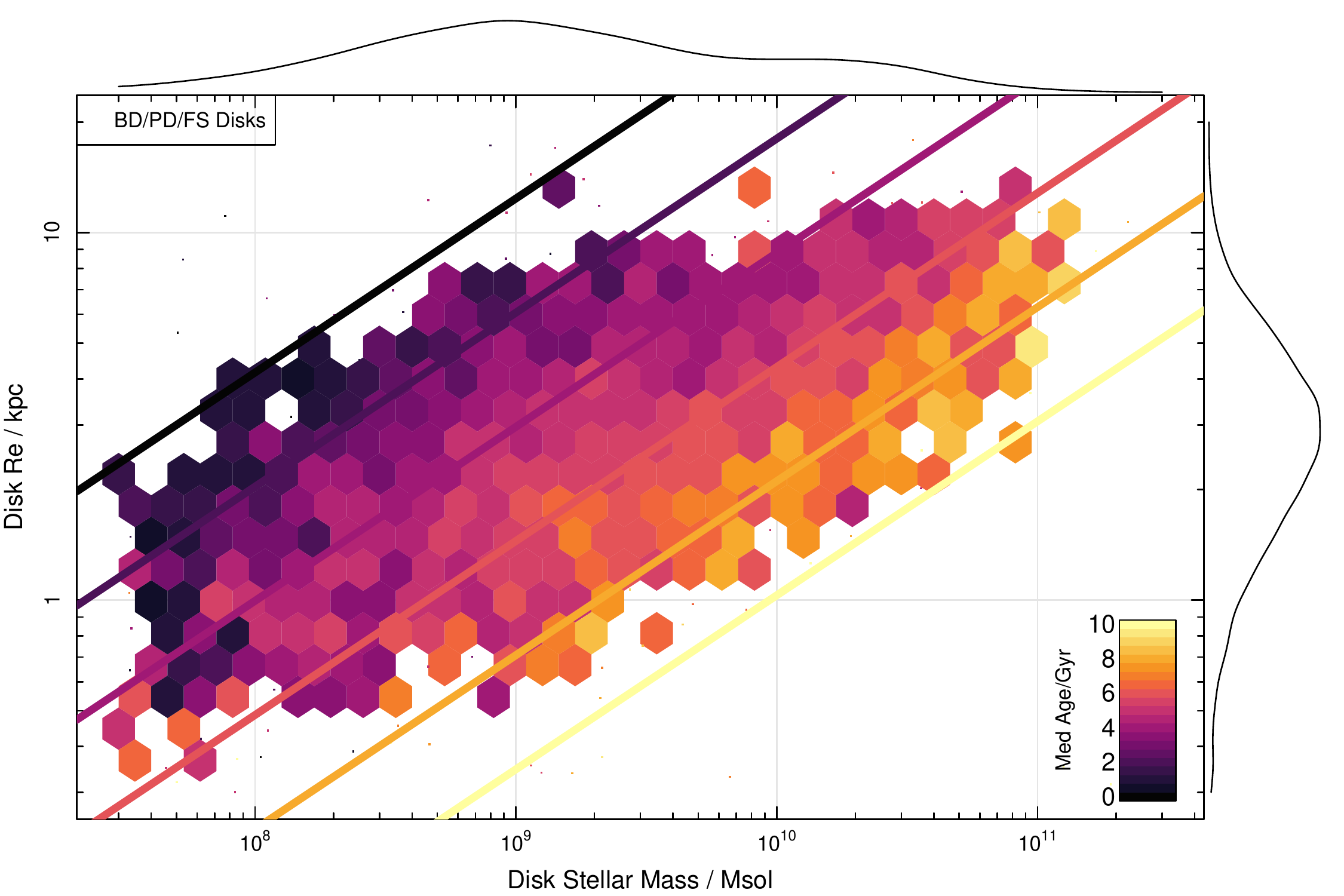}
\caption{Same as the bottom panel of \ref{fig:SMvRevAge_all}, but with the \hyperfit{} best fitting plane shown as parallel lines with 2 Gyr steps in age.}
\label{fig:SMvRevAge_all_Hyper}
\end{figure*}

For clarity, Figure \ref{fig:SMvRevAge_all_Hyper} presents the bottom panel of Figure \ref{fig:SMvRevAge_all} with the \hyperfit{} plane added as parallel lines of 2 Gyr steps in age (as reflected by their colour). This novel mass--size--age plane appears to do a reasonable job of describing the diversity of observations over this mass range. Importantly for the robustness of this result and the existence of a true mass--size--age plane, we find in Appendix \ref{app:Disk_Size} that even pure disks with no bulge show the same age structure. As such, our results are not attributable to being a complex systematic of our BD or PD models, and the plane is present even when fitting the simplest FS model with only one unified SFH and ZH.

There is little published literature that allows for a direct comparison of these results, which is not surprising since self-consistent stellar-population structural decomposition in the manner of \profuse{} is a novel approach. The IFS work of \citet{Scott:2017ud} is probably the most directly comparable work, which uses data from the Sydney-AAO Multi-object Integral \cite[SAMI;][]{Bryant:2015ug} survey along with the \profit{} derived galaxy sizes provided by GAMA \cite{Robotham:2017tl, Moffett:2019tw}. The top panel of Figure 9 in that work shows very similar global behaviour for the mass--size--age plane, where systematically the youngest galaxies are also the largest for all stellar masses. Figure 12 in that work also attempts to split the sample into disk-like (LHS) and spheroidal (RHS) samples, and broadly the result of larger younger disks is still clear. As in our work, the spheroidal population has less clear trends in the most comparable top-right panel.

\citet{McDermid:2015tk} investigated early type galaxies using data from the ATLAS3D IFS survey \citep{Cappellari:2011vq}, although including S0 galaxies with disks. Figure 6 in that work presents LOESS smoothed relationship of dynamical mass versus size as a function of stellar age and metallicity. Clear trends in age are present in the same sense as we find for our \profuse{} disks (younger galaxies being larger for a given dynamical mass). Whilst the dynamic range and sample size in \citet{McDermid:2015tk} is much smaller than we have here, the age-size dependency appears to be in qualitative agreement (2 Gyrs in age being a factor two difference in size). This agreement is perhaps surprising since we only see strong evidence for mass--size--age dependency for our disk sample, our unambiguously early-type class of preferred fits (the more massive FS spheroids) do not show such behaviour.

Other analogue works are those of \citet{Barone:2021td} and \citet{Maltby:2018uu}. The former of these works \citep{Barone:2021td} is focussed on the stellar population fitting of quiescent galaxies (a cut we do not make here), but broadly has the same outcome that younger stellar populations are the more extended in term of $R_e$. The latter work \citep{Maltby:2018uu} does not explicitly extract stellar ages, but does describe a consistent picture where disk-like galaxies with more current star formation (which will tend to make them `younger' by our mass-weighted age definition) are also the most extended. Comparable findings are also presented in \citet{Courteau:2007vv}, where rather than age cuts they find bluer later-type spirals (crudely analogous to being younger) are more extended for a given luminosity (analogous to our disk stellar mass).

The general picture of larger younger disks is consistent with the apparent growth in galaxies since $z\sim2.5$ as presented in \citet{Trujillo:2006uj, Trujillo:2007wr}. Those works found a factor $\sim3$ size growth for disk-like objects galaxies since $\sim10.5$ Gyrs look-back time, however they considered the sizes of galaxies at a given epoch which themselves would have a range of ages. In fact, cutting our sample to select disks with a mean stellar age older than 10.5 Gyrs returns a mass--size relationship that has the same slope (within error) but a factor 1.65 lower. This suggests a picture where disks that formed entirely at an early epoch must have grown by a factor 1.5--2 for our low redshift GAMA results to be consistent with apparent sizes of higher redshift disk-like galaxies. Notably this size evolution is not enough to make these disks as large as those formed more recently (i.e.\ they are still distinguishable in our mass--size--age plane). Whether this disk growth could be secular in nature is unknown, but there are far more mechanisms that allow galaxies to become larger over time than smaller \citep{Trujillo:2011wj}.

The most directly comparable theoretical work is \citet{Tonini:2017wl}, which is based on an extension to the Semi-Analytic Galaxy Evolution \citet[SAGE;][]{Croton:2016uh} model. Figure 9 in that work presents a similar mass--size--age plane, with similar trends of age across the size dimension (younger galaxies again being more massive). The apparent dynamic range of the effect is visually similar too, spanning disk ages from a few Gys at the largest extreme and 10 Gyrs for the most compact. Detailed comparisons to predictions from semi-analytic models \citep{Lagos:2018wc} and appropriate higher resolution hydrodynamical simulations \citep{Schaye:2015wm} are deferred to future work (Bellstedt et al., in prep.). Obvious avenues of interest given associated works are the degree to which disk sizes can evolve without any star formation (in a secular fashion), the role of galaxy-galaxy interactions, and the impact new star formation has on the distribution of stellar material already formed. Given the apparent lack of requirement for a BDD model to fit our GAMA sample (which can produce radial gradients in the SFH) it is also not yet clear how important inside-out growth mechanisms are to explain these results.

\subsubsection{Mass--Size--$Z$ Relationship}

\begin{figure}
 \includegraphics[width=\columnwidth]{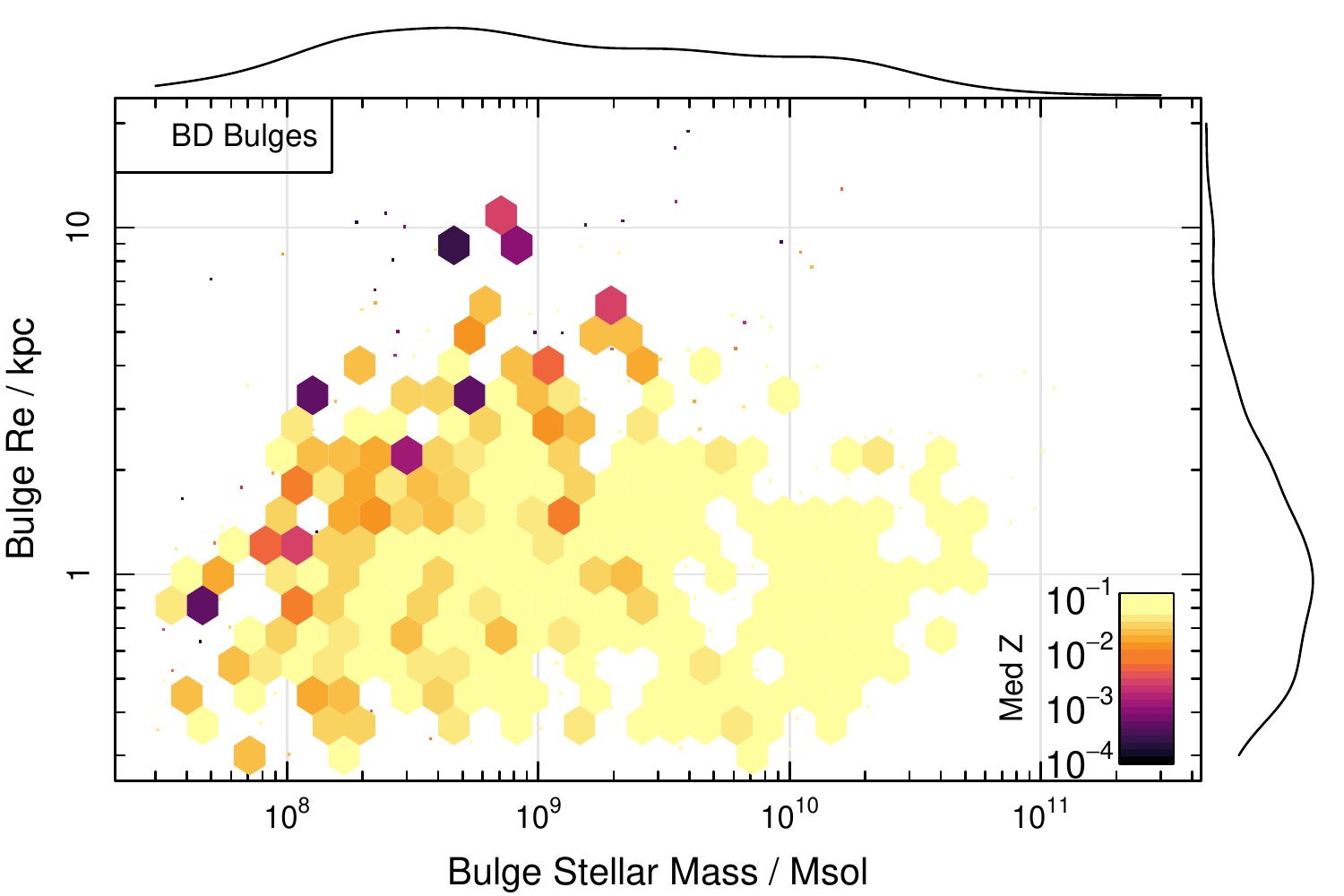} \
 \includegraphics[width=\columnwidth]{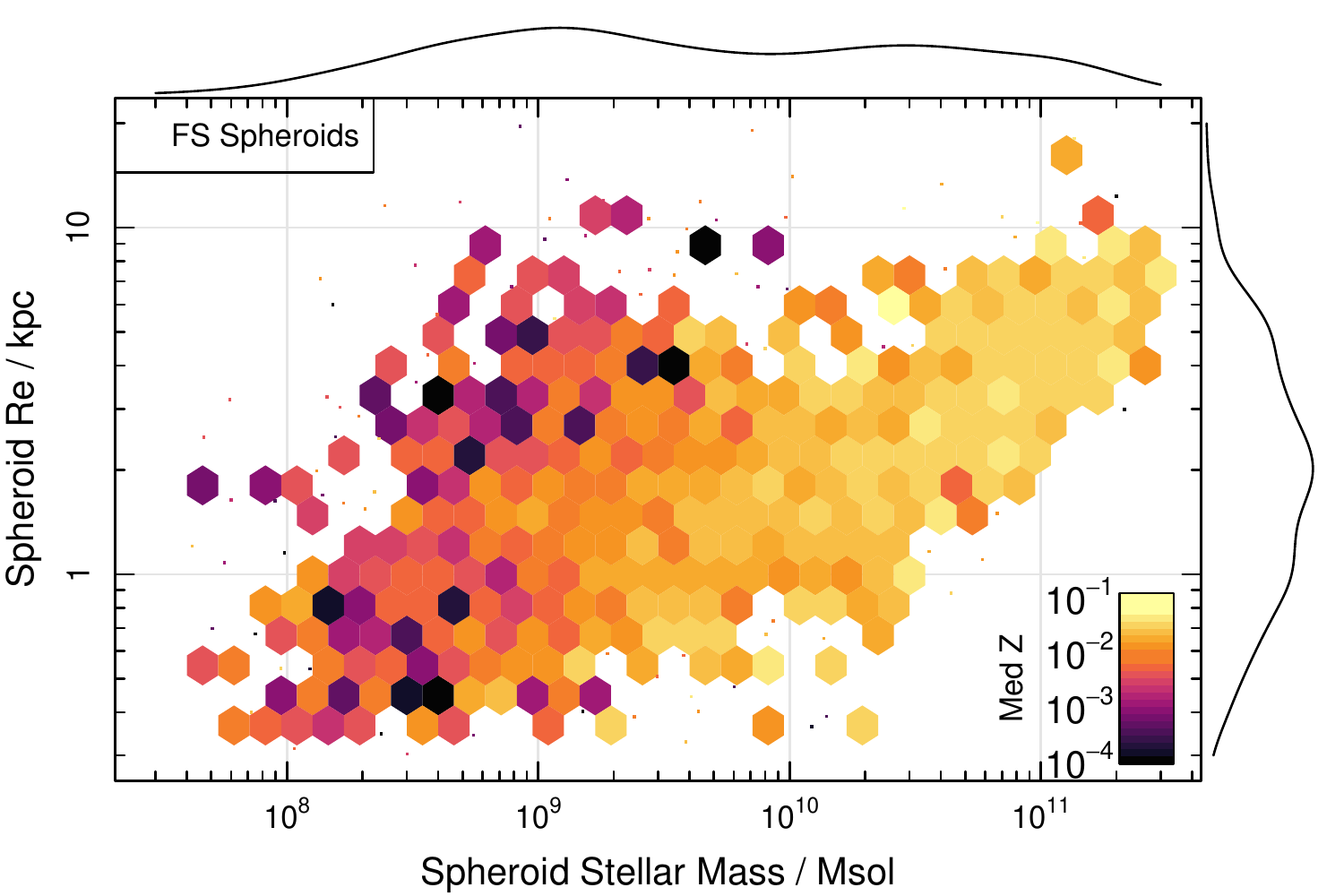} \
 \includegraphics[width=\columnwidth]{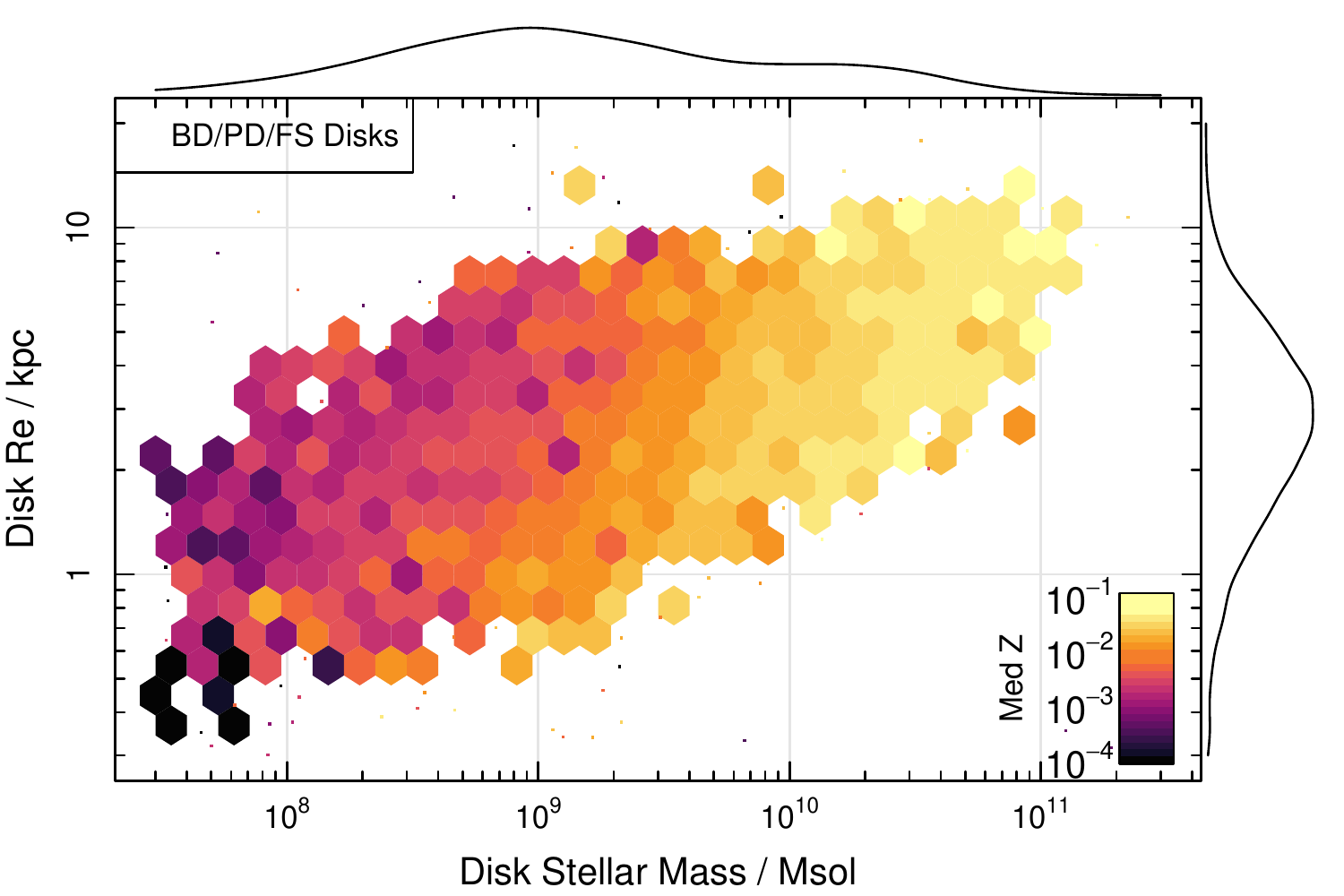}
 \caption{Component stellar mass versus size versus metallicity. Top panel shows BD/PD bulges and middle panel FS spheroids. The addition of metallicity does not seem to produce any discernible trends. Bottom panel shows disks. We see a some co-dependency between stellar mass, size and metallicity, but the horizontal stellar mass metallicity relationship clearly dominates.}
 \label{fig:SMvRevZ_all}
\end{figure}

Extending our analysis to metallicity, Figure \ref{fig:SMvRevZ_all} presents the mass--size relationship coloured by the median disk $Z_{\text{gas}}$ within a hexagonal cell. There are no strong trends seen for our bulge and spheroid populations, and the disks show weak co-dependence between stellar mass, size and $Z_{\text{gas}}$. The horizontal dependency is naturally strong \citep[due to the expected mass-metallicity relationship, e.g.][]{Lara-Lopez:2010tp}, but the lack of strong diagonal banding suggests there is no intrinsic co-dependent plane between the three parameters, instead the stellar mass is independently able to predict the size and metallicity.

This null result appears to be very consistent with \citet{Scott:2017ud}. In the middle-left panel (late-type galaxies, which we can take as analogues of our disks) of figure 11 we see strong mass--$Z$ trends, but little evidence of any vertical trends of Z varying with size in any systematic manner. In the middle-right panel (early-type galaxies, which we can take as analogues of our spheroids) of figure 11 there is again little evidence of an age--$Z$ trend, and only a small amount of mass--$Z$ dependency. The latter cannot be strongly concluded given the small amount of dynamic range in mass present for the early-type galaxies in that work.

\citet{McDermid:2015tk} find equally strong evidence for a mass--size--$Z$ plane as a mass--size--age plane. This would appear to be in some tension with this work, which may be related to the early-type galaxy focus of \citet{McDermid:2015tk}, and perhaps even the use of dynamical masses.

Given the above, there is no justification to attempt a four dimensional mass--size--age--$Z$ hyper plane, but in conclusion we do find compelling evidence for a true co-dependency between mass, size and age. The full implications of this novel three dimensional plane will be further explored in future work (Bellstedt et al., in prep), including the impact of our choice of stellar population library (BC03 in this work).

\section{Limitations and Future Paths}

Whilst we present \profuse{} as a step forward in physically motivated spatial decomposition for large photometric samples, there are a number of limitations with the current software. A few of the more obvious and physically relevant are listed and discussed in the following subsections.

\subsection{Radial Trends}

Whilst many authors suggest wavelength trends in multiple properties \citep{Kelvin:2014ul, Haussler:2013tz}, it is non-trivial to create a generative model that recreates such variations naturally. \prospect{} does support simple disk colour gradients via the option of fitting a coupled disk model (BDD above). In this case the disk geometries of two disks are all coupled except for $R_e$ which is allowed to vary (and optionally $n$). Each disk is also allowed to have completely distinct \prospect{} SEDs, i.e. different SFHs and ZHs etc (see Figure \ref{fig:BDDcartoon} for a schematic view of this model). The natural consequence of this is the possibility of radial colour gradients along the disk components either due to radial trends in SFH or ZH (and possibly even the dust screen). This is the simplest \profuse{} model that can naturally explain colour gradients along disks (see Appendix \ref{app:BDD} for a discussion of this).

\begin{figure}
 \includegraphics[width=\columnwidth]{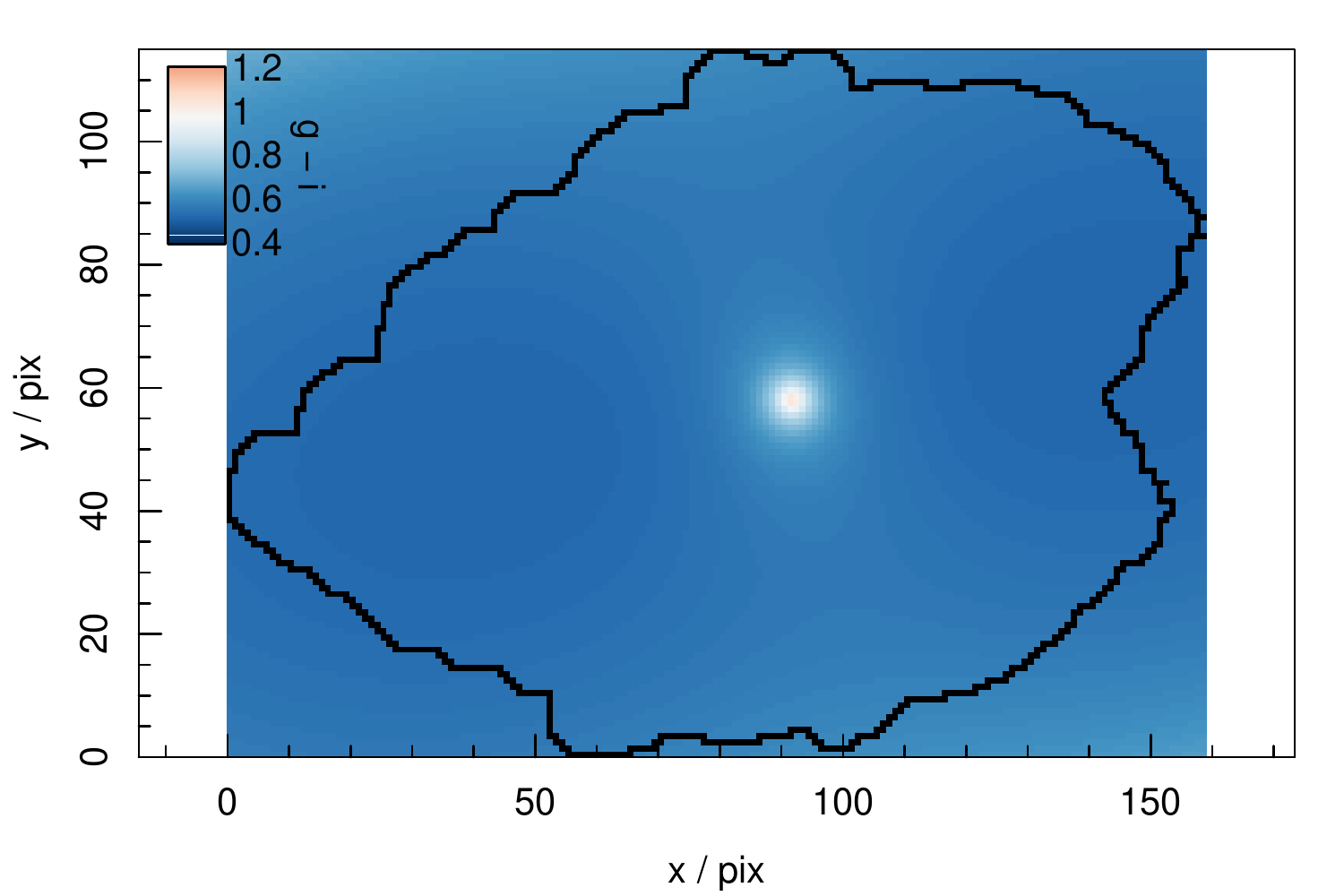}
 \caption{The model predicted observed $g-i$ colour per pixel (with PSF convolution, observed frame, and dust attenuated). The bluer disk light dominates everywhere, but as should be expected the redder bulge contribution increases towards the centre.}
 \label{fig:model_g-i}
\end{figure}

Whilst the above is an option (with additional fitting expense with a more flexible and possibly more degenerate model) it should be noted that even the simpler bulge-disk model discussed above produces natural colours gradients spatially just due to the extended interactions between the single SED (and therefore single colour) bulge and disk components. This is because differing scale lengths and geometries will naturally produce different observed colour trends over the full scale of the galaxy- the combined image will never be as simple as a distinct uniformly coloured bulge and disk. This can be seen clearly for our example galaxy (from Section \ref{sec:example}) in Figure \ref{fig:model_g-i}, where we show the observed $g-i$ colour across the full image. Whilst the bulge and the disk both have a fixed colour by construction, the combined model displays complex variations throughout. Particularly in the inner regions, the simple constant colour circular bulge and elliptical disk models show non-trivial interactions, especially along the minor-axis direction.

\subsection{Bulge-Disk Model Geometry}

For bulge-disk fitting, the elliptical disk and circular bulge geometry currently enforced is not so much a code limitation (it is not computationally expensive to allow for more complex geometries) but a pragmatic restriction to reduce the strong fit degeneracies that often occur with increased flexibility. The most serious consequence of the enforced circular bulge is only canonical `classical' bulges can be generated in \profuse, and the often non-spherical family of bluer pseudo-bulges cannot be well represented by the current model \citep[e.g. see][]{Athanassoula:2005wm}. Given the wealth of data available in the residuals it is relatively easy to identify these failing cases that truly do justify a different physical model, and at least these can be removed for relevant analysis. In practice for the low redshift GAMA sample we found few examples of galaxies that strongly compelled a more flexible pseudo-bulge like geometry that e.g.\ were truly better fitted with the more flexible BDD mode.

On top of the bulge restriction, it is clear that true elliptical galaxy geometries are compromised since they are not elegantly described by the sum of a circular de Vaucouleurs profile bulge and elliptical exponential disk. Again, it is easy to identify these compromised cases. In this situation they are generally better represented by a single component (i.e.\ only one \prospect{} SFH can be generated for the whole galaxy) elliptical free \sersic{} profile model (FS above), which is allowed in the current version of \profuse{}.

\subsection{Model Selection}

The general use case involves running at least the simplified bulge-disk model and the single component \sersic{} model and determining the most representative of the data from those (as we have done above in our initial analysis of GAMA with the BD and FS models). In a well behaved example the comparison of the DICs will often be enough, but some physical intuition is required to weed out situations where a model is numerically better (more `likely') but physically less meaningful.

Ironically, given the pursuit of ever deeper and better resolved data, the above model selection problem generally gets worse the better the data are. This is because we enter into the statistical regime of approximate models where in detail all models are `wrong'. Even in this case, it is reasonable, and in practice often possible, to attempt to select the most physically informative model. The GAMA data above at the typical redshift of $z\sim0.04$ are in a good regime of depth and resolution for the \profuse{} approach, where a significant fraction of our models produce very clean $data-model$ residuals. More thought and effort would need to be applied to solving this problem when the data resolve features our model cannot describe, e.g.\ the increase in the visually irregular structure of galaxies at higher redshift \citep{Driver:1995vv}.

\subsection{Bar Model}

Currently \profuse{} does not offer a bar model component, be that Ferrer or some sort of modified \sersic{} profile \citep[see][for a detailed discussion of the profiles available]{Robotham:2017tl}. This is a relatively simple addition longer term, and is likely to be one of the first extensions when expanding beyond this initial application to moderately resolved GAMA data. An interesting question for the bar component is whether its physical SED model should be directly coupled to the disk (where the only free \prospect{} component would be the fraction of mass in the bar) or left to have a completely independent SED model. Even if the SFH and ZH end up being coupled, there would be a case to allow for a different bar specific dust model rather than assuming a completely constant dust screen for the whole disk and bar.

Regarding the impact this lack of bar modelling has on the simpler model explored in this paper, it has been noted that, similar to spiral arms, the ellipse-averaged contribution of a bar is similar to a smooth disk profile. That is to say the average intrinsic radial profile (when computed within geometrically linked annuli) is often not hugely modified by the presence of spiral arms and/or a bar even though the azimuthal distribution of light is hugely modified \citep{Mollenhoff:2004vz, Laurikainen:2005uf}. The positive in terms of the \profuse{} modelling is that the average disk properties should therefore be well conserved (assuming it is reasonable to exactly couple the star formation history of the bar to that of the disk), whilst the clear negative is that no separate accounting of mass inferred in the bulge versus the canonical disk is possible. We make a more general observation that only extremely well-resolved galaxies would lend themselves to meaningful bar modelling, and by visual inspection only a few percent of the GAMA galaxy decompositions analysed for this work have convincingly clear bars that would lend themselves to non-degenerate fitting (where the `bar' component does not simply model irregular features in the bulge or disk).

\subsection{Uncoupled Screen Dust}

\begin{figure*}
 \includegraphics[width=2\columnwidth]{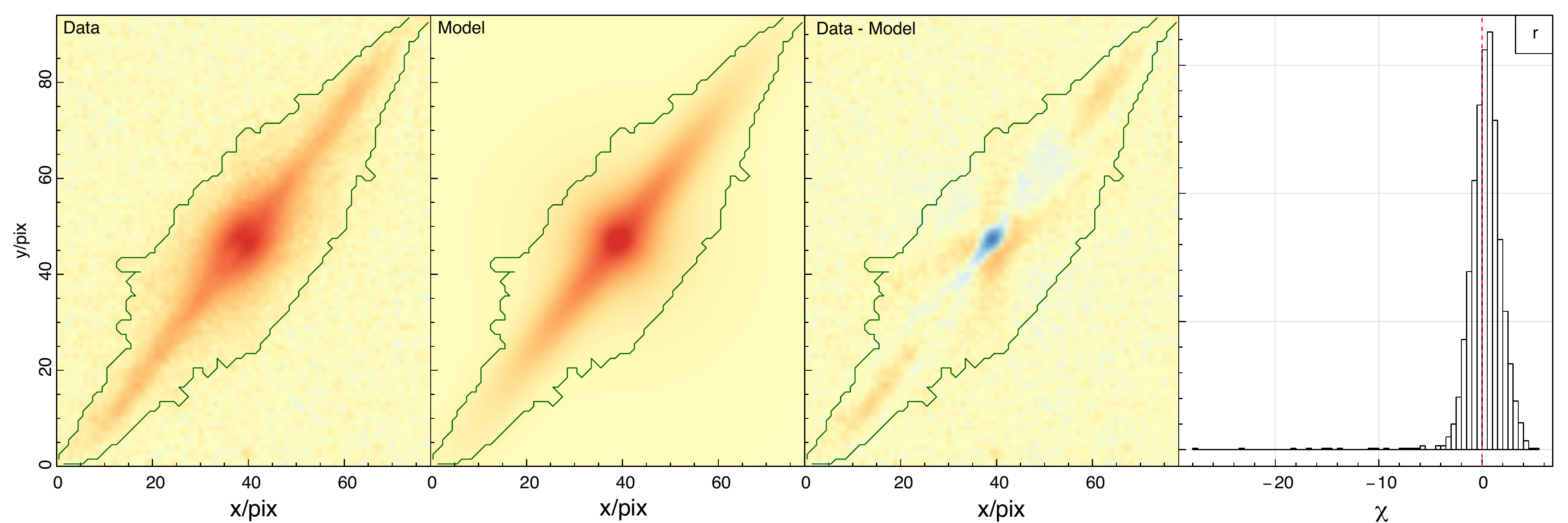}
 \caption{Example \profuse{} fit showing the $r$ band data (left panel), model (second panel), spatial residual (third panel) and error normalised histogram (right panel). The data show clear evidence of a thin dust lane what is not produceable by the \profuse{} model, it being limited to dust screens for each component. This is particularly clear in the third panel where we see a blue model over-subtraction in the inner part of the profile where the dust lane is observationally dimming the light.}
 \label{fig:dust_lane}
\end{figure*}

Currently the dust model for both the bulge and disk are entirely independent, and one does not know about the other. In practice this means bulge light that, given the specific geometry of the system, would be attenuated by the disk dust is currently only attenuated by the bulge dust. However, to do this meaningfully takes a path towards full 3D radiative transfer fitting. For now we note that it is relatively easy to identify these problem cases from their edge-on disk geometry (very high ellipticity disks). These also tend to be the systems with visually obvious dust lane features, which are also not well described by the simple bulge-disk \profuse{} model (e.g.\ see Figure \ref{fig:dust_lane}). We note positively that such extreme systems in a volume limited sample appear to be statistically rare, so few of our galaxy insights are significantly biased by this modelling limitation. For such cases, it is also possible to mask out dust lane features.

\subsection{Intra-Halo Light Profile}

A near term ambition will be to add the option of a projected intra-halo light (IHL) component to \prospect{}. This should allow us to probe the low surface brightness outskirts of galaxies, where potentially strong priors can be placed on the age of stellar populations in this component. Such analysis should be revealing since theory suggests many elements of halo assembly (including in particular merger activity) and even cosmology can be associated with the IHL \citep[e.g.][]{Power:2016tx, Drakos:2019tg, Drakos:2019to, Canas:2020wk}. The most appropriate profile to use for the projected IHL is a matter of much uncertainty since it is often hard to detected at all, let alone distinctly profile \citep{Merritt:2016uk}. Theoretically defining the optimal stellar populations to extract is part of planned future work (Proctor et al., in prep.).

\section{Conclusions}

Below we summarise the major outcomes from this work:

\begin{itemize}
\item We present \profuse{}, a natural extension and {\it fusion} of the structural analysis tool \profit{} and the SED fitting tool \prospect.
\item The current implementation of \profuse{} can fit physically self-consistent models of galaxies comprised of axis-symmetric bulges (unresolved PSF or \sersic{} profile) and disks (up to two \sersic{} profiles).
\item In this initial implementation we process the low redshift GAMA sample of \citet{Bellstedt:2020vq} that already has results from our previous \prospect{} fits. We process the sample of 6,664 galaxies in four different modes: single free \sersic{} [FS];  de Vaucouleurs bulge and exponential disk [BD]; PSF bulge and exponential disk [PD] and finally de Vaucouleurs bulge and double exponential disk [BDD]. Determining the BDD model was very rarely preferred (the quality of data does not reveal strong colour gradients in our disks), we focussed our analysis on the FS, BD and PD models.
\item We find globally consistent behaviour between our new GAMA \profuse{} fits and our published \prospect{} results, giving confidence that our new methodology is working in a comprehensible manner.
\item We find our disks are systematically (and almost always) younger than our bulges for galaxies preferring a BD or PD model.
\item In terms of mass-age-Z separation, our bulges appear to be tri-modal: we find a very metal rich old population of true bulges, and two populations for our preferred FS (spheroidal) galaxies. One is more massive older metal rich elliptical, and the other less massive younger metal poor spheroids. The latter appear to be analogues to the Little Blue Spheroid (LBS) population discussed in the literature.
\item The mass-age-Z separation of disks is simpler and effectively uni-modal: we find a smooth mass--$Z$ dependency.
\item Expanding the analysis to include the structural properties of \profuse{} we find the most compelling and novel results. Whilst our bulge and spheroid populations show little compelling requirement for a mass--size--age or mass--size--$Z$ plane, our disks display strong co-dependency in the mass--size--age plane. We find a strong trend in age along both the mass (as expected) and size dimensions, where younger system are larger for a given mass.
\end{itemize}

The most interesting result is certainly the extraction of a compelling mass--size--age plane for galaxy disks, and as much as can be compared this seems to be in agreement with previous works (although none are directly comparable given differences in data and approach). Further observational and theoretical work on the \profuse{} modelling of GAMA is planned (Bellstedt et al. in prep.), and future extensions to the DEVILS and WAVES surveys are planned. Longer term, multi-band space based missions (e.g\ James Webb, Vera Rubin, Euclid and Nancy Grace Roman observatories) offer the best prospects to extending our analysis to higher redshifts.

\section*{Acknowledgements}

ASGR acknowledges support via the Australian Research Council Future Fellowships Scheme (FT200100375).

This work made use of the following software:\citet{Robotham:2016ue,Robotham:2016vu,Robotham:2018aa,Robotham:2016ux,Robotham:2016wx,Robotham:2020wo}. This was all written in the \R{} language \citep{R-Core-Team:2021wf}.

GAMA is a joint European-Australasian project based around a spectroscopic campaign using the Anglo-Australian Telescope. The GAMA input catalogue is based on data taken from the Sloan Digital Sky Survey and the UKIRT Infrared Deep Sky Survey. Complementary imaging of the GAMA regions is being obtained by a number of independent survey programmes including GALEX MIS, VST KiDS, VISTA VIKING, WISE, Herschel-ATLAS, GMRT and ASKAP providing UV to radio coverage. GAMA is funded by the STFC (UK), the ARC (Australia), the AAO, and the participating institutions. The GAMA website is http://www.gama-survey.org/ .

Based on observations made with ESO Telescopes at the La Silla Paranal Observatory under programme ID 179.A-2004.

Based on observations made with ESO Telescopes at the La Silla Paranal Observatory under programme ID 177.A-3016.

This work was supported by resources provided by the Pawsey Supercomputing Centre with funding from the Australian Government and the Government of Western Australia.”

Thank you to Luke J.~M. Davies for assistance in creating Figure \ref{fig:BDDcartoon}. ASGR's original fever dream sketch was successfully translated into something comprehensible thanks to his efforts and Keynote prowess.

Thank you to the anonymous referee for their constructive comments on the original version of this manuscript. Figures \ref{fig:prospectVprofuseFS_SM} and \ref{fig:prospectVprofuseFS_SFR} in particular are much improved based on their suggestions.

\section*{Data and Code Availability}
 
All the \R{} packages discussed, including the new software \prospect, are already available publicly on primary author A.~Robotham's main GitHub accounts\footnote{https://github.com/asgr and https://github.com/ICRAR}. The main multi-band example data are included as part of the \profound{} package. The analysis of GAMA galaxies made use of publicly available GAMA data.

All GAMA data discussed in this work, including reprocessed images and comparison data, will be released as part of the comprehensive GAMA Data Release 4 (DR4; Driver et al, in press).



\bibliographystyle{mnras}
\bibliography{ProFuse} 




\appendix

\section{BD Versus BDD Mode}
\label{app:BDD}

\begin{figure}
 \includegraphics[width=\columnwidth]{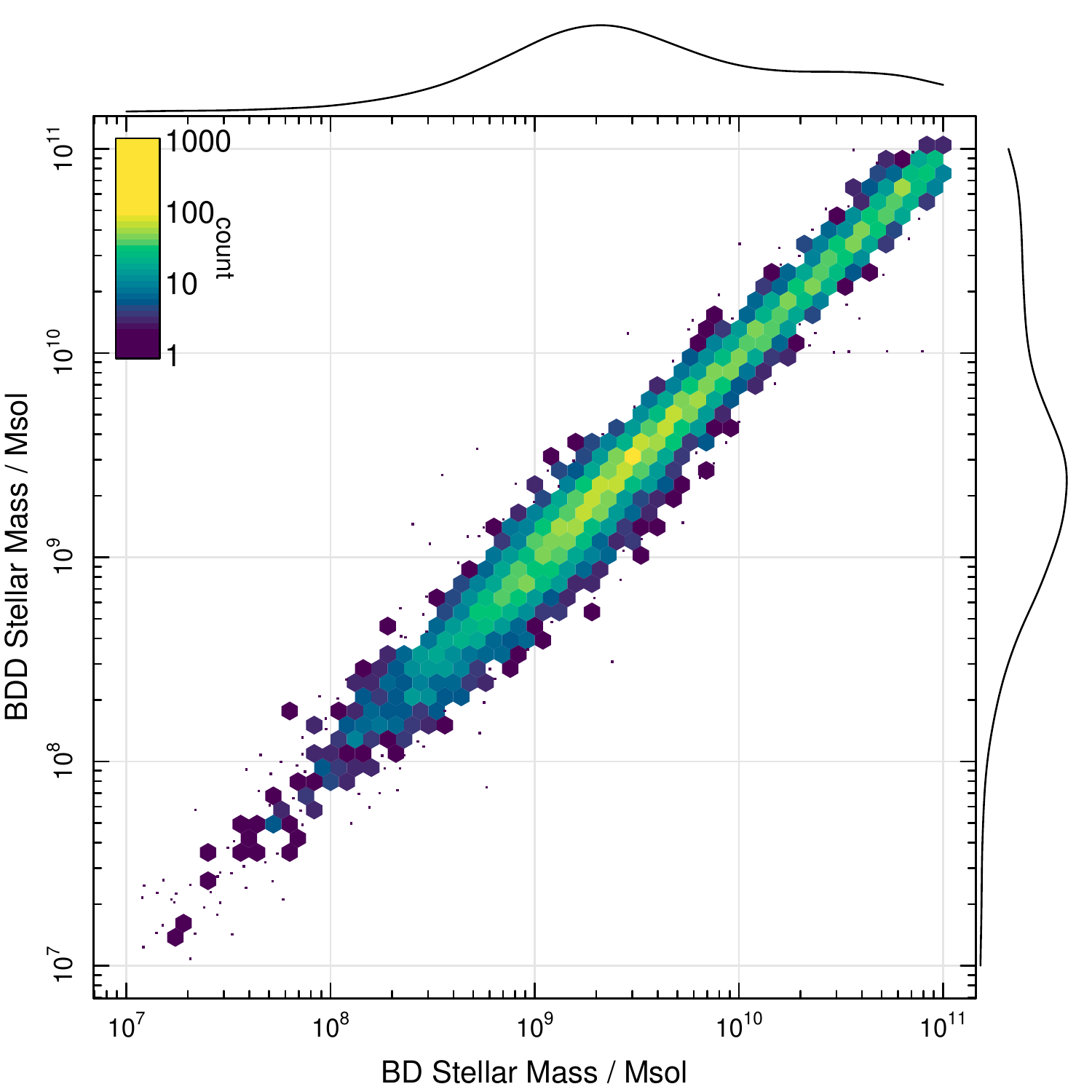}
 \caption{Comparison of disk stellar masses between BD and BDD modes. For the BDD we are showing the sum of both disk components. The disks being formed in the more complex BDD are representing almost exactly the same amount of stellar mass.}
 \label{fig:SM_BDvBDD}
\end{figure}

\begin{figure}
 \includegraphics[width=\columnwidth]{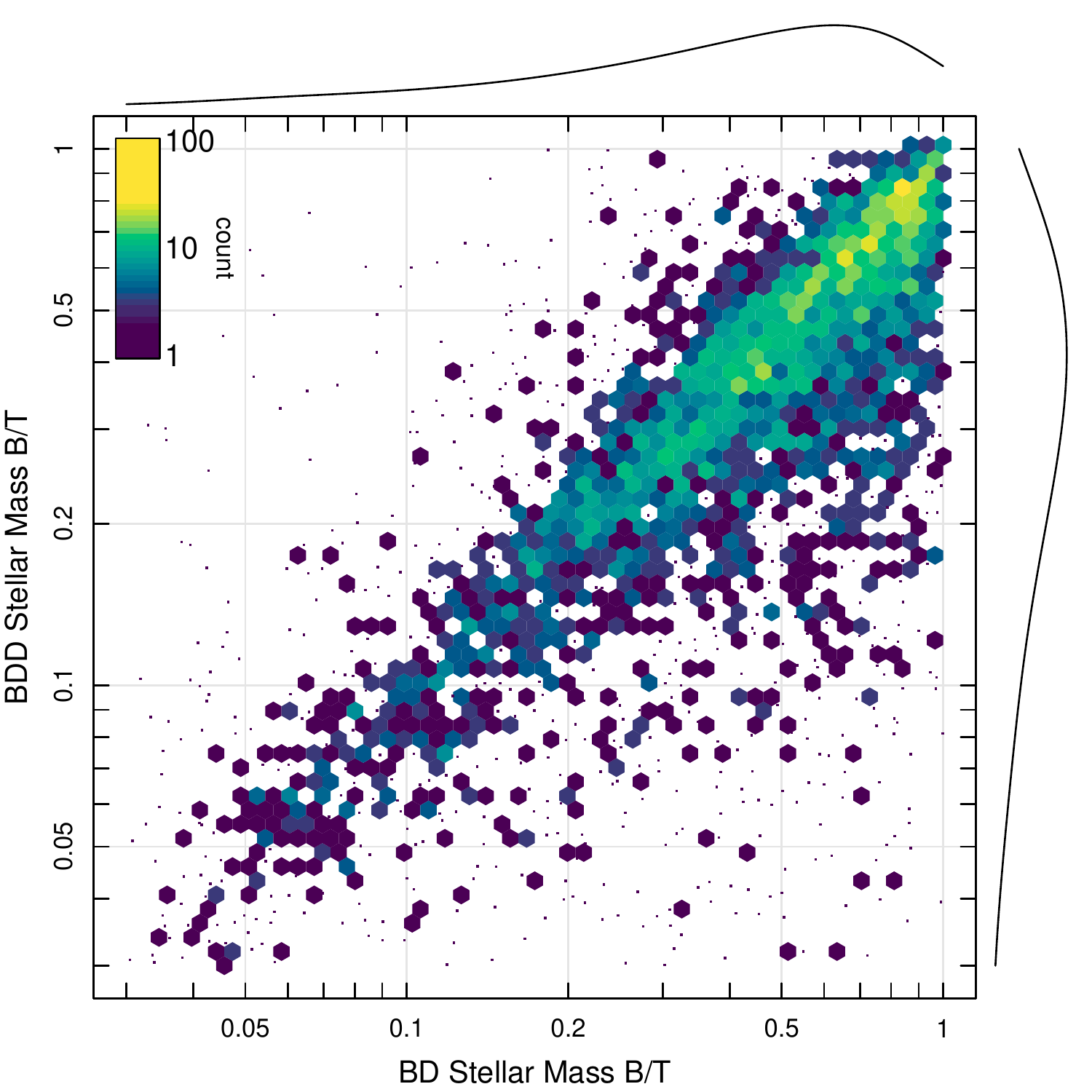}
 \caption{Comparison of bulge divided by total stellar masses between BD and BDD modes. Despite the significantly more complex double disk model, the amount of stellar mass being attributed to the bulge is broadly the same. This is significant since in principle the BDD mode can better describe a pseudo-bulge component (compact and disk-like), if this was often required.}
 \label{fig:B_T_BDvBDD}
\end{figure}

\begin{figure}
 \includegraphics[width=\columnwidth]{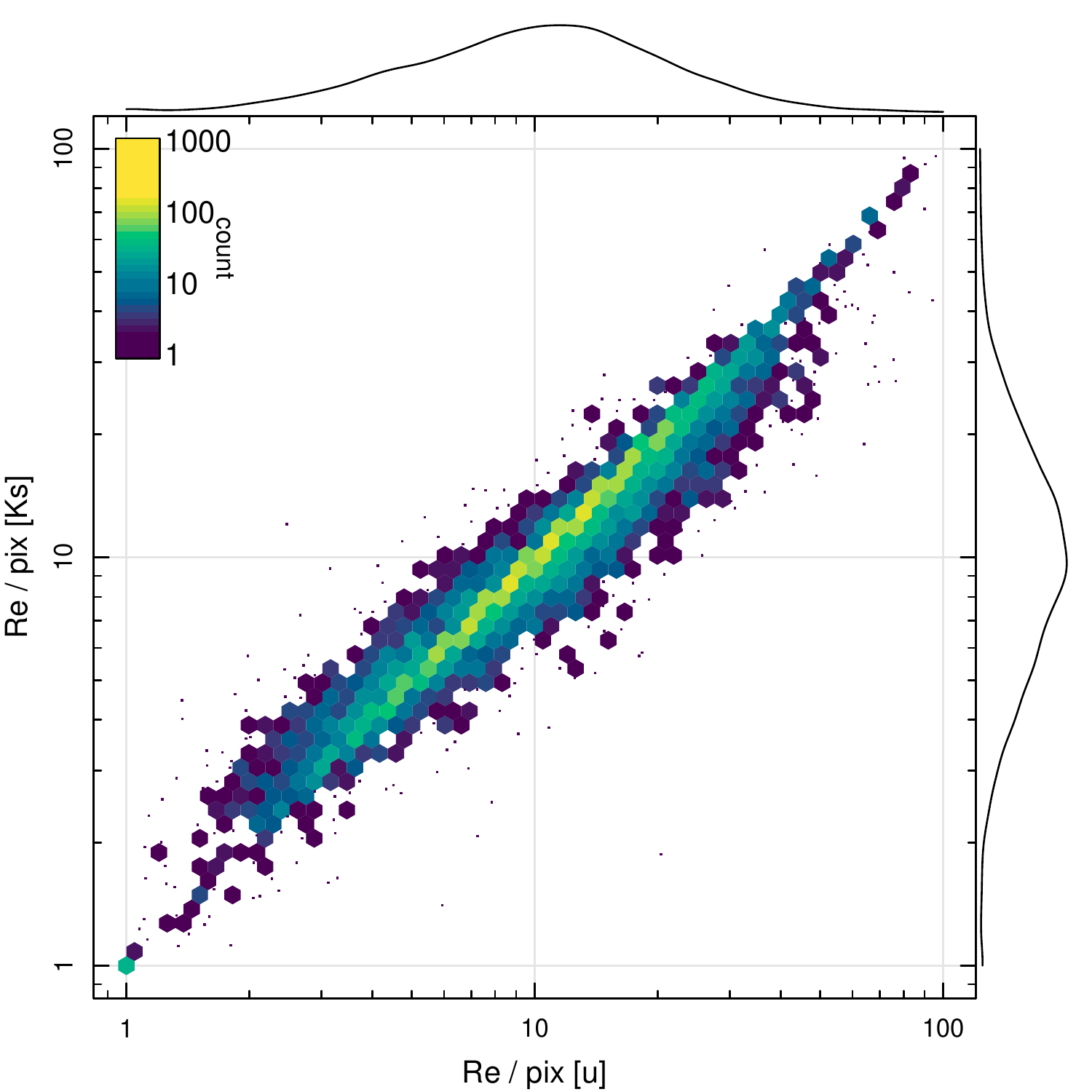}
 \caption{Comparison of inferred disk Re (from the mass-weighted average of both disk components) for the bluest ($u$) and reddest (Ks) bands used for this work. Over the full wavelength range we only see a variation of $\sim$10\% for a small fraction of the sample.}
 \label{fig:Re_uvKs_BDD}
\end{figure}

All 6,664 GAMA galaxies used for the low redshift analysis in this paper were also run with a bulge plus double disk BDD mode. In practice the amount of stellar mass attributed to the disk and the bulge to total ratio of the model remains broadly similar (see Figures \ref{fig:SM_BDvBDD} and \ref{fig:B_T_BDvBDD} respectively). In much better resolved galaxies such a complexity of operation might be warranted, since disk colour gradients in disks have been recovered in many galaxies \citep{MacArthur:2004ta, Bakos:2008ts}. However, at the quality of the GAMA data at a typical redshift of $\sim0.04$ strong disk colour gradients are rare. This is further demonstrated by Figure \ref{fig:Re_uvKs_BDD}, where we largely infer very similar sizes for the $u$ band and Ks band disks. The model is sufficiently flexible that strong colour gradients can be generated if necessary, so we conclude that the added complexity of the BDD mode over the BD (with the subsequent increase in computational cost) is not warranted for the GAMA sample analysed here.

\section{\profuse{} Fits of GAMA Galaxies at various Resolutions}
\label{app:resolve}

\begin{figure*}
 \includegraphics[width=18cm]{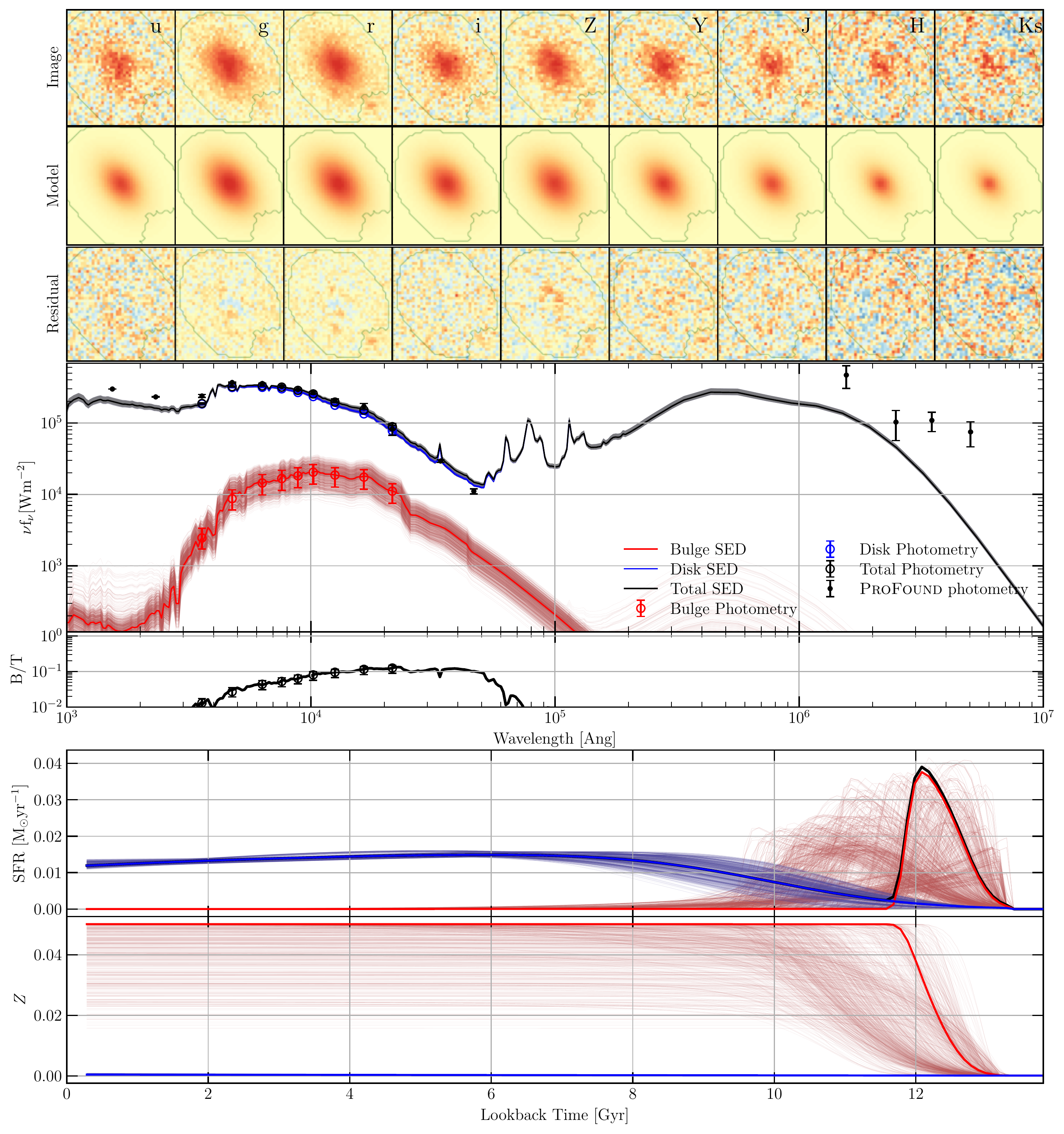}
 \caption{Full set of visual diagnostics for GAMA 7615 (an example of a marginally resolved object, approximately at the $20^{th}$ percentile of our sample). The preferred model BD is shown. Other information as per Figure \ref{fig:ProFuse_7688_Summary_BD}.}
 \label{fig:ProFuse_7615_Summary_BD}
\end{figure*}

\begin{figure*}
 \includegraphics[width=18cm]{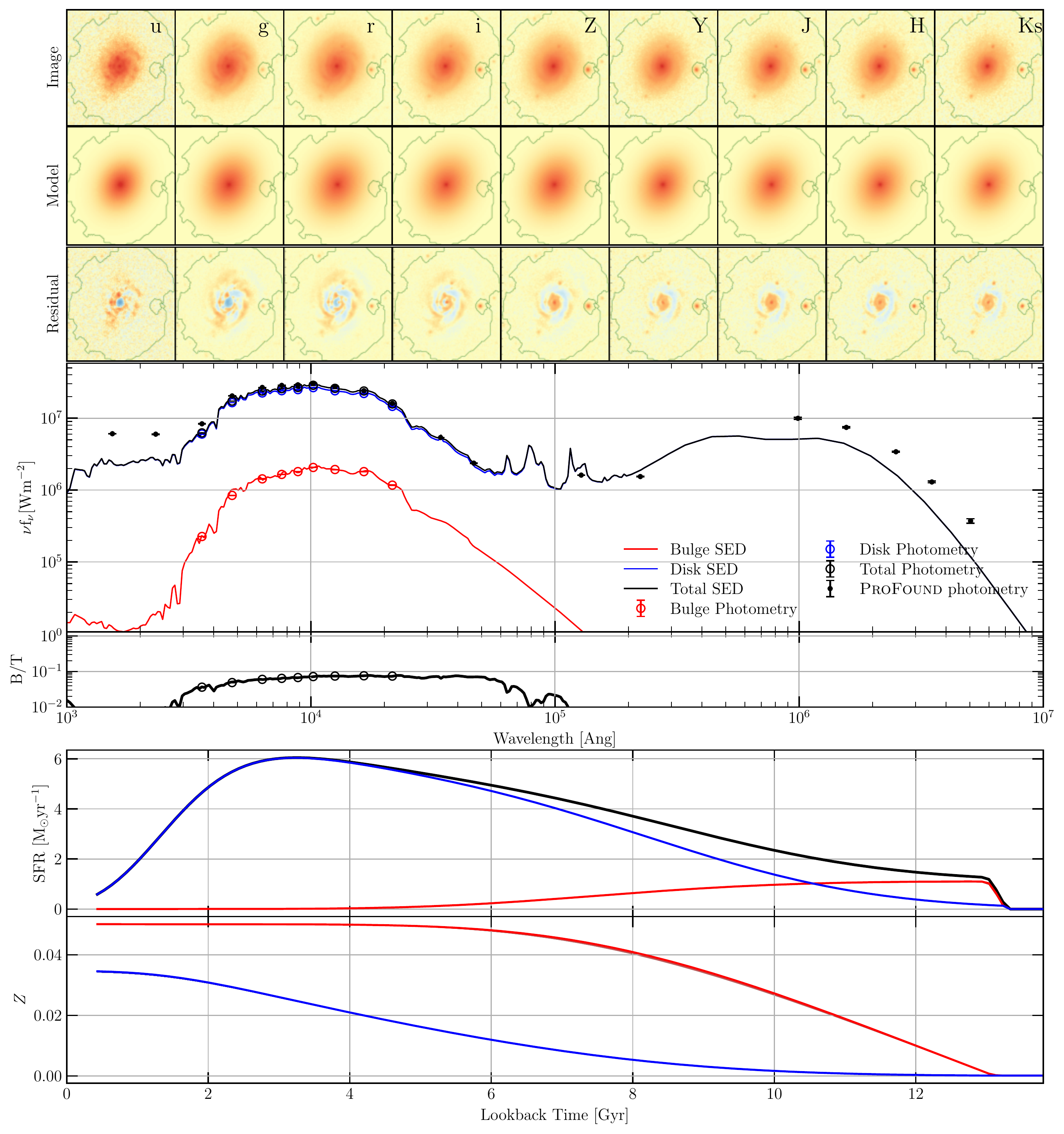}
 \caption{Full set of visual diagnostics for GAMA 3895257 (an example of a well-resolved object, approximately at the $80^{th}$ percentile of our sample). The preferred model BD is shown. Other information as per Figure \ref{fig:ProFuse_7688_Summary_BD}. The posterior sampling lines are present, but the fit is so well converged that they are difficult to discern below the expectation fit lines.}
 \label{fig:ProFuse_3895257_Summary_BD}
\end{figure*}

For reference we present examples of a marginally- and well-resolved GAMA galaxy from our sample of 6,664 used in this work. The marginally resolved source (GAMA 7615, see Figure \ref{fig:ProFuse_7615_Summary_BD}) is at approximately the $20^{th}$ percentile of the sample in terms of the intrinsic $R_e$ of our FS model. The well-resolved source (GAMA 3895257, see Figure \ref{fig:ProFuse_3895257_Summary_BD}) is at approximately the $80^{th}$ percentile of the sample in terms of the intrinsic $R_e$ of our FS model. 

\section{Two Component (BD/PD) and Single Component (FS) Disks}
\label{app:Disk}

In the main body of the paper the BD/PD disks were merged together with FS models, where disk-like models were taken as those with $n<1.5$ \citep[as suggested in][]{Trujillo:2006uj}. Whilst disks in these models broadly overlap, here we briefly explore whether there are discernible differences between our two component (BD/PD) and single component (FS) disks.

\subsection{Star Formation Properties}
\label{app:Disk_SFR}

\begin{figure}
  \includegraphics[width=\columnwidth]{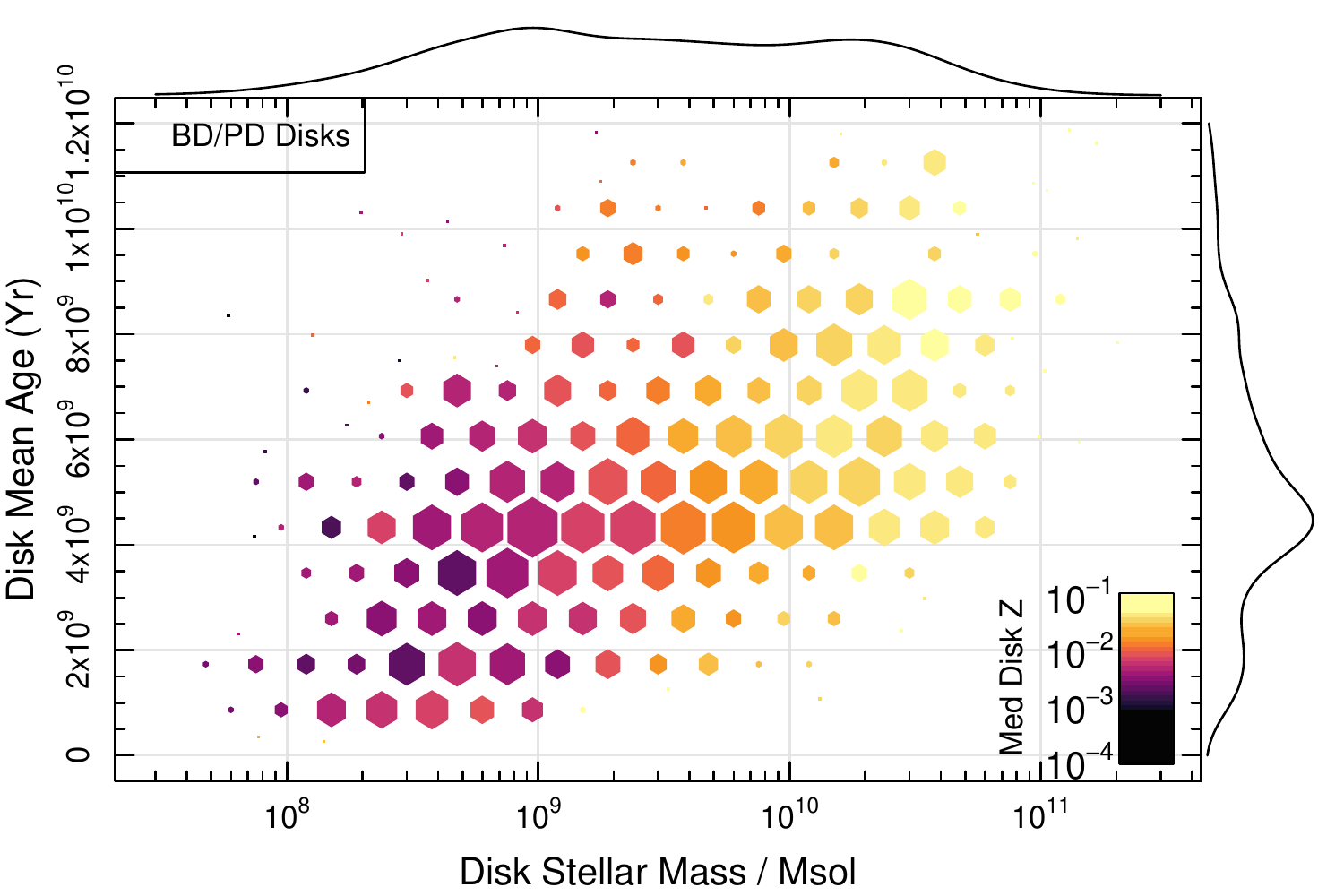} \
 \includegraphics[width=\columnwidth]{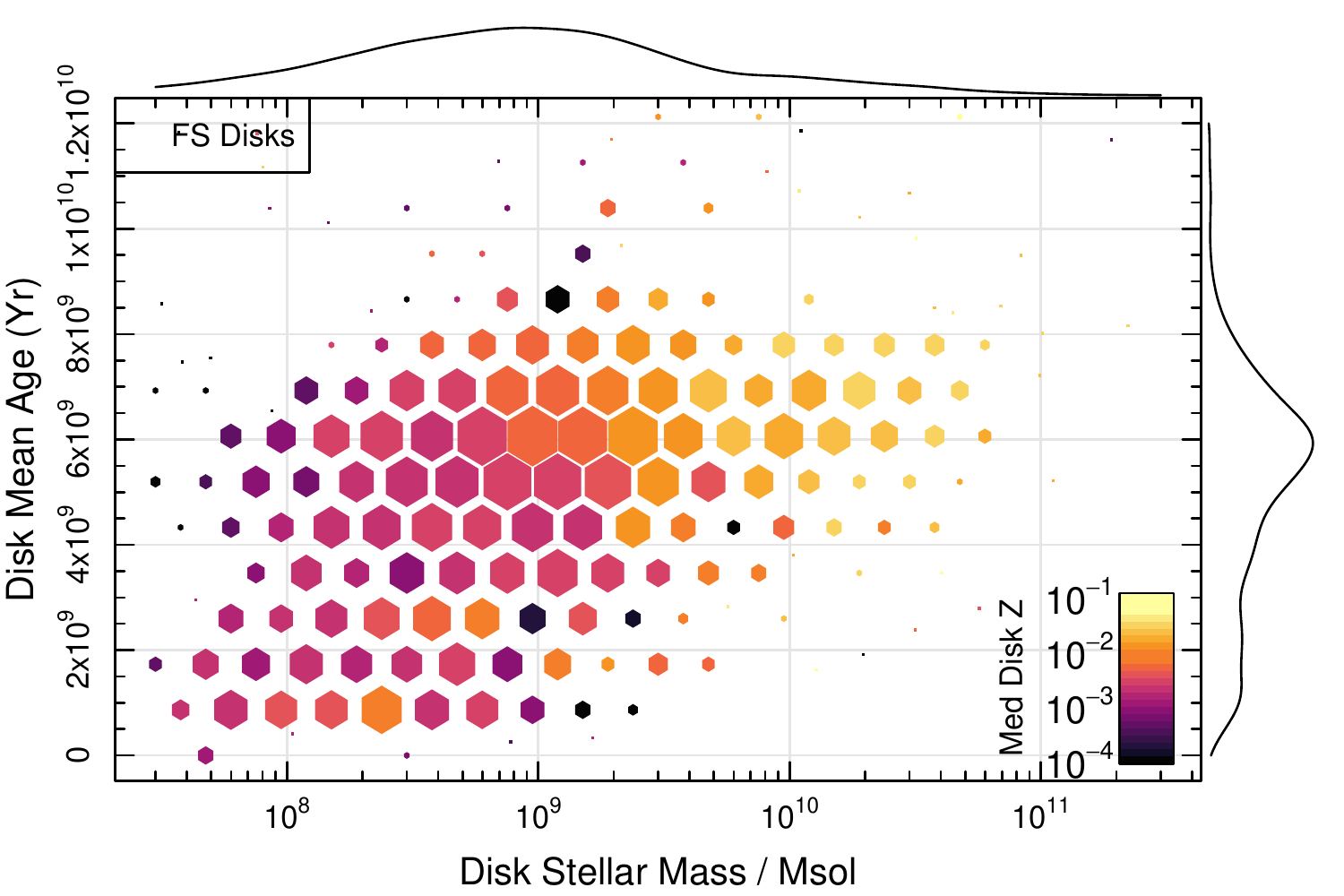}
 \caption{Component stellar mass versus age versus metallicity. Top panel shows BD/PD disks, bottom panel shows FS disks ($n<1.5$). Size scaling reflects the logarithmic counts in cells.}
 \label{fig:SMvAgevZ_disk}
\end{figure}

Figure \ref{fig:SMvAgevZ_disk} presents the age versus metallicity properties for disks, split into the two component BD/PD model disks (top panel) and pure FS model disks (bottom panel). All disks display very similar properties in the stellar mass range of overlap, where (not surprisingly) we find that our pure disk FS model (no bulge) systems are predominantly found at lower stellar mass, whilst the bulge-disk systems span up to the most massive galaxies in GAMA.

\subsection{Structural Properties}
\label{app:Disk_Size}

\begin{figure}
  \includegraphics[width=\columnwidth]{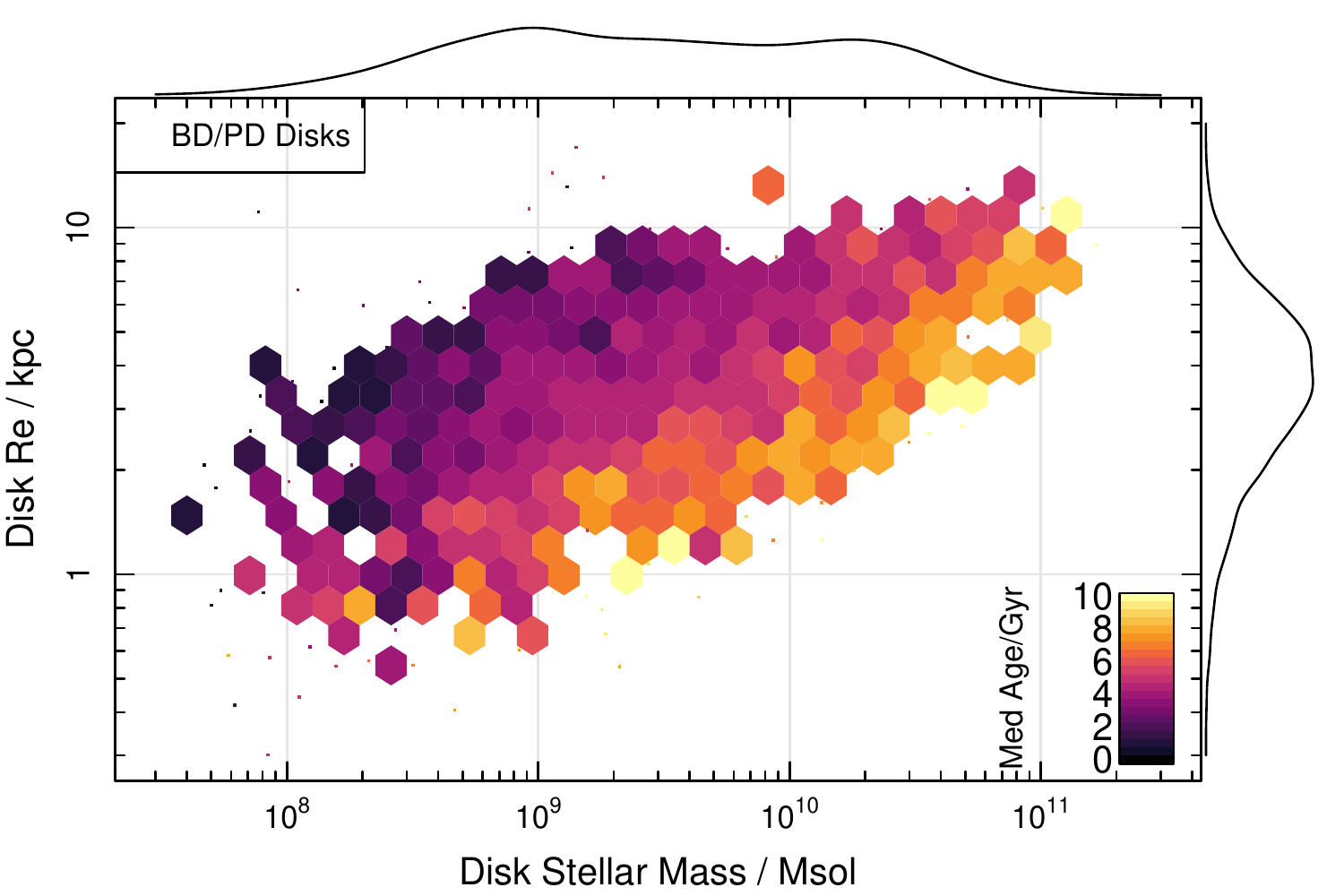}
 \includegraphics[width=\columnwidth]{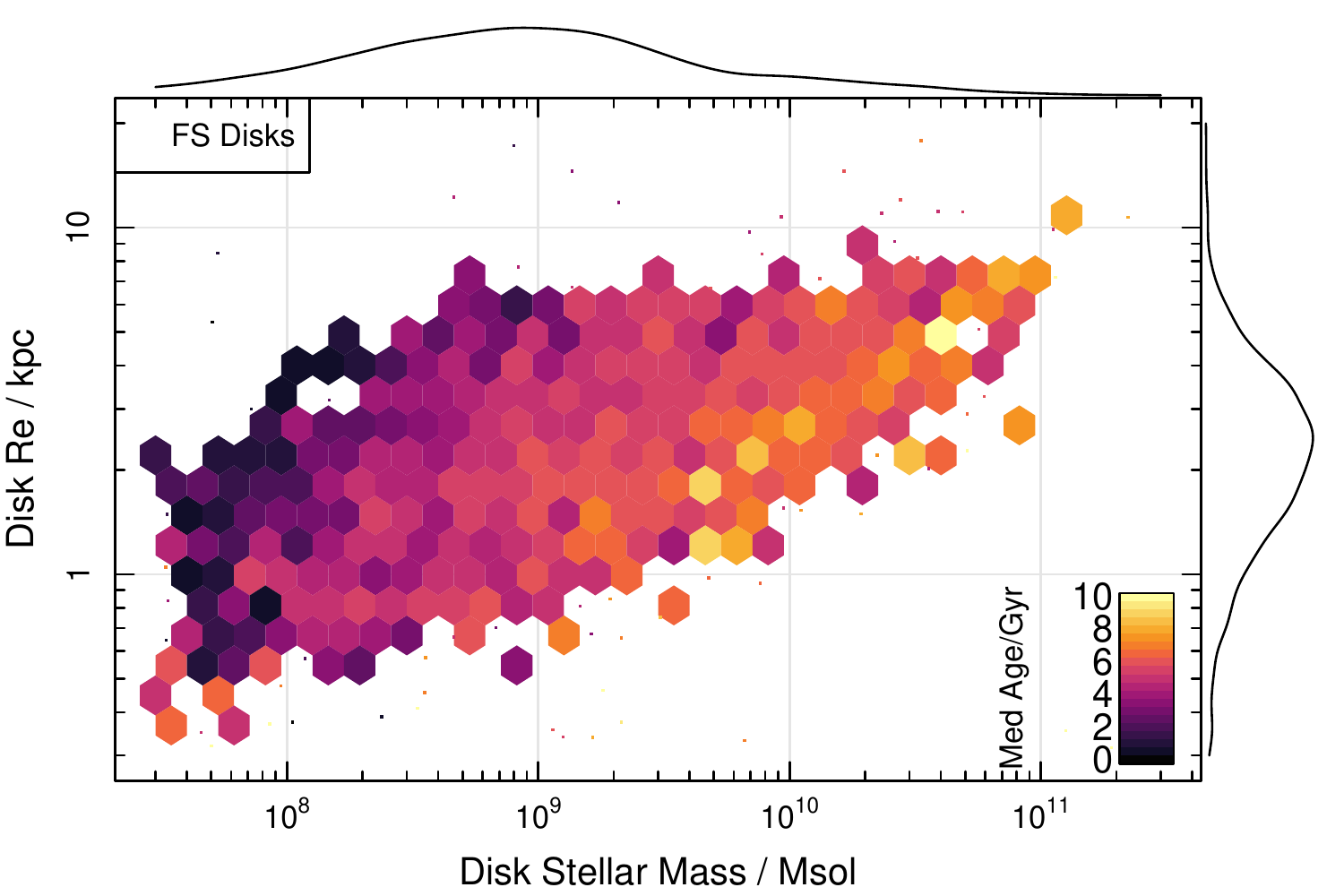}
 \caption{Component stellar mass versus size versus age. Top panel shows BD/PD disks, bottom panel shows FS disks ($n<1.5$). Size scaling reflects the logarithmic counts in cells.}
 \label{fig:SMvRe_disk}
\end{figure}

Figure \ref{fig:SMvRe_disk} presents the component mass--size relationships for disks, split into the two component BD/PD model disks (top panel) and pure FS model disks (bottom panel). The BD/PD two component and FS pure disks create very similar planes with metallicity, but span slightly different mass regimes (as seen in the main body of the paper). The main conclusion to be drawn is combining them together into a single population is generally reasonable when considering the mass--size plane.

\section{Other Potential Third Dimensions to Mass--Size Plane}
\label{app:Diskother}

\begin{figure}
  \includegraphics[width=\columnwidth]{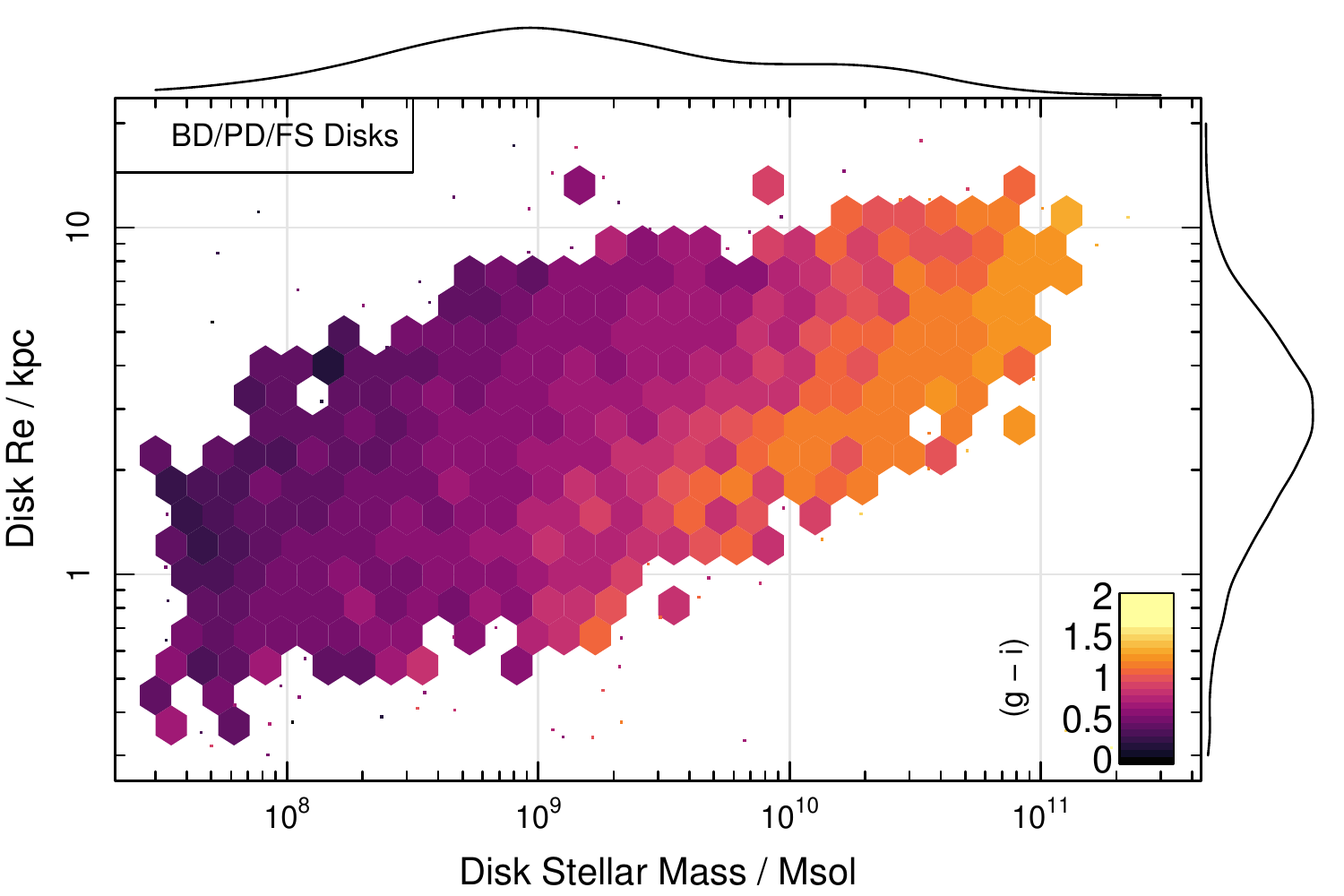} \
 \includegraphics[width=\columnwidth]{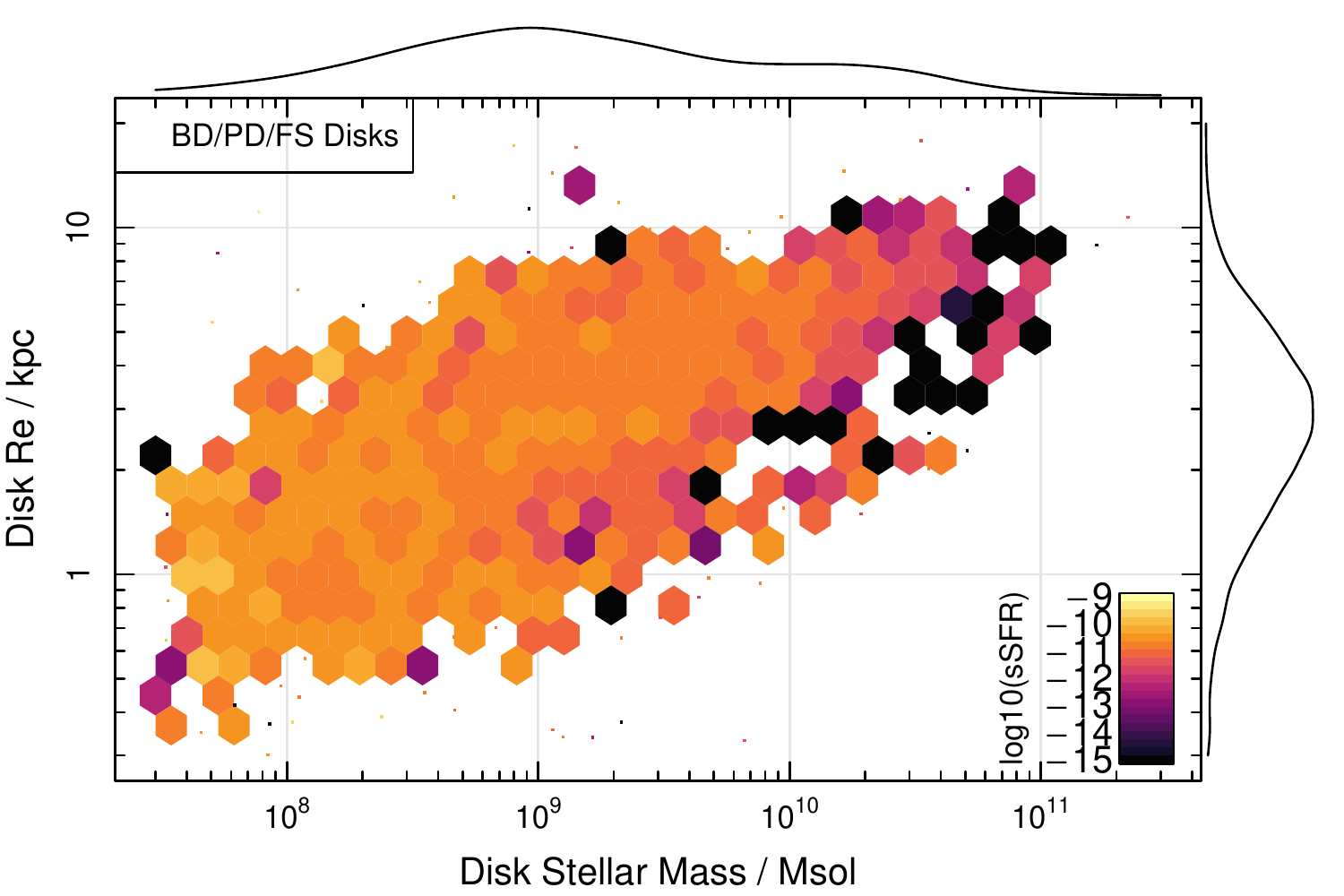}
 \caption{Other mass--size plane third dimensions explored for disks. Top panel shows the observed frame $g-i$ colour of the disk, and the specific SFR (sSFR) of the disk.}
 \label{fig:SMvAgevother_disk}
\end{figure}

Figure \ref{fig:SMvAgevother_disk} presents other third dimensions of interest when exploring extensions to the mass--size plane for disks. The top panel shows the simple $g-i$ colour, with a strong colour mass dependence and some evidence of banding. However, it is much more vertical than we see for our age axis in Figure \ref{fig:SMvRevAge_all_Hyper}, which is much like we found when investigating the dependency on metallicity. The bottom panels shows the sSFR of the disk, and this is even less convincing with only a small residual trend of the most massive disks to have lower sSFR. The current star formation of a disk appears to be a poor predictor for its size given a disk stellar mass.

\bsp	
\label{lastpage}
\end{document}